\documentclass[prd,aps,showpacs,onecolumn,preprintnumbers,nofootinbib,amssymb]{revtex4}
\UseRawInputEncoding
\usepackage{graphicx}
\usepackage{epsf}
\usepackage{amsmath}
\usepackage{epstopdf}
\usepackage{bm}
\usepackage{color}
\usepackage{tabularx}
\usepackage{enumitem}
\usepackage{float}
\usepackage{array,booktabs}
\usepackage{footnote}
\usepackage{threeparttable}
\usepackage{amssymb,epsf}
\usepackage{latexsym}
\usepackage{epsfig}
\usepackage{multirow}
\usepackage{bm}
\usepackage{caption}
\usepackage{xcolor}
\usepackage[colorlinks=true,linkcolor=blue]{hyperref}
\def\e3p{$\eta \rightarrow 3 \pi$}
\def\CL{\,\,\, \longrightarrow^{^{^{\hskip -0.55cm C.L.}}} \,\,\,}

\begin{document}

\title{%
\hfill{\normalsize\vbox{%
 }}\\
{Chiral Nonet Mixing in $\pi \eta$ Scattering}}

\author{Amir H. Fariborz \footnote{Email: fariboa@sunypoly.edu}\\}

\affiliation{Department of Mathematics and Physics, State University of New York, Polytechnic Institute, Utica, NY 13502, USA}

\author{Soodeh Zarepour \footnote{Email: szarepour@phys.usb.ac.ir}\\} 
\affiliation{Department of Physics, University of Sistan and Baluchestan, Zahedan, Iran}

\author{Esmaiel Pourjafarabadi \footnote{Email: epourjafar@shirazu.ac.ir}} 
   
\author{S. Mohammad Zebarjad \footnote{Email: mzebarjad@shirazu.ac.ir}}

\affiliation{Department of Physics, Shiraz University, Shiraz 71454, Iran}

\date{\today}

\begin{abstract}
The generalized linear sigma model for mixing among two- and four-quark components of scalar (and psedudosclar) mesons below and above 1 GeV is applied to the $\pi\eta$ channel in which the isovector scalars $a_0(980)$ and $a_0(1450)$ are probed.    In the leading order,  the model parameters have been previously fixed by various low-energy experimental data,  and then applied to $\pi\pi$ and $\pi K$ channels in which the properties of the light and broad  $\sigma$ and $\kappa$ mesons are extracted in agreement with estimates reported in the literature.   With the same parameters fixed in the leading order, in the present work  the prediction of the model for the  $\pi\eta$ scattering amplitude in the elastic region is  given and unitarized with the K-matrix method. The poles of the unitarized scattering amplitude, which determine the mass and decay width of $a_0(980)$ and $a_0(1450)$ are computed.  It is found that the model predicts an isovector scalar state below 1 GeV,  with  mass 984 $\pm$ 6 MeV and decay width 108 $\pm$ 30 MeV which is a clear signal for the $a_0(980)$.     The $a_0$ pole extracted in this work,    further supports the plausibility of the mixing  patterns for scalar mesons predicted by this model according to which there is a significant underlying mixing among scalars below and above 1 GeV, with those below 1 GeV being  generally of four-quark nature while those above 1 GeV being overall closer to quark-antiquark states.     Predictions for various  scattering lengths as well as for properties of $a_0(1450)$ are also presented.

\end{abstract}


\maketitle
\section{introduction}

Although the perturbative implementation of the fundamental theory of strong interactions (QCD) breaks down at low energies wherein the light hadrons reside, nevertheless,   pioneering works have opened the path of significant progress in uncovering the strong interaction phenomena in this important low-energy QCD region.
Historically, linear sigma model \cite{LsM_1960}, nonlinear realization of spontaneous symmetry breaking \cite{NLR_1969} and  Nambu-Jona-Lasinio approach to dynamical chiral symmetry breaking \cite{NJL_61} have provided powerful platforms for understanding the general characteristics of strong interactions.
Lattice QCD \cite{lattice} has spearheaded  an ambitious path of working directly with the  fundamental QCD degrees of freedom,  while  chiral perturbation theory \cite{ChPT},  
and its extensions such as chiral unitary approach \cite{ChUA1}-\cite{ChUA9} (for a recent review see \cite{ChUA9}) and inverse amplitude method \cite{IAM1}-\cite{IAM4}, have provided  practical   frameworks  for computing physical quantities in terms of a systematic energy expansion.  Most (if not all) models and approaches that are currently used to explore low-energy QCD have been inspired by these pioneering works, and, in one way or another,  solicit the  general guiding principles of low-energy QCD including the chiral symmetry and its breakdown, U(1)$_{\rm A}$  and trace anomalies and various assumptions about the QCD vacuum.

The physics of light pseudoscalar mesons, the Goldstone bosons of strong interaction,  have been fairly well understood and their quark substructure have generally followed the basic quark-antiquark model.    The scalar mesons on the other hand, have continuously challenged the conventional wisdom of the quark model  and have effectively turned into the predicaments  of low-energy QCD.   Their light mass and inverted mass spectrum deviates from what one would expect solely from a quark-antiquark spectroscopy.    Many approaches have been put forward for analyzing the physics of scalar mesons.  In the seminal approach of the MIT bag model \cite{Jaf}  the light scalars are considered to be diquark-antidiquark states which provides an explanation for their unusually  low and inverted mass spectrum.    Many other investigators have tackled scalar mesons from different perspectives  \cite{Jaf}-\cite{07_FJS3}.  For comprehensive reviews see \cite{07_KZ,Pelaez_Review}.

Scalars above 1 GeV are generally expected to be closer to  quark-antiquark states, however,  when their mass spectrum and decay properties are carefully scrutinized \cite{Mec}, it can be seen that their substructure show deviations from  pure quark-antiquark combinations.   This observation then triggers the question of whether the physics of scalars below and above 1 GeV are correlated, and if so,  can a mixing among different  quark-antiquark and four-quark components  (as well as mixing of these components with glue in the case of isosinglet states) can account for some of the unusual properties of these scalar mesons.
This calls for a global treatment of all scalar states below 2 GeV within a single framework.    This global study has been the platform of the framework developed in \cite{global} (and references therein) upon  which the present work is built.

In the global picture  of Ref. \cite{global},  a generalized linear sigma model (GLSM) which is formulated in terms of two scalar nonets and two pseudoscalar nonets (a two- and a four-quark nonet) is developed and the underlying mixings among the scalars and among pseudoscalars is studied.   The framework employs chiral fields which allow a straightforward development of a general chiral invariant Lagrangian as well as a piece that, while preserves chiral symmetry,  breaks U(1)$_{\rm A}$ in a manner that mocks up the axial anomaly of QCD.  In addition, terms that resemble quark mass terms and explicitly break chiral symmetry are added.   Spontaneous  chiral symmetry breaking is then invoked when isosinglet quark-antiquarks and four-quarks develop nonzero vacuum expectation values.    The chiral invariant part of the Lagrangian (as well as the part that breaks the chiral symmetry) can contain a large (or, in principle, an infinite) number of terms.   Therefore, for the model to be practical, an approximation scheme needs to be defined that allows, according to a cogent criterion,  limiting these terms at a given order,  with the hope that the predictions can then be systematically improved at higher orders.   The criterion introduced in \cite{global} is that the terms in the Lagrangian be evaluated according to their total number of quark and antiquark lines and that to consider terms with a large  number of lines to be less important compared to those with
fewer lines.    This approximation scheme,  allows a semi-quantitative organization of the Lagrangian.     At the order where only terms of at most eight quark and antiquark lines are retained,  the Lagrangian parameters were all determined in \cite{global} by various fits to low-energy data.     Consequently, at this order a detailed analysis of two- and four-quark components of both scalars (as well as pseudoscalars) were also accomplished  in the work of \cite{global} and it was observed
that,  while scalar mesons below 2 GeV are distinctively mixed,  light scalars below 1 GeV are mostly of  two quark-two antiquark
 nature while those above 1 GeV are closer to quark-antiquarks (in the same setting,  the light pseudoscalars below 1 GeV emerge dominantly of quark-antiquark structure, as expected from established phenomenology).

 Within the same set of parameters at this order, several  predictions of the model were studied in follow up works, including the prediction of $\pi\pi$ scattering amplitude in \cite{mixing_pipi}.  In that work, K-matrix method was used  for unitarization of the scattering amplitude up to about 1 GeV.  The poles of the unitarized scattering amplitude were determined which in turn give the mass and the decay   width of the isosinglet scalars.  In this approach the first pole has the characteristics of $f_0(500)$ with mass and width:

\begin{eqnarray}
m[f_0(500)]  &=& 477 \pm 8 \, {\rm MeV}, \nonumber \\
\Gamma[f_0(500)] &=& 398 \pm 107 \, {\rm MeV}.
\end{eqnarray}
in agreement with PDG \cite{pdg}:
\begin{eqnarray}
m[f_0(500)] &=& 400-550 \, {\rm MeV}  \hskip 0.5cm ({\rm PDG}) \nonumber \\
\Gamma [f_0(500)] &=& 400-700  \, {\rm MeV} \hskip 0.5cm ({\rm PDG})
\end{eqnarray}
The main advantage of the K-matrix unitarization method is that it does not introduce any additional parameters and therefore provides a simple way of estimating the final-state interactions of the pions in $\pi\pi$ scattering.

In the same order of the model, the  $I=1/2$, $J=0$, $\pi K$ scattering amplitude was studied in \cite{mixing_piK} and a close agreement with experiment was observed up to about 1 GeV.     Since all the parameters of the generalized linear sigma model of Ref. \cite{global} have been previously fixed in this leading order,  the analysis of the $\pi K$ scattering was another prediction of the model and provided further test of the mixing patterns predicted in \cite{global}.     In the work of \cite{mixing_piK},  the predictions of the model for the poles of the K-matrix unitarized scattering amplitude were also determined.   The mass and decay width of the first pole found in \cite{mixing_piK}, correspond to $K_0^*(800)$ (or kappa meson) with
\begin{eqnarray}
 m[K_0^*(800)] &=& 670-770 \, {\rm MeV}, \nonumber \\
\Gamma[K_0^*(800)] &=& 640-750  \, {\rm MeV}.
\end{eqnarray}
is consistent with the averaged values reported by PDG \cite{pdg}:
\begin{eqnarray}
 m[K_0^*(800)] &=& 682 \pm 29 \, {\rm MeV} \hskip 0.5cm ({\rm PDG}) \nonumber \\
 \Gamma[K_0^*(800)] &=& 547 \pm 24  \, {\rm MeV} \hskip 0.5cm ({\rm PDG})
\end{eqnarray}

In addition to the model predictions for the $\pi\pi$ and $\pi K$ scattering amplitudes,  the  decay $\eta'\rightarrow \eta\pi\pi$ has been recently  investigated in \cite{LsM_mmp_eta3p} within the same leading order of the model discussed above.      It is found that the prediction for the partial decay width  that includes the effect of the final-state interaction of pions agrees with the experiment up to about 1\%, and that the model is able to give a reasonable prediction of the energy dependencies of the normalized decay amplitude squared.  This further promotes the prediction of the model for the underlying mixing of two- and four-quark components of the scalar mesons.

In this work,  we apply the generalized linear sigma model of Ref. \cite{global} (in the leading order with the same set of parameters used in the $\pi\pi$ and  $\pi K$ scatterings as well as $\eta'\rightarrow \eta\pi\pi$ decay, discussed above)  to study $\pi\eta$ scattering in which the $a_0(980)$ is probed.    This will complete the probe of the scalar meson nonet below 1 GeV within this   mixing model.
Unlike $\pi\pi$ and $\pi K$ scatterings,  there is still a lack of  experimental data on $\pi\eta$ scattering. Nevertheless, this process has been studied by many investigators using different approaches and from different perspectives, such as computation of $\pi\eta$ scattering amplitude  at the next-to-leading order of chiral perturbation theory \cite{ChPT_bernard,ChPT_Novotny}; nonlinear chiral Lagrangian study of $\pi\eta$ scattering and its pertinence to the light scalar meson nonet \cite{pieta};   effects of vacuum fluctuations of quark condensates probed in $\pi\eta$ scattering \cite{ChPT_kolesar};   the $\pi^0 \eta$ rescattering effects  in
$\gamma\gamma\rightarrow \pi^0 \eta$ data of Belle Collaboration and probing the tetraquark nature of $a_0(980)$ in this analysis \cite{belledata}; form factor computation of isotriplet scalar currents from S-wave $\pi\eta$ scattering phase shift \cite{Albaladejo_ph}; probe of isotriplet scalars in coupled channel $\pi\eta$, $K {\bar K}$, $\pi\eta'$ analysis in lattice QCD \cite{pieta_lattice}; chiral study of the $a_0(980)$ resonance and $\pi\eta$ scattering phase shifts in unitarized chiral perturbation theory \cite{guo}; and the S-wave $\pi\eta$ scattering and the properties of $a_0$ resonances from photon-photon scattering \cite{Lu}.

Sec. II provides the basic set up and notation followed by  the prediction of the single nonet SU(3) linear sigma
model (SNLSM) for the $\pi \eta$ scattering amplitude in Sec. III.  The generalized linear sigma model is reviewed in Sec. IV and its predictions for the $\pi \eta$ scattering amplitude are given in Sec. V together with a comparison  with the single nonet results as well as the results obtained by  other investigators.    Sec. VI gives a summary of the results and the conclusions as well as directions for future studies.


\section{Basic set up and notation}

\begin{figure}[!htb]
\centering
\includegraphics[scale=0.6]{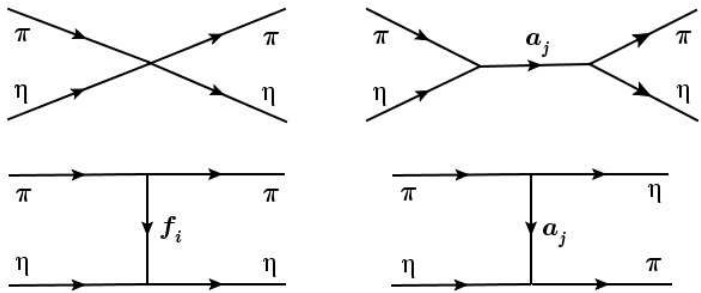}
\caption{Tree-level Feynman diagrams representing the $\pi \eta$ scattering.    The parameters $i = 1 \cdots n_f$ and $j = 1 \cdots n_a$,  where $n_f$ and $n_a$ represent the number of isosinglet and isotriplet scalars respectively, which in the single (double) nonet model are equal to two (four) and one (two). }
\label{F_FD}
\end{figure}

Motivated by large $N_c$ approximation to QCD,  we only consider the contribution of tree diagrams to the scattering amplitude. 
The generic tree-level Feynman diagrams for this scattering are displayed in Fig. \ref{F_FD}.  These include a four-point interaction diagram (contact diagram) together with diagrams representing the contributions of the isovector and isosinglet  scalar mesons. In the single (double) nonet model there are two (four) isosinglet scalars and one (two) isotriplet scalars contributing to these diagrams.

There are no vector meson exchanges in $\pi\eta$ scattering at tree level. The tree level invariant amplitude ($I=1$ projection) is
\begin{equation}
A(s,t,u)=-\gamma_{\pi \eta}^{(4)} +
\sum _{i=1} ^{n_f}\frac{{{2 \sqrt{2}}}\,\gamma_{f_i\pi\pi} \gamma_{f_i\eta \eta}}{m_{f_i}^2-t}+\sum _{j=1}^{n_a} \gamma_{a_j \pi \eta}^2\left[\frac{1}{m_{a_{j}}^2-s}+\frac{1}{m_{a_{j}}^2-u}\right],
\label{Astu_template}
\end{equation}
where in the single (double) nonet model $n_f$ is two (four) and $n_a$ is one (two), and the coupling constants are defined as
\begin{eqnarray}
-{\cal L} &=&
\frac{1}{4}\gamma_{\pi \eta}^{(4)}\eta \eta\, \mbox{\boldmath ${\pi}$} \cdot
{\mbox{\boldmath ${\pi}$}} +
 \frac{\gamma_{f_i \pi \pi}}{\sqrt 2}
f_i \mbox{\boldmath ${\pi}$} \cdot
{\mbox{\boldmath ${\pi}$}}
+ \gamma_{f_i \eta \eta} f_i  \eta  \eta
+ \gamma_{a_j \pi\eta} {\bf a_j} \cdot  \mbox{\boldmath
${\pi}$}  \eta
+ \cdots.
\label{gamma_temp_2}
\end{eqnarray}
The ``bare'' $J=0$ partial wave amplitude (s-wave) is obtained from
\begin{equation}
T_{0}^{I\,B} = \frac{\rho(s)}{2}\int_{-1}^{1} d\cos\theta P_0 (\cos\theta)  A^I(s,t,u),
\end{equation}
with $\rho(s)=q/(8\pi \sqrt(s))$, where q is the center of mass momentum
\begin{equation}
q= \frac{1}{2\sqrt{s}}\sqrt{(s-(m_{\pi}+m_{\eta})^2)(s-(m_{\pi}-m_{\eta})^2)}.
\end{equation}
Performing the partial wave projection we find the "bare" $I=1$, $J = 0$ amplitude
\begin{eqnarray}\label{pietatem}
T_{0}^{1\,B}= \frac{q(s)}{16 \pi \sqrt{s}}\Bigg[&&- 2 \gamma_{\pi \eta}^{(4)} +\sum_{j=1}^{n_a} \gamma_{a_j \pi \eta}^2\left( \frac{1}{2q^2}\ln \left(\frac{(B_{\eta})_j+1}{(B_{\eta})_j-1}\right)+\frac{2}{m_{a_j}^2-s}\right)\nonumber\\
&&+\sum_{i=1}^{n_f} \frac{{{\sqrt{2}}}}{q^2} \gamma_{f_i \eta \eta} \gamma_{f_i \pi \pi}\ln \left(1+\frac{4 q^2}{m_{f_i}^2}\right)\Bigg],
\end{eqnarray}
where $(B_{\eta})_j$ is defined as
\begin{equation}
(B_{\eta})_j = \frac{1}{2q^2}\left[m_{a_j}^2-m_{\pi}^2-m_{\eta}^2+2 \sqrt{(m_{\pi}^2+q^2)(m_{\eta}^2+q^2)}\right].
\end{equation}
Here $s$, $t$ and $u$ are the usual Mandelstam variables
\begin{eqnarray}
t &=& -2q^2(1-\cos \theta), \nonumber \\
u &=& m_{\eta}^2 + m_{\pi}^2 - 2 \sqrt{(m_{\pi}^2 + q^2)(m_{\eta}^2 + q^2)} -2 q^2 \cos \theta,
\end{eqnarray}
where $\theta$ is the scattering angle. In this work,  Eq. (\ref{pietatem}) is our reference  equation for the scattering amplitude.

As in the case of $\pi\pi$ and $\pi K$ scatterings, we use K-matrix unitarization method to unitarize the $\pi \eta$ scattering amplitude:
\begin{equation}\label{T01_unitary}
T_0^1=\frac{T_{0}^{1\,B}}{1-iT_{0}^{1\,B}}.
\end{equation}
This is what we take as our physical amplitude and will compare with the experimental data. The physical masses (${\widetilde m}_j$) and physical decay widths (${\widetilde \Gamma}_j$) are determined from the poles in the unitarized amplitude (as before, $j=1 \cdots n_a$).   Solving for the roots ($z_j$) of the denominator of (\ref{T01_unitary})
\begin{equation}
1-i T_0^{1 B} = 0 \Rightarrow z_j = {\widetilde m}_j^2 - i {\widetilde m}_j {\widetilde \Gamma}_j.
\label{poles_eq}
\end{equation}
In general,  some of the poles may not be physical (for example, being below the threshold).


\section{$\pi \eta$ scattering in single nonet linear sigma model}

For the purpose of comparison with the generalized linear sigma model predictions for $\pi \eta$ scattering, in this section we give the prediction of the single nonet three flavor linear sigma model for this scattering  \cite{LsM}.        Using the  $ 3\times3$ chiral field

\begin{equation}
M = S +i\phi,
\label{sandphi}
\end{equation}
where $S=S^{\dagger}$ and $ \phi=\phi^{\dagger}$ representing  scalar and  pseudoscalar chiral nonets,  respectively. Under chiral transformation of the left-handed and right-handed quark fields, $q_L\rightarrow U_L q_L$, $q_R\rightarrow U_R q_R$, and consequently:
\begin{equation}
M \rightarrow U_L M U_R^{\dagger}.
\label{sm}
\end{equation}
The  Lagrangian density takes the general form
\begin{equation}
{\cal L} = - \frac{1}{2} {\rm Tr}
\left( \partial_\mu M \partial_\mu M^{\dagger}
\right) - V_0 \left( M \right) - V_{SB},
\label{LsMLag}
\end{equation}
where the potential $V_0$ is in general a function of 
$\rm{SU(3)_L} \times {\rm SU(3)_R} \times {\rm U(1)_V}$ invariants
\begin{eqnarray}
I_1&=&{\rm Tr} (M M^{\dagger}), \hspace{1cm} I_2={\rm Tr}(MM^{\dagger}MM^{\dagger}),\nonumber \\
I_3&=&{\rm Tr}\big((M M^{\dagger})^3\big)…,\hspace{.6 cm} I_4=6(\det M + \det M^{\dagger} ).
\label{SNLSM_Is}
\end{eqnarray}
Among these invariants, $I_4$ is the only one that is not invariant under ${\rm U(1)_A}$.     The minimum  symmetry
breaker $V_{SB}$ is:
\begin{equation}
V_{SB}=-2(A_1 S_1^1 + A_2 S_2^2 + A_3 S_3^3),
\end{equation}
The vacuum expectation values are:
\begin{equation}
\langle S_a^b\rangle= \alpha_a \delta_a^b.
\label{SNLSM_VEV}
\end{equation}
The decay constants can be derived in terms of these parameters
\begin{equation}
F_{\pi}=\alpha_1+\alpha_2 ,  \hspace{1cm}  F_K=\alpha_1+\alpha_3,
\label{SNLSM_SB}
\end{equation}
where in the isospin invariant limit
\begin{equation}
A_1=A_2 ,  \hspace{1cm} \alpha_1=\alpha_2.
\end{equation}
The stable point of the potential is found from 
\begin{equation}
\left< \frac{\partial V}{\partial S_a^a}\right>=0.
\end{equation}
The quantity
\begin{equation}
V_4\equiv\left< \frac{\partial V_0}{\partial I_4}\right>.
\end{equation}
contributes to the $\eta'$ mass. 
We can determine the free  parameters $A_1$ , $A_3$ , $\alpha_1$ , $\alpha_3$ and $V_4$ using the following experimental
inputs:
\begin{eqnarray}
m_{\pi} &=& 137\hspace{.1cm} {\rm MeV},\hspace{.2cm} m_K = 495 \hspace{.1cm}{\rm MeV},\nonumber \\
m_{\eta} &=& 547 \hspace{.1cm} {\rm MeV},\hspace{.2cm} m_{\eta^{\prime}} = 958 \hspace{.1cm} {\rm MeV},\nonumber \\
F_\pi &=& 131 \hspace{.1cm} {\rm MeV}.
\label{inputs}
\end{eqnarray}
In this framework, chiral symmetry, the choice of symmetry breakers as welll as the U(1)$_{\rm A}$ anomaly  determine the pseudoscalar masses, whereas not all scalar masses are predicted \cite{LsM}.   With the  parameters determined in \cite{LsM}, the $I$=1, $J$=0, $\pi \eta$ scattering amplitude can be calculated using (\ref{pietatem}).   The coupling constants  are computed from the ``generating equations'' that express the symmetry of the Lagrangian  (\ref{LsMLag}) \footnote{A computational algorithm for this calculation is given in \cite{LsM_Maple}}:
\begin{eqnarray}
\gamma^{(4)}_{\pi\eta}&=&\sum_{a,b} \left\langle \frac{\partial^4 V}{\partial \phi_1^2 \partial\phi_2^1\partial\phi_a^a\partial\phi_b^b}\right\rangle_0 (R_{\phi})_{2}^a(R_{\phi})_{2}^b,\nonumber\\
\gamma_{f_i\eta\eta}&=&\frac{1}{2}\sum_a\left\langle \frac{\partial^3 V}{\partial S_a^a \partial \phi_b^b \partial \phi_c^c}\right\rangle _0 (R_s)_{i+1}^a (R_\phi)_2^b (R_\phi)_2^c,\nonumber\\
\gamma_{f_i \pi\pi}&=&{1\over \sqrt{2}}\,\sum_a \left\langle \frac{\partial^3 V}{\partial S_a^a \partial\phi_1^2 \partial\phi_2^1}\right\rangle_0 (R_s)_{i+1}^a, \nonumber\\
\gamma_{a_0 \pi\eta}&=&\sum_a \left\langle \frac{\partial^3 V}{\partial S_1^2 \partial\phi_a^a\partial\phi_2^1}\right\rangle_0 (R_{\phi})_{2}^a,\nonumber\\
\end{eqnarray}
where the ``bare'' couplings and the rotation matrices ($R_s$ and $R_\phi$) are given in Appendix A.    Here $f_1=\sigma$ and $f_2=f_0(980)$.

\begin{figure}[!htbp]
	\begin{center}
		\epsfxsize = 1.5 cm
		\includegraphics[keepaspectratio=true,scale=0.75]{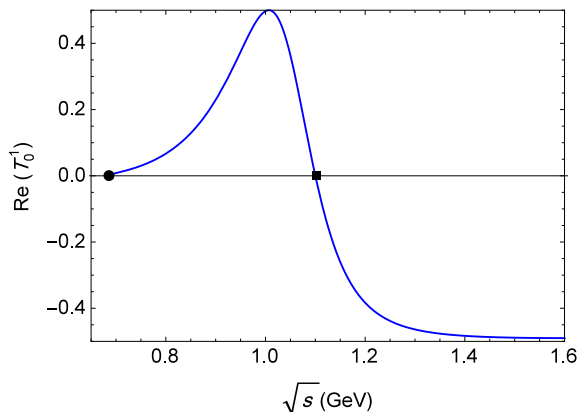}
		\caption{Prediction of the single nonet linear sigma model (SNLSM) for the real part of $T_0^1$.   The zero of this function (square) is at the location of ``bare'' isovector scalar mass. }
		\label{F_ReT01_sn}
	\end{center}
\end{figure}

\begin{figure}[!htbp]
	\begin{center}
		\epsfxsize = 7 cm
		\includegraphics[keepaspectratio=true,scale=0.75]{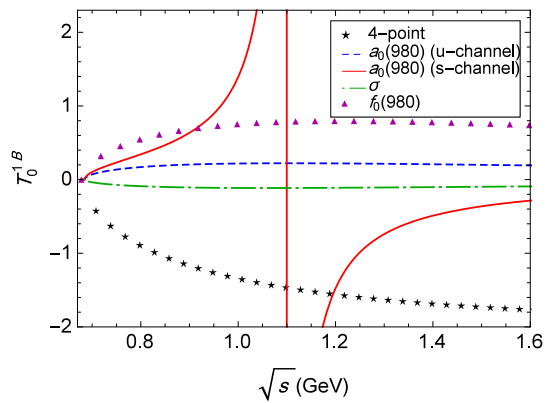}
		\caption{Individual contributions to the $\pi\eta$ scattering in the single nonet model. Up to about 1 GeV,  the  large contribution of the 4-point term and  the contributions of scalar mesons are partially balanced.}
\label{F_T01_indi}
			\end{center}
\end{figure}

Using the inputs (\ref{inputs}) together with the results of the best fit to $\pi\pi$ scattering amplitude of Ref. \cite{LsM},
the bare $I$=1, $J=0$, $\pi \eta$ scattering amplitude is computed from Eq. (\ref{pietatem}) and K-matrix unitarized according to (\ref{T01_unitary}).  The real part of the  $T_0^1$   is plotted in Fig. \ref{F_ReT01_sn}.   The amplitude vanishes around 1.1 GeV which, in this model,   is the location of the bare mass of the isovector scalar meson and is much larger than the mass of $a_0(980)$.   Also,  the decay width of this isotriplet state to $\pi\eta$ comes out around 0.381 GeV \cite{LsM} which is much larger than the decay width of $a_0(980)$.   Using (\ref{poles_eq}),  the pole of the K-matrix unitarized scattering amplitude gives the physical mass and decay width of this state \cite{LsM}:
\begin{eqnarray}
{\widetilde m}_a &=& 1.013 \hskip 0.1cm {\rm GeV} \nonumber \\
{\widetilde \Gamma}_a  &=& 0.241 \hskip 0.1cm {\rm GeV}
\end{eqnarray}
which are closer to the properties of $a_0(980)$, but still are not within the experimental ranges \cite{pdg}.   This clearly shows the shortcoming of the single nonet model, which, as we will see,  can be remedied by allowing the lowest and  the next-to-lowest scalar meson nonets to mix.    The individual contributions to the real part of the amplitude are plotted in Fig. \ref{F_T01_indi}.   The figure shows that, below 1 GeV,  the individual contributions balance the large four-point contribution, but above 1 GeV [that here lacks the contribution of $a_0(1450)$] this is not the case which further highlights the importance of
$a_0(1450)$ and chiral mixing.

\begin{figure}[!htbp]
\begin{center}
\epsfxsize = 4 cm
 \includegraphics[keepaspectratio=true,scale=0.75]{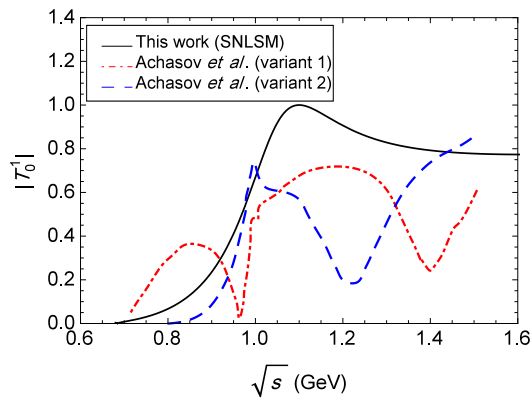}
\caption{Comparing the prediction of the single nonet linear sigma model (SNLSM) for the modulus of $T_0^1$ with the prediction of  Ref. \cite{belledata}, where up to about 1 GeV,  a qualitative agreement  with ``variant 2'' is seen.}
\label{F_modT01_sn}
\end{center}
\end{figure}

The modulus of $T_0^1$  is plotted in Fig. \ref{F_modT01_sn} and compared with the work of Ashasov and Shestakov for this quantity obtained from Belle data for $\gamma\gamma\rightarrow \pi^0\eta$ \cite{belledata} process. Up to about 1 GeV,  our result  agrees better with ``variant 2'' of Ref. \cite{belledata} (at least in mathematical form) in comparison with the large disagreements of  the two variants of that reference.   Above 1 GeV, the effect of $a_0(1450)$ kicks in but  this state is absent here in the single nonet approach, hence as seen in the figure, expectedly,  there is no agreement with either variants.  We will see later that when the chiral nonet mixing within the generalized linear sigma model is considered,   in which the underlying mixing of $a_0(980)$ and $a_0(1450)$ is naturally built in,  the (qualitative) agreement with the ``variant 2'' of Ref. \cite{belledata} extends to about 1.5 GeV, accentuating the importance of chiral nonet mixing model as the centerpiece of the present work.

\begin{figure}[!htbp]
	\begin{center}
		\epsfxsize = 7 cm
		\includegraphics[keepaspectratio=true,scale=0.75]{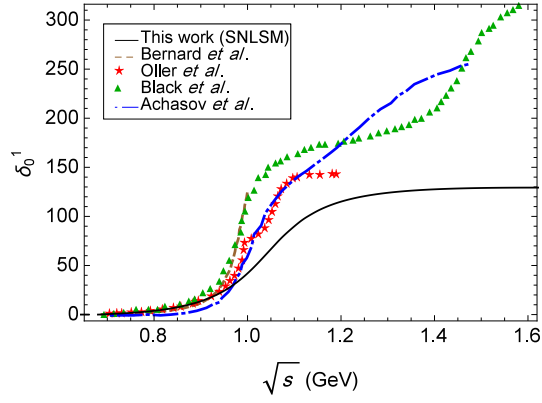}
		\caption{Phase shift computed from the K-matrix unitarized s-wave amplitude of $\pi\eta \rightarrow \pi \eta$ scattering in SNLSM is compared with the predictions by Bernard et al \cite{ChPT_bernard}, Oller et al \cite{oller2},  Black et al \cite{pieta} and Achasov et al  \cite{belledata}.  A qualitative  agreement is seen up to about  $1$ GeV. }
		\label{pieta_sn_phaseshift}
	\end{center}
\end{figure}

The   absence of chiral mixing is also manifested in the phase shift
\begin{equation}\label{phshi}
\sin(2\delta_l^I)=\frac{2 {\rm Re} T_I^l}{\mid 1+ 2 i T_l^I\mid},
\end{equation}
which can be compared with other model predictions.  The prediction for the $l=0$ phase shift in the single nonet linear sigma model is plotted in Fig. \ref{pieta_sn_phaseshift} which shows a qualitative agreement with other approaches such as the chiral unitary approach  \cite{ChPT_bernard,oller2},  the ``variant 2''  of  Ref. \cite{belledata} and the nonlinear chiral Lagrangian of Ref. \cite{pieta} up to about 1 GeV.  Here we have ignored the inelastic effects  which seems reasonable up to roughly about $1$ GeV, but the effects of inelastic channels, namely $\pi\eta \rightarrow K \bar{K}$ and $\pi\eta \rightarrow \pi \eta'$ are expected to become important above 1 GeV.

In the linear sigma model the scalar and  pseudoscalars fileds (as well as, when relevant,  other spin multiplets) are explicitly kept in the Lagrangian as opposed to be integrated out.  In studies of Goldstone boson scatterings the amplitudes are studied in the resonance region (away from threshold) with direct inclusion of resonances over a broad range of energy roughly up to 1 or 1.5  GeV.    In order to be able to compare with the leading order of  ChPT, one has to zoom in near the threshold by taking the scalar masses to infinity (see Appendix C). In order to roughly compare with higher orders of ChPT,  one can examine the 1/$m_i^2$ corrections ($m_i$  being the scalar masses).    An example of this type of comparison (for the case of pi pi scattering)  is given in \cite{LsM_scatt_length}.
Although  the linear sigma model is not expected to be very accurate near the threshold region and is designed to cover a larger energy range,   nevertheless it is still useful to check its accuracy (or its lack thereof) near the threshold.
For this purpose,  we compute the scattering lengths which probe the low-energy dynamics,  and compare them with the results of other models.  It is common to define the scattering lengths by an expansion near the threshold of the form
\begin{equation}\label{scatlengthexpan}
t_0^{I\, B}=\frac{\sqrt{s}}{2}  (a_0^I+\frac{q^2}{m_\pi^2}b_0^I + \frac{q^4}{m_\pi^4}c_0^I+\cdots), \quad {\rm for} \: q\rightarrow 0,\, s\rightarrow(m_\pi+m_\eta)^2,
\end{equation}
where the lower case bare amplitude is
\begin{equation}
t_0^{I\, B}=\frac{\sqrt{s}}{2q}T_0^{I\, B}=\frac{1}{32\pi}\int_{-1}^{1} d\cos\theta P_0(\cos\theta)A^I(s,t,u).
\end{equation}
Therefore
\begin{equation}
T_0^{I\, B}= q (a_0^I+\frac{q^2}{m_\pi^2}b_0^I + \frac{q^4}{m_\pi^4}c_0^I+\cdots), \quad {\rm for} \: q\rightarrow 0,\, s\rightarrow(m_\pi+m_\eta)^2.
\label{T0B_expand}
\end{equation}
In the units of $({\rm pion\, scattering\, wavelength)}^{2l+1}$, the S-wave scattering lengths are
\begin{eqnarray}
a_0^I &=& m_\pi \lim_{q\rightarrow 0} \frac{2}{\sqrt{s}}\, t_0^{I\, B}= m_\pi \lim_{q\rightarrow 0} \frac{1}{q}\, T_0^{I\, B},\nonumber\\
b_0^I &=&\frac{ m_\pi^3}{2!} \lim_{q\rightarrow 0} \frac{2}{\sqrt{s}}\, \frac{\partial^2 t_0^{I\, B}}{\partial q^2}=\frac{ m_\pi^3}{2!} \lim_{q\rightarrow 0} \frac{1}{q}\, \frac{\partial^2 T_0^{I\, B}}{\partial q^2},\nonumber\\
c_0^I &=& \frac{m_\pi^5}{4!} \lim_{q\rightarrow 0} \frac{2}{\sqrt{s}}\, \frac{\partial^4 t_0^{I\, B}}{\partial q^4}= \frac{m_\pi^5}{4!} \lim_{q\rightarrow 0} \frac{1}{q}\, \frac{\partial^4 T_0^{I\, B}}{\partial q^4}.
\end{eqnarray}
The effect of K-matrix unitarization on the scattering lengths can be obtained by substitution of  (\ref{T0B_expand}) into (\ref{T01_unitary})
\begin{eqnarray}
{\rm Re}(T_0^I)= q (\widetilde{a}_0^I+\frac{q^2}{m_\pi^2}\widetilde{b}_0^I + \frac{q^4}{m_\pi^4}\widetilde{c}_0^I+\cdots)= \frac{q(a_0^I+\frac{b_0^I}{m_\pi^2}q^2+\frac{c_0^I}{m_\pi^4}q^4+\cdots )}{1+q^2(a_0^I+\frac{b_0^I}{m_\pi^2}q^2+\frac{c_0^I}{m_\pi^4}q^4+\cdots )^2}.\nonumber\\
\end{eqnarray}
Therefore
\begin{eqnarray}
\widetilde{a}_0^I&=&a_0^I,\nonumber\\
\widetilde{b}_0^I&=&b_0^I-(a_0^{I})^3 m_\pi^2,\nonumber\\
\widetilde{c}_0^I&=&c_0^I-3\, (a_0^{I})^2 b_0^I\, m_\pi^2.
\end{eqnarray}

\begin{table}[!htbp]
\centering
\caption{The s-wave scattering lengths of $I=1$, $\pi\eta$ scattering in SNLSM (first row), GLSM (second row),
 ${\cal O} p^4$ ChPT \cite{ChPT_bernard} (third row); nonlinear chiral Lagrangian of Ref. \cite{pieta} (fourth row);  non-relativistic effective field theory \cite{kubis}
 (fifth row); estimates obtained at  ${\cal O} p^4$ in \cite{kubis} with low-energy constants taken from \cite{bijnens} and \cite{amoros} respectively (sixth and seventh rows);  estmiates  of Ref. \cite{belledata} using Belle data (eighth and ninth rows); and two-channel unitarity model of \cite{Albaladejo_ph} (last two rows).}
\renewcommand{\tabcolsep}{0.4pc} 
\renewcommand{\arraystretch}{1.5} 
\begin{tabular}{cccc}
\noalign{\hrule height 1pt}
\noalign{\hrule height 1pt}
& $a_0^1$ & $b_0^1$ & $c_0^1$ \\
\noalign{\hrule height 1pt}
This work, SNLSM (Sec. III) & $1.63 \times 10^{-2}$ & $1.59\times 10^{-2}$ & $2.86\times 10^{-3}$ \\
\noalign{\hrule height 0.1pt}
This work, GLSM (Sec. IV) & $(2.6-4.0)\times 10^{-2}$ & $(2.4-4.4)\times 10^{-2}$ & $(1.6-5.2)\times 10^{-3}$  \\
\noalign{\hrule height 0.1pt}
Bernard \emph{et al} \cite{ChPT_bernard}  & $7.3\times 10^{-3}$ & $\rule{0.7cm}{0.15mm}$ & $\rule{0.7cm}{0.15mm}$ \\
\noalign{\hrule height 0.1pt}
Extracted from Black \emph{et al } \cite{pieta}& $3.1 \times10^{-2}$ & $1.8 \times 10^{-2}$ & $0.85 \times 10^{-3}$\\
\noalign{\hrule height 0.1pt}
\multirow{3}{*}{Kubis \emph{et al} \cite{kubis}}& $(-0.2\pm 7.7)\times10^{-3}$ & $\rule{0.7cm}{0.15mm}$ & $\rule{0.7cm}{0.15mm}$ \\
& $(1.57\pm 2.39)\times10^{-2}$ & $(9.9 \pm 22.9) \times 10 ^{-3}$ & $\rule{0.7cm}{0.15mm}$\\
& $(0.98 \pm 1.58)\times 10^{-2}$ & $(0.4 \pm 18.8) \times 10 ^{-3}$ & $\rule{0.7cm}{0.15mm}$ \\
\noalign{\hrule height 0.1pt}
\multirow{2}{*}{Achasov \emph{et al} \cite{belledata}}  & $0.98 \times 10^{-2}$ (variant 1)& $\rule{0.7cm}{0.15mm}$& $\rule{0.7cm}{0.15mm}$\\
& $0.66 \times 10^{-2}$ (variant 2)& $\rule{0.7cm}{0.15mm}$ & $\rule{0.7cm}{0.15mm}$ \\
\noalign{\hrule height 0.1pt}
\multirow{2}{*}{Albaladejo \emph{et al} \cite{Albaladejo_ph}} & $0.67 \times 10^{-2}$ (large $L_4$, $L_6$) & $-15.0 \times 10^{-3}$ (large $L_4$, $L_6$) & $\rule{0.7cm}{0.15mm}$\\
& $1.62 \times 10^{-2}$ (small $L_4$, $L_6$) & $10.6 \times 10^{-3}$ (small $L_4$, $L_6$) & $\rule{0.7cm}{0.15mm}$\\
\noalign{\hrule height 1pt}
\noalign{\hrule height 1pt}
\end{tabular}\\[2pt]
\label{sn_sl}
\end{table}

The predictions of the SNLSM for the s-wave scattering lengths of $I=1$, $\pi\eta$ scattering are given in the first row of
Table \ref{sn_sl} and compared with those of the generalized linear sigma model \cite{global} (second row --  will be discussed in
Sec. IV);  ${\cal O} p^4$ chiral perturbation theory results \cite{ChPT_bernard} (third row); estimates  extracted from the work of
\cite{pieta} within a nonlinear chiral Lagrangian (fourth row);  non-relativistic effective field theory \cite{kubis}
(fifth row); estimates obtained at  ${\cal O} p^4$ in \cite{kubis} with low-energy constants taken from \cite{bijnens} and \cite{amoros} respectively (sixth and seventh rows);  estmiates  of Ref. \cite{belledata} using Belle Collaboration data (eighth and ninth rows); and the  work of \cite{Albaladejo_ph} that models a two-channel unitarity that matches  ${\cal O} p^4$ chiral expansion (last row).     A close agreement of SNLSM predictions with the estimates of \cite{Albaladejo_ph} and \cite{kubis}, and a qualitative agreement with other works is evident.

\section{Brief review of the generalized linear sigma model}


The model is constructed in terms of 3$\times$3 matrix
chiral nonet fields:
\begin{equation}
M = S +i\phi, \hskip 2cm
M^\prime = S^\prime +i\phi^\prime,
\label{sandphi1}
\end{equation}
where $M$ and $M'$ transform in the same way under
chiral SU(3) transformations 
\begin{eqnarray}
M &\rightarrow& U_L\, M \, U_R^\dagger,\nonumber\\
M' &\rightarrow& U_L\, M' \, U_R^\dagger,
\end{eqnarray}
but transform differently under U(1)$_A$
transformation properties
\begin{eqnarray}
M &\rightarrow& e^{2i\nu}\, M,  \nonumber\\
M' &\rightarrow& e^{-4i\nu}\, M'.
\label{U1A}
\end{eqnarray}
$M$ and $M'$ respectively represent the 
quark-antiquark and the  two-quark two-antiquark chiral nonets.
In this framework the type of the four-quark content of  $M'$ is not determined and therefore we consider it to be a linear combination of  diquark-antidiquark and  molecular structure.  The way that the model  distinguishes  two-quark from four-quark is through the U(1)$_A$ transformation (\ref{U1A}).

We can write down the  Lagrangian density 
\begin{equation}
{\cal L} = - \frac{1}{2} {\rm Tr}
\left( \partial_\mu M \partial_\mu M^\dagger
\right) - \frac{1}{2} {\rm Tr}
\left( \partial_\mu M^\prime \partial_\mu M^{\prime \dagger} \right)
- V_0 \left( M, M^\prime \right) - V_{SB},
\label{mixingLsMLag}
\end{equation}
where $V_0(M,M^\prime)$ is constructed out of SU(3)$_{\rm L} \times$ SU(3)$_{\rm R}$  but not necessarily U(1)$_{\rm A}$) invariants.
Clearly, there  are many such terms, even when we consider the renormalizable potential.  However, for practical purposes, we define an approximation strategy  that limits the number of terms at each level of calculation.
In \cite{07_FJS1} an evaluation of Lagrangian was examined   in terms of the number of the quarks and antiquarks in each term.
The  leading order corresponds
to eight or fewer quark and antiquark lines:
\begin{eqnarray}
V_0 =&-&c_2 \, {\rm Tr} (MM^{\dagger}) +
c_4^a \, {\rm Tr} (MM^{\dagger}MM^{\dagger})
\nonumber \\
&+& d_2 \,
{\rm Tr} (M^{\prime}M^{\prime\dagger})
     + e_3^a(\epsilon_{abc}\epsilon^{def}M^a_dM^b_eM'^c_f + {\rm H. c.})
\nonumber \\
     &+&  c_3\left[ \gamma_1 {\rm ln} (\frac{{\rm det} M}{{\rm det}
M^{\dagger}})
+(1-\gamma_1){\rm ln}\frac{{\rm Tr}(MM'^\dagger)}{{\rm
Tr}(M'M^\dagger)}\right]^2.
\label{SpecLag}
\end{eqnarray}
With the  exception of the  last two terms (which generate the axial anomaly)
all other terms are invariant under U(1)$_{\rm A}$.  Terms that violate OZI rule are not included.
The symmetry breaking consistent with the QCD mass term
is:
\begin{equation}
V_{SB} = - 2\, {\rm Tr} (A\, S),
\label{vsb}
\end{equation}
where $A={\rm diag} (A_1,A_2,A_3)$ and the diagonal elements are proportional to the light quark
current masses.
The model allows for both quark-antiquark as well as the four-quark condensates:
$\alpha_a=\langle S_a^a \rangle$ and 
$\beta_a=\langle {S'}_a^a \rangle$, respectively.
In the limit of isospin 
symmetry $A_1 = A_2$ and:
\begin{equation}
\alpha_1 = \alpha_2  \ne \alpha_3, \hskip 2cm
\beta_1 = \beta_2  \ne \beta_3.
\label{ispinvac}
\end{equation}
The ``minimum" conditions are:
\begin{equation}
\left< \frac{\partial V_0}{\partial S}\right> + \left< \frac{\partial
V_{SB}}{\partial
S}\right>=0,
\quad \quad \left< \frac{\partial V_0}{\partial S'}\right>
=0.
\label{mincond}
\end{equation}
The parameter space of the model in this order contains the six coupling constants  in Eq. (\ref{SpecLag}), the two quark mass parameters ($A_1=A_2,A_3$) and the four condensates  ($\alpha_1
=\alpha_2,\alpha_3,\beta_1=\beta_2,\beta_3$).   These twelve parameters reduce to eight when we use the four minimum equations. Then we input the following five  experimental inputs:
\begin{eqnarray}
 m[a_0(980)] &=& 980 \pm 20\, {\rm MeV},
\nonumber
\\ m[a_0(1450)] &=& 1474 \pm 19\, {\rm MeV},
\nonumber \\
 m[\pi(1300)] &=& 1300 \pm 100\, {\rm MeV},
\nonumber \\
 m_\pi &=& 137 \, {\rm MeV},
\nonumber \\
F_\pi &=& 131 \, {\rm MeV}.
\label{inputs1}
\end{eqnarray}
reducing the unknown parameters to three. Note that $m[\pi(1300)]$ has a large uncertainty which in turn will be reflected in our model predictions.  
The sixth input is the light
``quark mass ratio" $A_3/A_1$ which is varied over its range and reduces the unknown parameters to two.   We expect that the model should not predict the pole mass $a_0(980)$ too different from its Lagrangian mass and we will test this in Sec. V.    

The remaining two parameters ($c_3$ and $\gamma_1$) only affect the isosinglet pseudoscalars (whose properties also
depend on the ten parameters discussed above).    However, there are several choices for determination of these two parameters depending on how the  four isosinglet pseudoscalars predicted in this model are matched to many experimental candidates below 2 GeV.   The two lightest predicted by the model ($\eta_1$ and $\eta_2$)  are identified with $\eta(547)$ and $\eta'(958)$ with masses:
\begin{eqnarray}
m^{\rm exp.}[\eta (547)] &=& 547.853 \pm
0.024\, {\rm
	MeV},\nonumber \\
m^{\rm exp.}[\eta' (958)] &=& 957.78 \pm 0.06
\, {\rm
	MeV}.
\end{eqnarray}
For the two heavier ones ($\eta_3$ and $\eta_4$),   there are six ways that they can be identified with the four experimental candidates above 1 GeV:  $\eta(1295)$,  $\eta(1405)$,  $\eta(1475)$, and $\eta(1760)$ with masses,
\begin{eqnarray}
m^{\rm exp.}[\eta (1295)] &=& 1294 \pm 4\, {\rm
	MeV},\nonumber \\
m^{\rm exp.}[\eta (1405)] &=& 1409.8 \pm 2.4 \,
{\rm
	MeV},
\nonumber \\
m^{\rm exp.}[\eta (1475)] &=& 1476 \pm 4\, {\rm
	MeV},\nonumber \\
m^{\rm exp.}[\eta (1760)] &=& 1756 \pm 9 \,
{\rm
	MeV}.
\end{eqnarray}
This led to six scenarios considered in detail in \cite{global}.
The two experimental inputs for determination of the two parameters $c_3$ and $\gamma_1$ are taken to be the trace and the determinant of the isosinglet pseudoscalar $4\times 4$ square mass matrix ($M_\eta^2$), i.e.
\begin{eqnarray}
{\rm Tr}\, \left(  M^2_\eta  \right) &=&
{\rm Tr}\, \left(  {M^2_\eta}  \right)_{\rm exp},
\nonumber \\
{\rm det}\, \left( M^2_\eta \right) &=&
{\rm det}\, \left( {M^2_\eta} \right)_{\rm exp}.
\label{trace_det_eq}
\end{eqnarray}

Moreover,  for each of the six scenarios,  $\gamma_1$ is found from a quadratic equation, and as a result, there are altogether twelve possibilities for determination of $\gamma_1$ and $c_3$.    Since only Tr and det of experimental masses are imposed for each of these twelve possibilities, the resulting  $\gamma_1$ and $c_3$ do not necessarily recover the exact individual experimental masses,  therefore the best overall agreement between the predicted masses (for each of the twelve possibilities) were examined in \cite{global}.   Quantitatively,  the
goodness of each solution was measured by the smallness of
the following quantity:
\begin{equation}
\chi_{sl} =
\sum_{k=1}^4
{
	{\left| m^{\rm theo.}_{sl}(\eta_k)  -
		m^{\rm exp.}_{s}(\eta_k)\right|}
	\over
	m^{\rm exp.}_{s}(\eta_k)
},
\label{E_chi_sl}
\end{equation}
in which $s$ corresponds to the scenario
(i.e. $s= 1 \cdots 6$) and
$l$ corresponds to the solution number
(i.e. $l=$ I, II).   The quantity $\chi_{sl}
\times 100$ gives the overall percent
discrepancy between our theoretical prediction
and experiment.   For the six scenarios and
the two solutions for each scenario,
$\chi_{sl}$ was analyzed  in ref. \cite{global}.
For the third scenario (corresponding to identification of $\eta_3$ and $\eta_4$ with experimental candidates $\eta(1295)$ and $\eta(1760)$) and  solution I the best agreement with the mass spectrum of the eta system was obtained (i.e. $\chi_{3\rm{I}}$ was the smallest).      Disfavoring $\eta(1405)$ and $\eta(1475)$ as the $\eta_3$ and $\eta_4$ is consistent with speculations that these two state are pseudoscalar glueballs \cite{07_KZ}.
Furthermore,   all six scenarios were examined in the analysis of $\eta'\rightarrow\eta\pi\pi$ decay in \cite{LsM_mmp_eta3p} and it was found that the best overall result (both for the partial decay width of $\eta'\rightarrow \eta\pi\pi$ as well as the energy dependence of its squared decay amplitude) is obtained for scenario ``3I'' consistent with the analysis of ref. \cite{global}. In this work,  we use the result of ``3I'' scenario.

Given these inputs there are a very large number of
predictions. At the level of the quadratic terms in the
Lagrangian, we predict all the remaining masses
 and decay constants as well
as the angles describing the mixing between each of
($\pi,\pi'$),
($K,K'$), ($a_0,a_0'$), ($\kappa,\kappa'$) multiplets
and each of the 4$\times$4
isosinglet mixing matrices
 (each formally described by six angles).

Consequently,  all twelve parameters of the model (at the present order of approximation) are evaluated by the method discussed above using four minimum equations and eight experimental inputs.   The uncertainties of the experimental inputs result in uncertainties on the twelve model parameters which in turn result in uncertainties on  physical quantities that are computed in this model.    In the work of Ref. \cite{global} all rotation matrices describing the underlying mixing among two- and four-quark components for each spin and isospin states are computed. Tables \ref{pseudotable} and \ref{scalartable} give the outcome of the computations for masses and quark contents of both pseudoscalars and scalars below and above 1 GeV. For the study of $\pi \eta$ scattering, we need the following rotation matrices:
\begin{eqnarray}
\left[
\begin{array}{cc}
\pi^+(137)\\
\pi^+(1300)
\end{array}
\right]
&=&
R_\pi^{-1}
\left[
\begin{array}{cc}
\phi_1^2\\
{\phi'}_1^2
\end{array}
\right],
\hskip 1cm
\left[
\begin{array}{cc}
a_0^+(980)\\
a_0^+(1450)
\end{array}
\right]
=
L_a^{-1}
\left[
\begin{array}{cc}
S_1^2\\
{S'}_1^2
\end{array}
\right],\nonumber\\
\left[
\begin{array}{cc}
f_1\\
f_2\\
f_3\\
f_4
\end{array}
\right]
&=&
L_0^{-1}
\left[
\begin{array}{cc}
f_a\\
f_b\\
f_c\\
f_d
\end{array}
\right],
\hskip 1cm
\left[
\begin{array}{cc}
\eta_1\\
\eta_2\\
\eta_3\\
\eta_4
\end{array}
\right]
=
R_0^{-1}
\left[
\begin{array}{cc}
\eta_a\\
\eta_b\\
\eta_c\\
\eta_d
\end{array}
\right],
\label{E_SRot}
\end{eqnarray}
where $R_\pi^{-1}$ and  $L_a^{-1}$ are the rotation matrices for  $I=1$  pseudoscalars and scalars respectively; $f_i, i=1\cdots 4$ are four of the physical isosinglet scalars below 2 GeV (in this model $f_1$ and $f_2$ are clearly identified with $f_0(500)$ and $f_0(980)$ and the two heavier states resemble two of the heavier isosinglet scalars above 1 GeV); and
\begin{eqnarray}
f_a&=&\frac{S^1_1+S^2_2}{\sqrt{2}} \hskip .7cm
\propto \hskip .5cm n{\bar n},
\nonumber  \\
f_b&=&S^3_3 \hskip 1.6 cm \propto \hskip .5cm s{\bar s},
\nonumber    \\
f_c&=&  \frac{S'^1_1+S'^2_2}{\sqrt{2}}
\hskip .5 cm \propto \hskip .5cm ns{\bar n}{\bar s},
\nonumber   \\
f_d&=& S'^3_3
\hskip 1.5 cm \propto \hskip .5cm nn{\bar n}{\bar n},
\label{f_basis}
\end{eqnarray}
where the non-strange ($n$) and strange ($s$) quark content
for each basis state has been listed at the end of
each line above.

Similarly,    $\eta_i, i=1\cdots4$ are four of the physical isosinglet pseudoscalars below 2 GeV (where $\eta_1$ and $\eta_2$ are identified with $\eta(547)$ and $\eta'(958)$ and the two heavier states are identified with two of the heavier isosinglet pseudoscalars above 1 GeV),  and

\begin{eqnarray}
\eta_a&=&\frac{\phi^1_1+\phi^2_2}{\sqrt{2}} \hskip .7cm
\propto \hskip .5cm n{\bar n},
\nonumber  \\
\eta_b&=&\phi^3_3 \hskip 1.6 cm \propto \hskip .5cm s{\bar s},
\nonumber    \\
\eta_c&=&  \frac{\phi'^1_1+\phi'^2_2}{\sqrt{2}}
\hskip .5 cm \propto \hskip .5cm ns{\bar n}{\bar s},
\nonumber   \\
\eta_d&=& \phi'^3_3
\hskip 1.5 cm \propto \hskip .5cm nn{\bar n}{\bar n}.
\label{eta_basis}
\end{eqnarray}

\begin{table}[!htbp]
	\renewcommand\thetable{II}
	\centering
	\caption{
		Mass and four-quark percentage of the pseudoscalar mesons below and above 1 GeV predicted by the leading order of the GLSM \cite{global}.   In first column the masses in the square brackets are inputs and other values are model predictions. The corresponding experimental values reported in PDG \cite{pdg}  are displayed in the second column (for the mass of pion and kaon the average of their charged and neutral masses are extracted from PDG); the last column gives the estimate of the four-quark content of these states.  The model predictions have a range of variation that stem from two of the model inputs with large uncertainties ($m[\pi(1300)]$=1.22$-$1.38 GeV and $A_3/A_1$=27$-$30).  Each predicted quantity in columns one and three  is the average of that quantity over its range of variation and its  uncertainty is one standard deviation around the average.  All displayed masses are in MeV.   The predicted properties of states above 1 GeV are expected to improve by inclusion of higher order terms in the potential as well as the scalar and pseudoscalar glueballs.
	}
	\renewcommand{\tabcolsep}{0.4pc} 
	\renewcommand{\arraystretch}{1.5} 
	\begin{tabular}{cccc}
		\noalign{\hrule height 1pt}
		\noalign{\hrule height 1pt}
		State	& Mass (MeV) & Experiment & Four-quark percentage   \\
		\noalign{\hrule height 1pt}
		$\pi(137)$	 & [137] & 137.2734 $\pm$ 0.0005 & 14 $\pm$ 1  \\
		\noalign{\hrule height 1pt}
		$\pi(1300)$	 & [$1220-1380$] & 1300 $\pm$ 100 & 86 $\pm$ 1\\
		\noalign{\hrule height 1pt}
		$K(496)$  & 502 $\pm$ 9	& 495.64 $\pm$ 0.02 & 12 $\pm$ 2 \\
		\noalign{\hrule height 1pt}
		$K(1460)$  & 1275 $\pm$ 48	& $-$  & 87 $\pm$ 2 \\
		\noalign{\hrule height 1pt}
		$\eta(547)$  & 542 $\pm$ 7	& 547.862 $\pm$ 0.017  & 9 $\pm$ 2 \\
		\noalign{\hrule height 1pt}
		$\eta'(958)$  & 972 $\pm$ 20	& 957.78 $\pm$ 0.06 & 17 $\pm$ 3 \\
		\noalign{\hrule height 1pt}
		$\eta (1295)$  & 1293 $\pm$ 41	& 1294 $\pm$ 4 & 78 $\pm$ 6 \\
		\noalign{\hrule height 1pt}
		$\eta (1760)$  & 1749 $\pm$ 22	& 1751 $\pm$ 15  & 96 $\pm$ 1 \\
		\noalign{\hrule height 1pt}
	\end{tabular}\\[2pt]
	\label{pseudotable}
	\end{table}

\begin{table}[!htbp]
	\renewcommand\thetable{III}
	\centering
	\caption{Mass and four-quark percentage of the scalar mesons below and above 1 GeV predicted by the leading order of the GLSM \cite{global}.   The first column gives the Lagrangian mass (with the exception of those  in the square brackets which are taken as inputs,  all other values are model predictions).   The second column provides the physical mass extracted from the poles of the relevant K-matrix unitarized scattering amplitudes;   the third column provides the corresponding experimental values reported in PDG \cite{pdg}; and the last column gives the estimate of the four-quark content of these states.  The model predictions have a range of variation that stem from two of the model inputs with large uncertainties ($m[\pi(1300)]$=1.22$-$1.38 GeV and $A_3/A_1$=27$-$30).  Each predicted quantity in columns one, two and four is the average of that quantity over its range of variation and its  uncertainty is one standard deviation.  All displayed masses are in MeV.   The predicted properties of states above 1 GeV are expected to improve by inclusion of higher order terms as well as the scalar and pseudoscalar glueballs. }
	\renewcommand{\tabcolsep}{0.4pc} 
	\renewcommand{\arraystretch}{1.5} 
	\begin{tabular}{ccccc}
		\noalign{\hrule height 1pt}
		\noalign{\hrule height 1pt}
		State	& Mass  & Physical mass  &  Experiment &Four-quark percentage   \\
		\noalign{\hrule height 1pt}
		$a_0(980)$	 & [980] & 985 $\pm$ 5  & 980 $\pm$ 20 & 59 $\pm$ 11  \\
		\noalign{\hrule height 1pt}
		$a_0(1450)$	 & [1474] & 1083 $\pm$ 33 & 1474 $\pm$ 19  & 40 $\pm$ 11\\
		\noalign{\hrule height 1pt}
		$K_0^*(800)$  & 1113 $\pm$ 32& 748 $\pm$ 9	& 682 $\pm$ 29 & 81 $\pm$ 8 \\
		\noalign{\hrule height 1pt}
		$K_0^*(1430)$  & 1570 $\pm$ 30  & 1118 $\pm$ 37	& 1425 $\pm$ 50 & 19 $\pm$ 8 \\
		\noalign{\hrule height 1pt}
		$f_0(500)$  & 659 $\pm$ 30	&  477 $\pm$ 3  & 400-550 & 48 $\pm$ 6 \\
		\noalign{\hrule height 1pt}
		$f_0(980)$  & 1145 $\pm$ 41	& 1065 $\pm$ 33 &  990 $\pm$ 20 &  89 $\pm$ 5 \\
		\noalign{\hrule height 1pt}
		$f_0(1370)$  & 1507 $\pm$ 6	&  1157 $\pm$ 34  & 1200-1500 & 45 $\pm$ 4 \\
		\noalign{\hrule height 1pt}
		$f_0(1710)$  & 1713 $\pm$ 33	&  1691 $\pm$ 25 & 1704 $\pm$ 12 & 17 $\pm$ 7 \\
		\noalign{\hrule height 1pt}
	\end{tabular}\\[2pt]
	\label{scalartable}
\end{table}


\section{Generalized linear sigma model prediction of $\pi \eta$ scattering Amplitude}

Using the potential defined in Eq. (\ref{SpecLag}), we compute the Feynman diagrams of Fig. \ref{F_FD} which  include a four-point contact term, contribution of two isotriplet scalars in the $s$- and $u$-channels, as well as contribution of four isosinglet scalars in the $t$-channel.   This leads to computing the coupling constants in our reference equation for the amplitude [Eq. (\ref{pietatem})] as follows:
\begin{eqnarray}
\gamma_{\pi \eta}^{(4)} &=& \sum _{A,B,M,N}\left\langle \frac{\partial ^4 V}{\partial(\phi_1^2)_A\partial(\phi_2^1)_B\partial \eta_M\partial \eta_N}\right\rangle  (R_{\pi})_{A1}
 (R_{\pi})_{B1} (R_{0})_{M1} (R_{0})_{N1},\nonumber\\
\gamma_{f_{i} \pi \pi} &=& \frac{1}{\sqrt{2}}\sum _{A,B,K}\left\langle \frac{\partial ^3 V}{\partial(\phi_1^2)_A\partial(\phi_2^1)_B\partial f_K}\right\rangle (R_{\pi})_{A1} (R_{\pi})_{B1} (L_{0})_{Ki},\nonumber\\
 \gamma_{f_{i} \eta \eta} &=& \frac{1}{2}\sum _{K,M,N}\left\langle \frac{\partial ^3 V}{\partial f_K\partial \eta_M\partial \eta_N}\right\rangle (L_0)_{Ki} (R_{0})_{M1} (R_{0})_{N1},\nonumber\\
 \gamma_{a_j \pi \eta} &=& \sum _{A,B,M}\left\langle \frac{\partial ^3 V}{\partial(S_1^2)_A\partial(\phi_2^1)_B\partial \eta_M}\right\rangle  (L_{a})_{Aj}
 (R_{\pi})_{B1} (R_{0})_{M1}.
\end{eqnarray}
where $A$ and $B$ can take values of 1 and 2 (with 1 referring to nonet $M$ and 2 referring to nonet $M'$) and $K$ is a placeholder for  {\it a},{\it b},{\it c} and {\it d} that
represent the four bases in Eq. (\ref{f_basis}) and $M$ and $N$ are placeholders for  {\it a},{\it b},{\it c} and {\it d} that
represent the four bases in Eq. (\ref{eta_basis}).  $L_0$, $L_a$, $R_0$ and $R_\pi$  are the rotation matrices defined in previous section.  The bare coupling constants  are all given in Appendix B.     Consistency of these couplings are checked in Appendix C by recovering the current algebra result for the $\pi\eta$ scattering.

\subsection{The scattering amplitude and phase shift}
\begin{figure}[!htb]
\begin{center}
\vskip 1cm
\epsfxsize = 7.5cm
\includegraphics[keepaspectratio=true,scale=0.65]{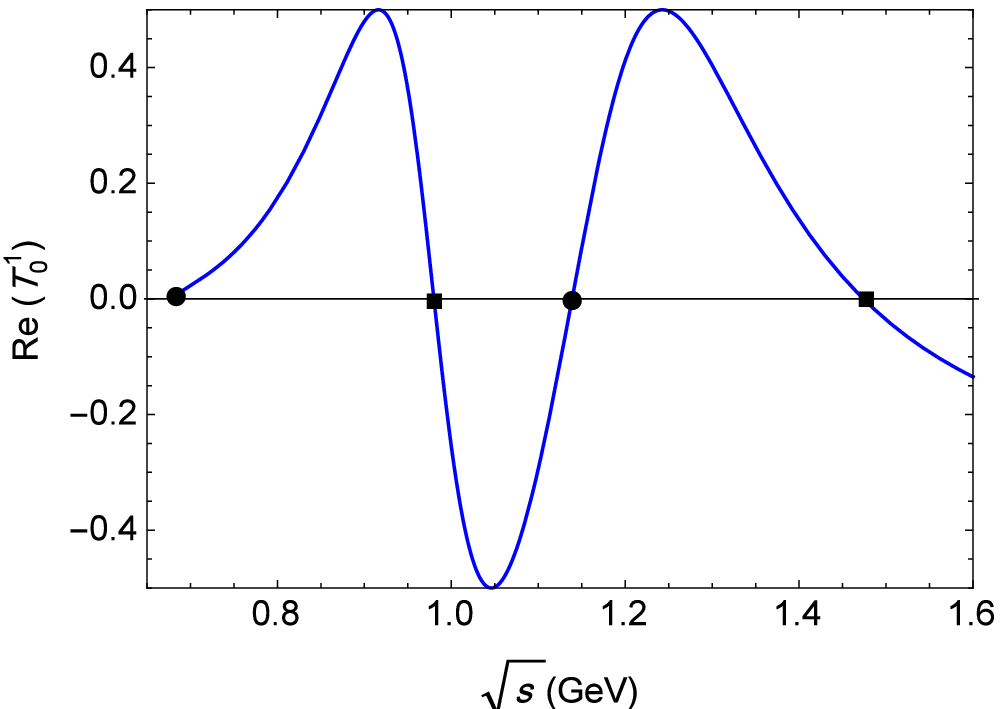}
\hskip 1cm
\epsfxsize = 7.5cm
\includegraphics[keepaspectratio=true,scale=0.65]{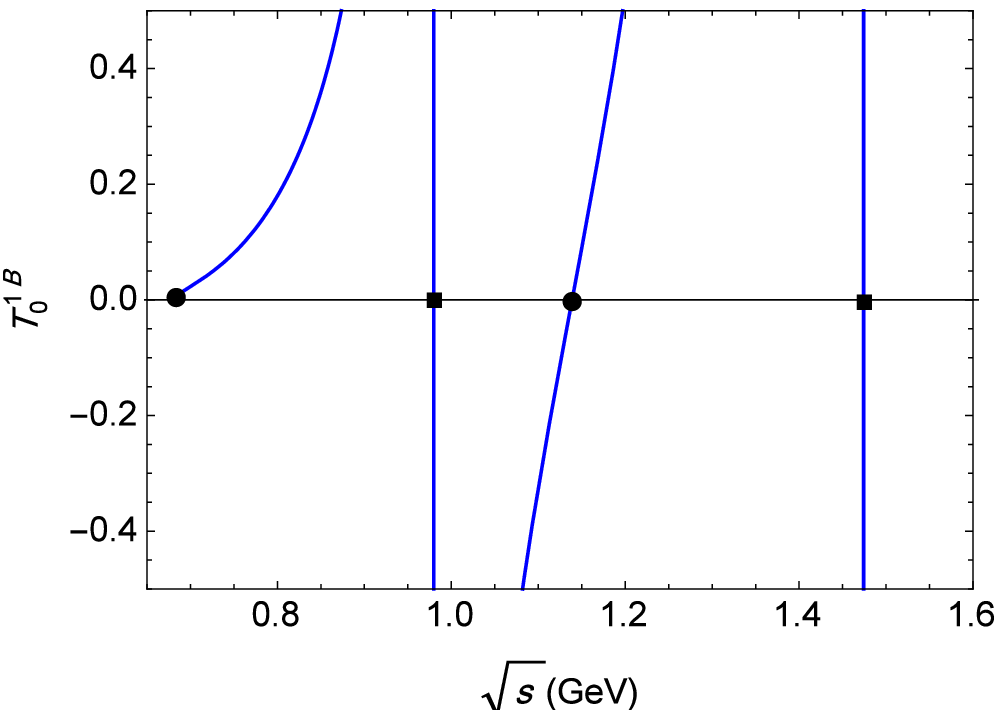}
\caption{Real part of the $I=1$, $J=0$, $\pi \eta$ scattering amplitude computed in GLSM.     The bare amplitude (left) contains zeros (circles) and poles (squares) at which the unitarized amplitude (right) also vanishes.}
\label{eta_generic_T10}
\end{center}
\end{figure}

\begin{figure}[!htbp]
\begin{center}
\epsfxsize = 4 cm
 \includegraphics[keepaspectratio=true,scale=0.67]{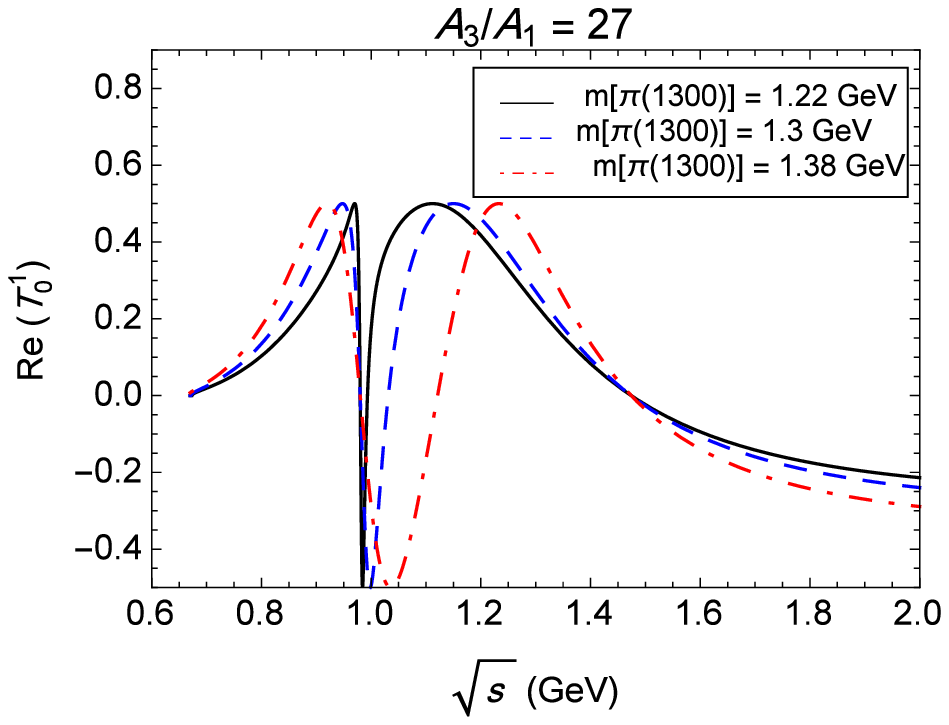}
\hskip .1 cm
\epsfxsize = 4 cm
\includegraphics[keepaspectratio=true,scale=0.67]{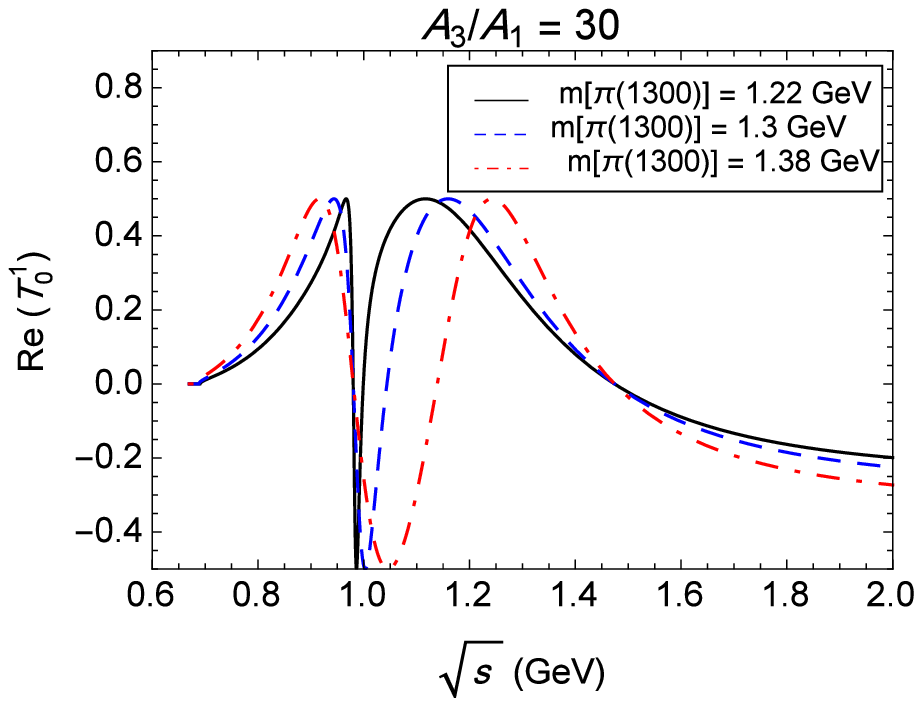}
\caption{Real part of the unitarized $\pi \eta$ scattering amplitude computed in GLSM for $A_3/A_1 =27 $ (left) and  $A_3/A_1 =30 $ (right) for three different choices of  $m[\pi(1300)]$ (with scenario $3I$ discussed in Sec. IV).}
\label{t01a3pa1}
\end{center}
\end{figure}

\begin{figure}
\begin{center}
\epsfxsize = 3 cm
\includegraphics[keepaspectratio=true,scale=0.67]{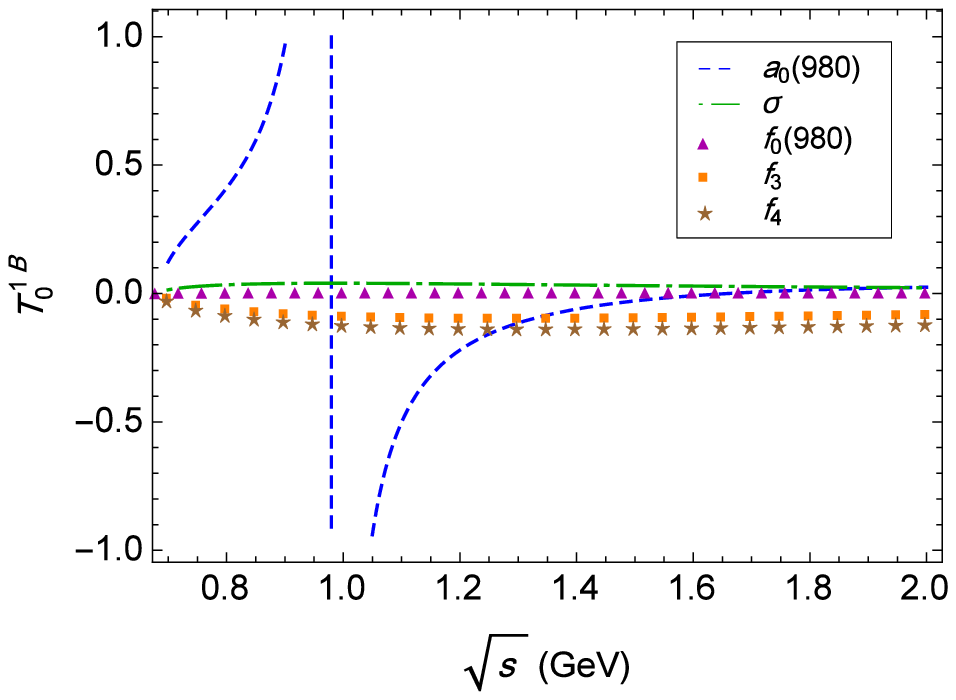}
\hskip .1 cm
\epsfxsize = 3 cm
\includegraphics[keepaspectratio=true,scale=0.67]{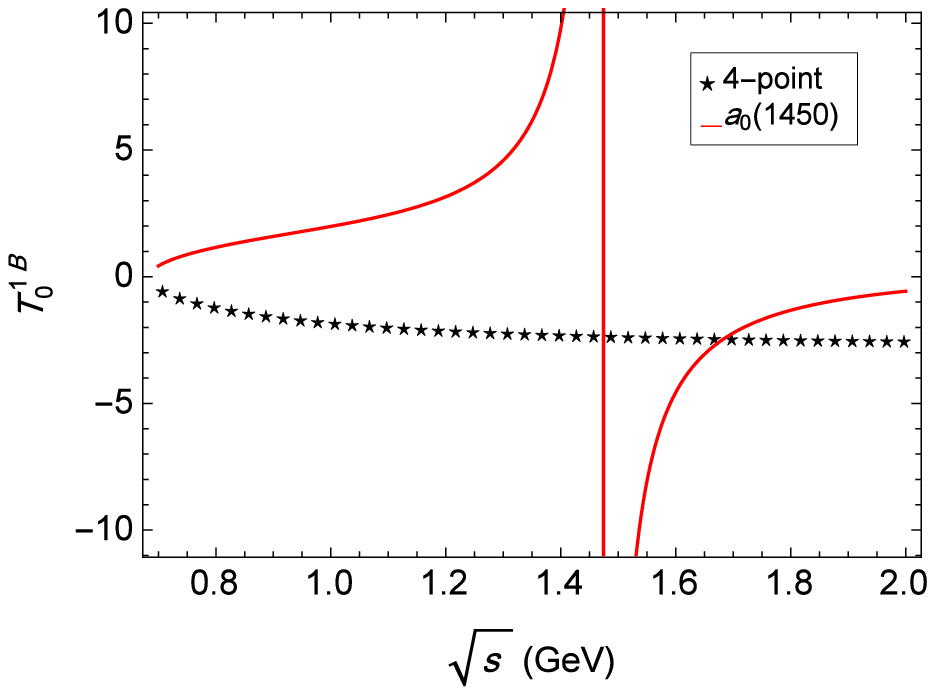}
\caption{Individual contributions to real part of the unitarized $\pi \eta$ scattering amplitude for  $A_3/A_1 =30$ and $m[\pi(1300)]=1.38$ GeV.   (Note that the contributions are plotted into two figures to avoid overcrowding  and to be able to show contributions that are of considerable size difference.) }
\label{F_GLSM_ind}
\end{center}
\end{figure}

We start with the  real part of the partial wave scattering amplitude and compare the bare amplitude (\ref{pietatem}) with its  K-matrix unitarized amplitude for a typical input of the model parameters.  Illustrated in Fig. \ref{eta_generic_T10}  is the bare amplitude (left),  given for direct comparison  side by side to its K-matrix unitarized amplitude (right).   The zeros in the unitarized amplitude stem from two sources in the bare amplitude:  The poles and the zeors in the bare amplitude (respectively shown with squares and circles  in Fig. \ref{eta_generic_T10}).       For several values of $A_3/A_1$ and $m[\pi(1300)]$,  the real part of the amplitudes are shown in Fig. \ref{t01a3pa1}.     The individual contributions to the bare amplitude are shown in Fig. \ref{F_GLSM_ind}.

\begin{figure}[!htbp]
\begin{center}
\epsfxsize = 3 cm
\includegraphics[keepaspectratio=true,scale=0.57]{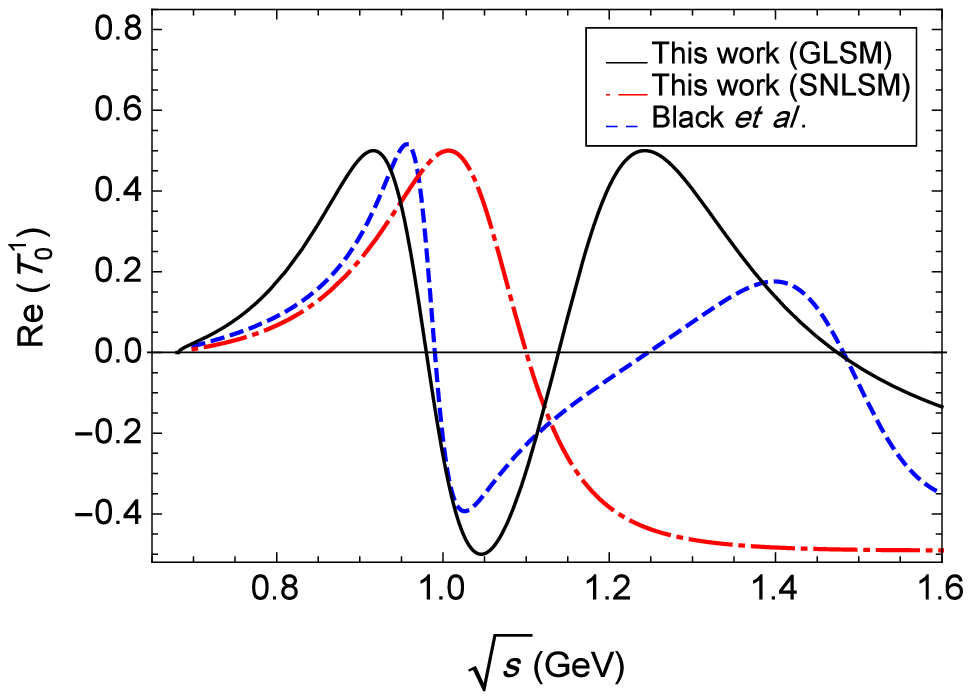}
\hskip 0.1cm
\epsfxsize = 3 cm
\includegraphics[keepaspectratio=true,scale=0.57]{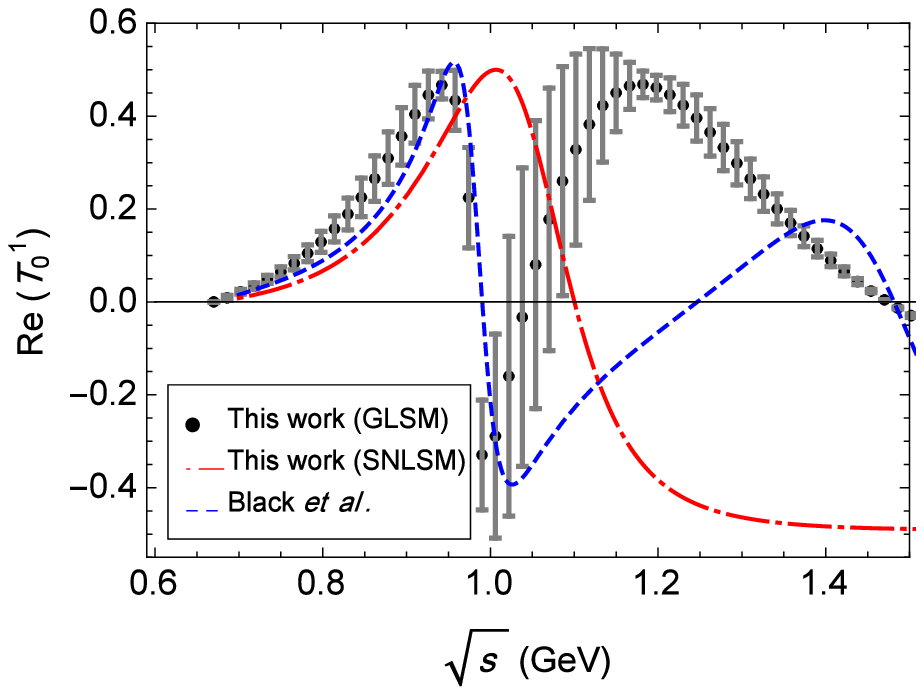}

\epsfxsize = 3 cm
\includegraphics[keepaspectratio=true,scale=0.57]{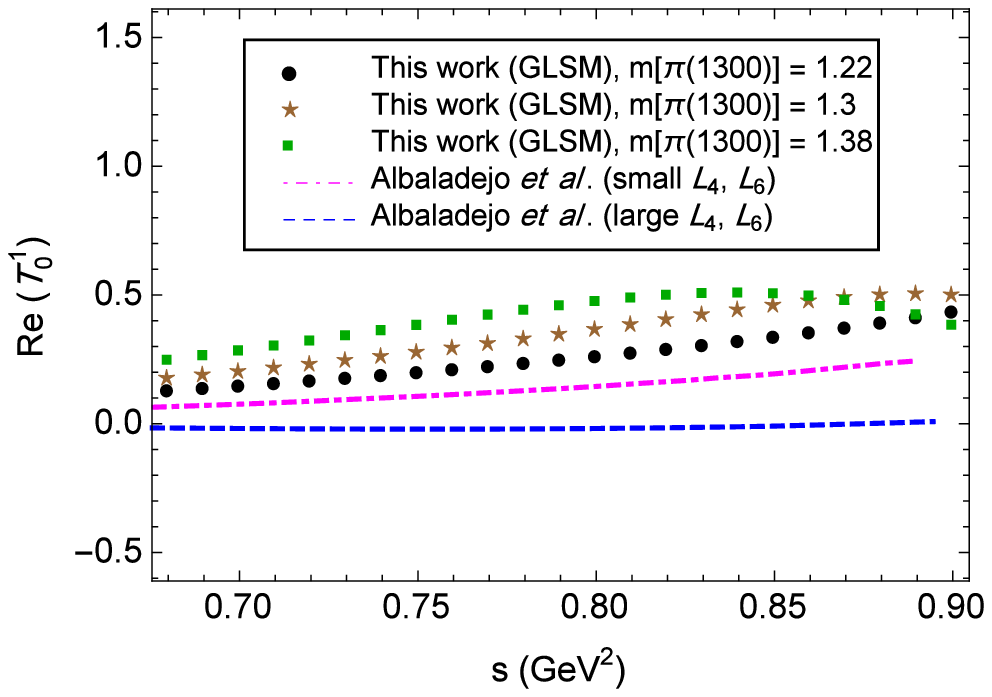}
\hskip 0.1cm
\epsfxsize = 3 cm
\includegraphics[keepaspectratio=true,scale=0.57]{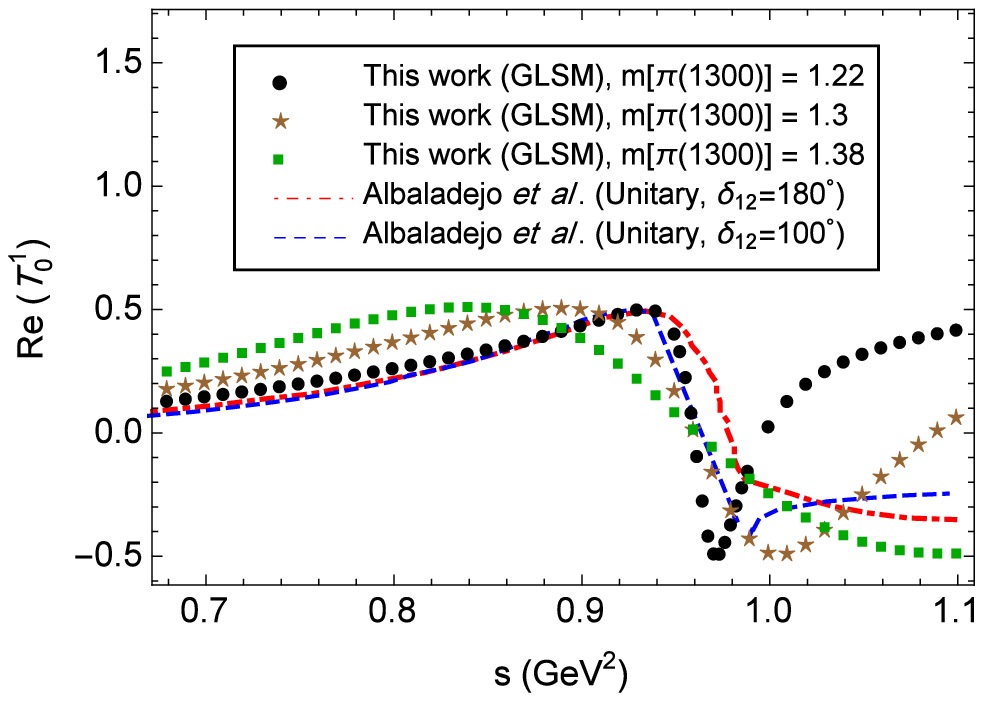}

\epsfxsize = 3 cm
\includegraphics[keepaspectratio=true,scale=0.57]{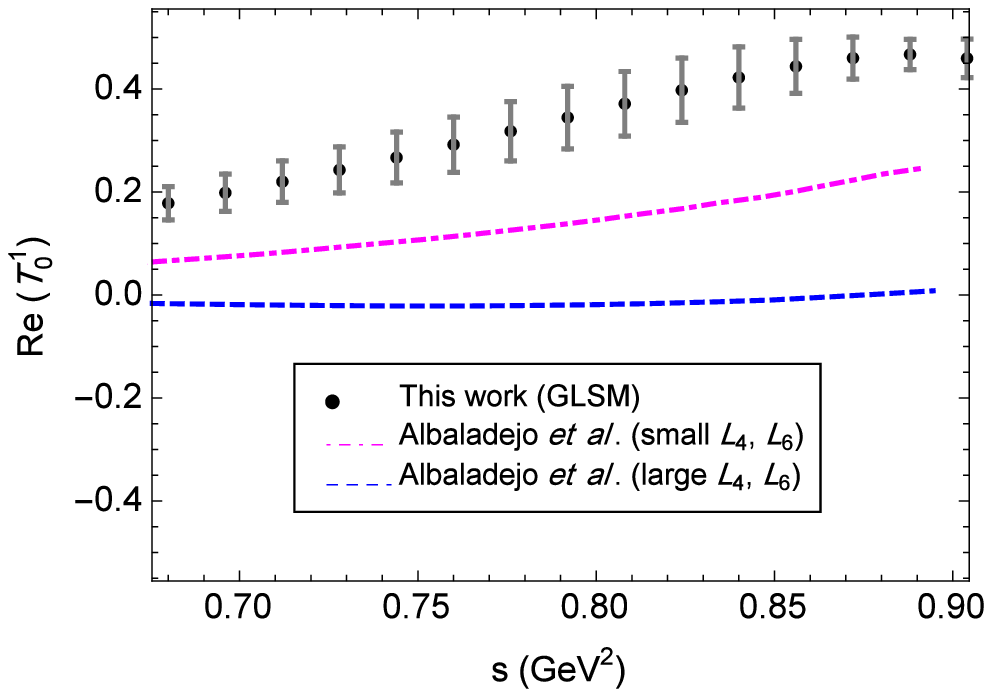}
\hskip 0.1cm
\epsfxsize = 3 cm
\includegraphics[keepaspectratio=true,scale=0.57]{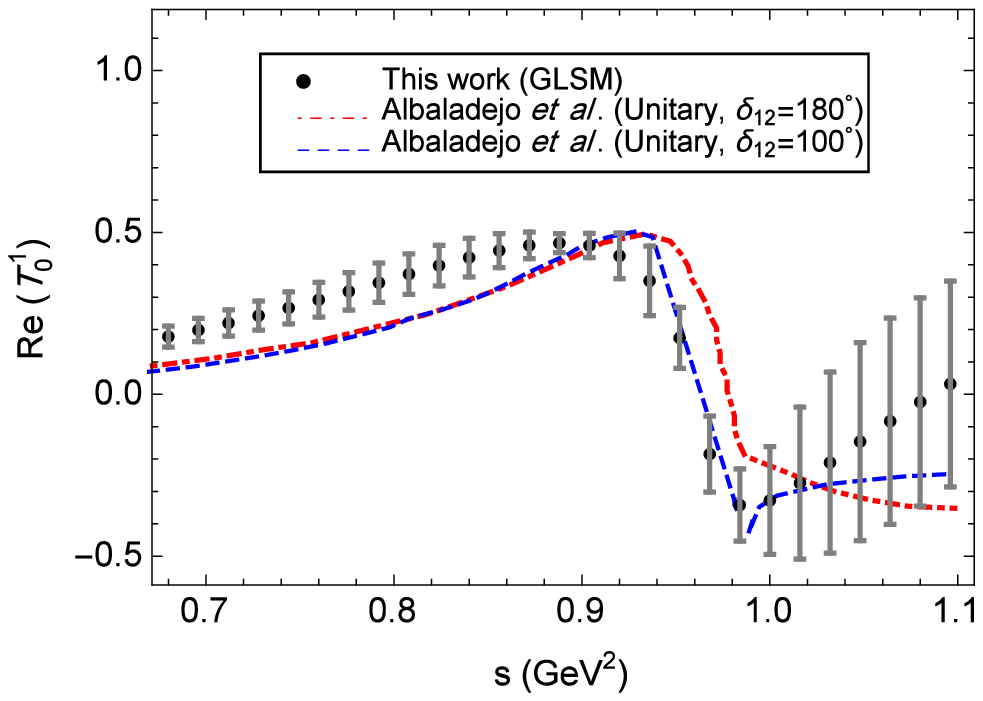}
\caption{ The real part of the predicted K-matrix unitarized $\pi \eta$ scattering amplitude computed  in GLSM (for a typical value of $m[\pi(1300)]$ and $A_3/A_1$) is compared with  single nonet linear sigma model (SNLSM) and the nonliear sigma model (NLSM) of Ref. \cite{pieta} (top left).  In the top right figure,  the averages (dots) and standard deviations (error bars) of the prediction of GLSM resulted from variation of  $m[\pi(1300)]$ and $A_3/A_1$ show much less sensitivity below 1 GeV.    Up to about 1 GeV,  the results qualitatively agree, however above 1 GeV the GLSM and NLSM only have a similar functional behavior (increasing to a maximum and decaying with energy), while the SNLSM lacks any structure and flattens to a constant value.   Also shown are comparisons with predictions of Ref. \cite{Albaladejo_ph}  using the next-to leading order of ChPT (middle left)
and  two-channel unitary amplitudes (middle right),  where limited qualitative agreements with the latter can be seen. The effects of coupled variations of $m[\pi(1300)]$ and $A_3/A_1$ (in ranges $1.22-1.38$ GeV and $27-30$, respectively) are plotted in the bottom two Figs. and compared with those of \cite{Albaladejo_ph}.  These coupled variations do not appreciably add to the uncertainties depicted in the middle two figures.
}
\label{ReT_GLSM_NLSM}
\end{center}
\end{figure}

The real part of the amplitude was also computed in the work of Ref. \cite{pieta} within a nonlinear chiral Lagrangian model with explicit inclusion of intermediate resonances.   The real part of the scattering amplitude found in this work  within GLSM (for the specific choice of $m[\pi(1300)]=1.38$ GeV and $A_3/A_1=30$) is compared with the prediction of \cite{pieta} in Fig. \ref{ReT_GLSM_NLSM} (top left).    Also shown (top right) is the effect of variation of $m[\pi(1300)]$ and $A_3/A_1$ on the prediction of GLSM with circles and error bars being the prediction averages and one standard deviation around the averages respectively.   Up to 1 GeV,  the variations mildly overlap with the work of \cite{pieta},  but
above 1 GeV only their functional form is similar.    Also shown are comparisons of this result (for specific inputs: $A_3/A_1=30$ and $m[\pi(1300)]=1.22,1.3,1.38$ GeV) with those of \cite{Albaladejo_ph} in next-to leading order in ChPT (middle left) and the two-channel unitarity amplitude (middle right). Within the uncertainties of both GLSM and the two models of Ref. \cite{Albaladejo_ph} there are some limited qualitative overlaps.    However,  comparatively,  the  GLSM results are in a better agreement with the two-channel unitarity model of \cite{Albaladejo_ph},   and clearly less consistent with  the  large  $L_4$ and $L_6$ ChPT (compared to the small $L_4$ and $L_6$ scenario) which is also less favored according to the work of Ref.  \cite{Albaladejo_ph}. 
Quantitatively, \footnote{ As a quantitative measure of the disagreement between two functions $f_{1}(x)$ and $f_{2}(x)$, we define 
			\begin{equation}
			\Delta (x) = \frac{{\frac{{\left| {{f_1}(x) - {f_2}(x)} \right|}}{2}}}{{\frac{{\left| {{f_1}(x)} \right| + \left| {{f_2}(x)} \right|}}{2}}} \times 100 = \frac{{\left| {{f_1}(x) - {f_2}(x)} \right|}}{{\left| {{f_1}(x)} \right| + \left| {{f_2}(x)} \right|}} \times 100,
						\end{equation}
This gives a measure of the percent disagreement between the two functions (compared to their average). 
The absolute values in the denominator avoids division by zero.  The mean disagreement is:
\begin{equation} 
{\bar \Delta} = {1\over {x_2-x_1}}\,\int_{x_1}^{x_2} \Delta(x)\, dx
\label{disag} 
\end{equation}.  
		}
	 the mean disagreements between GLSM predictions with inputs $m\left[ {\pi \left( {1300} \right)} \right] = 1.22$, $1.3$, and $1.38$ GeV, and  the predictions of ChPT \cite{Albaladejo_ph} respectively are (see footnote):  $19\% $, $30\%$ and $41\%$ (with their small $L_4$ and $L_6$);   $97\% $, $ 97\% $ and $ 98\% $  (with their large ${L_4},{L_6}$).  Similar comparisons with the unitary model of \cite{Albaladejo_ph} respectively give: $ 22\% $, $ 21\% $
and $27\%$ (with their  ${\delta _{12}} = {180^o}$) and $ 26\% $, $ 27\% $ and $ 38\% $ (with their ${\delta _{12}} = {100^o}$).
Therefore, our predictions are in general more consistent with the unitary model of  \cite{Albaladejo_ph} than their ChPT driven results.
The uncertainties around the central value of $m[\pi (1300)]$ mass for the real part of $J=0$, $I=1$ scattering amplitude average around $17\%$, while this is $66\%$ for the predictions of ChPT \cite{Albaladejo_ph}  and $16\%$ for their prediction of unitary approach \cite{Albaladejo_ph}.  The effects of combined variations of $m[\pi(1300)]$ and $A_3/A_1$ (in ranges $1.22-1.38$ GeV, and $27-30$, respectively) are given in the last two sub-figures of Fig. \ref{ReT_GLSM_NLSM} (with circles and error bars being the averages and standard deviations, respectively),   and show that effectively such coupled variations do not add additional uncertainties to the GLSM predictions.

\begin{figure}[!htbp]
	\begin{center}
		\epsfxsize = 4 cm
		\includegraphics[keepaspectratio=true,scale=0.72]{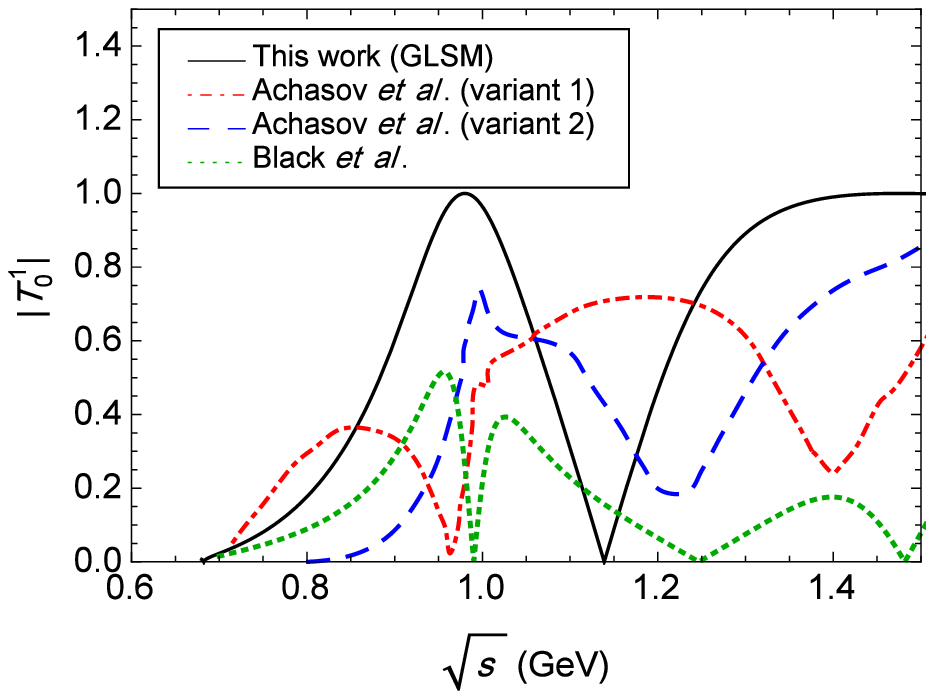}
		\epsfxsize = 4 cm
			\includegraphics[keepaspectratio=true,scale=0.72]{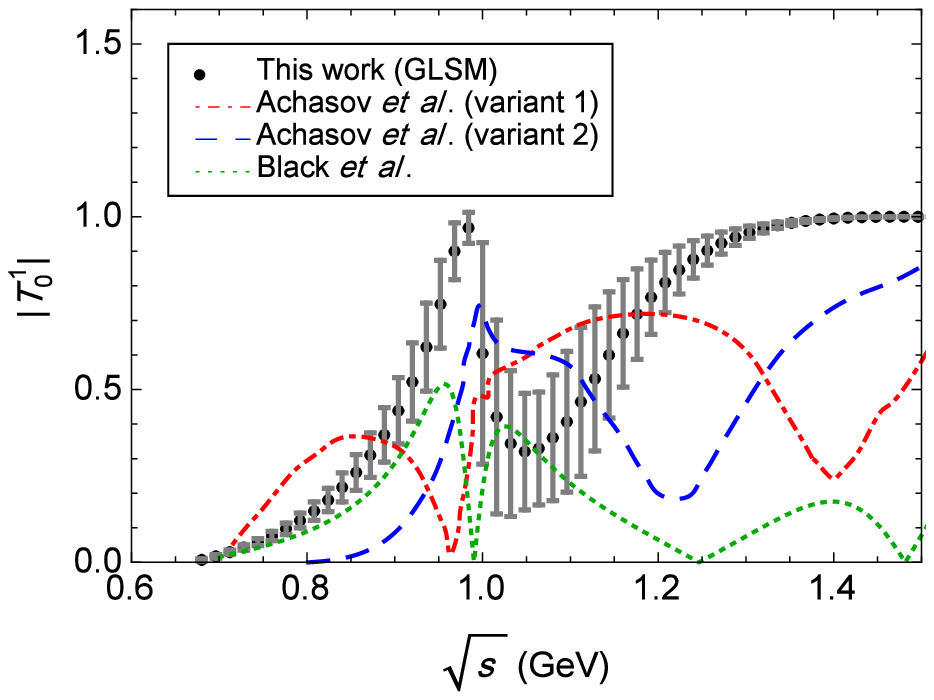}
		\caption{Comparing the prediction of the GLSM (left) for the modulus of $T_0^1$ with the predictions by Achasov et al \cite{belledata} and by Black et al \cite{pieta}. The behavior of ``variant 2'' of \cite{belledata} (dashed curve) is qualitatively closer to that of  GLSM (with inputs $A_3/A_1=30$ and $m[\pi(1300)]=1.38$ GeV).  
	The right figure gives the averages (dots) and standard deviations (error bars) of the prediction of GLSM (resulted from variation of  $m[\pi(1300)]$ [1.22 GeV,1.38 GeV] and $A_3/A_1$ [27, 30]) showing that,  as expected,  the model uncertainty grows above 1 GeV.		
						    }
		\label{modT_GLSM}
	\end{center}
\end{figure}

\begin{figure}[!htbp]
\begin{center}
\epsfxsize = 1 cm
 \includegraphics[keepaspectratio=true,scale=0.64,width=5.75cm]{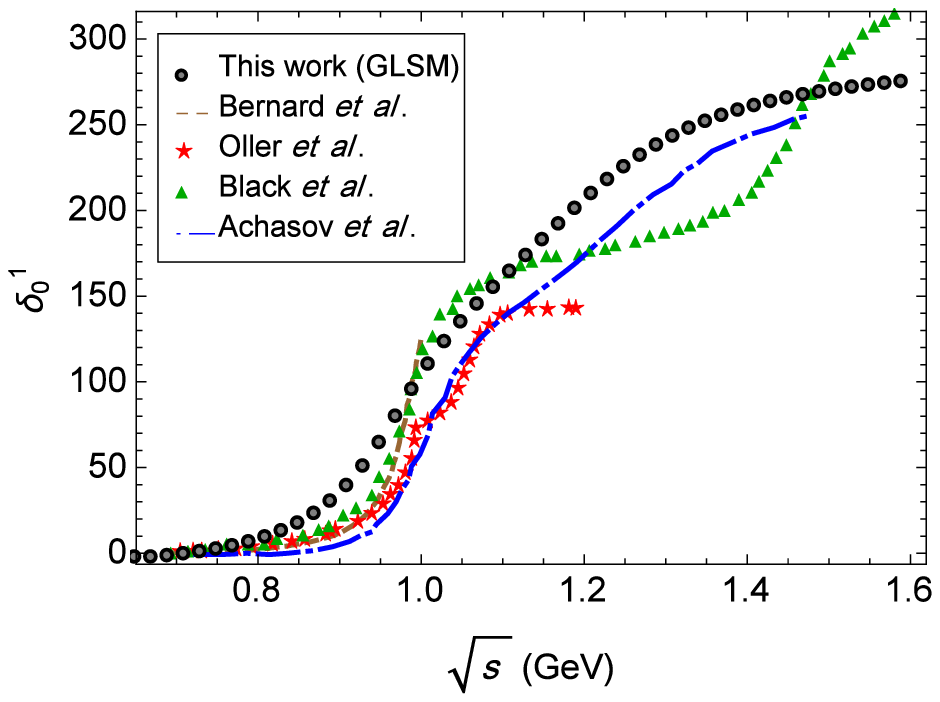}
\epsfxsize = 1 cm
\includegraphics[keepaspectratio=true,scale=0.64,width=5.75cm]{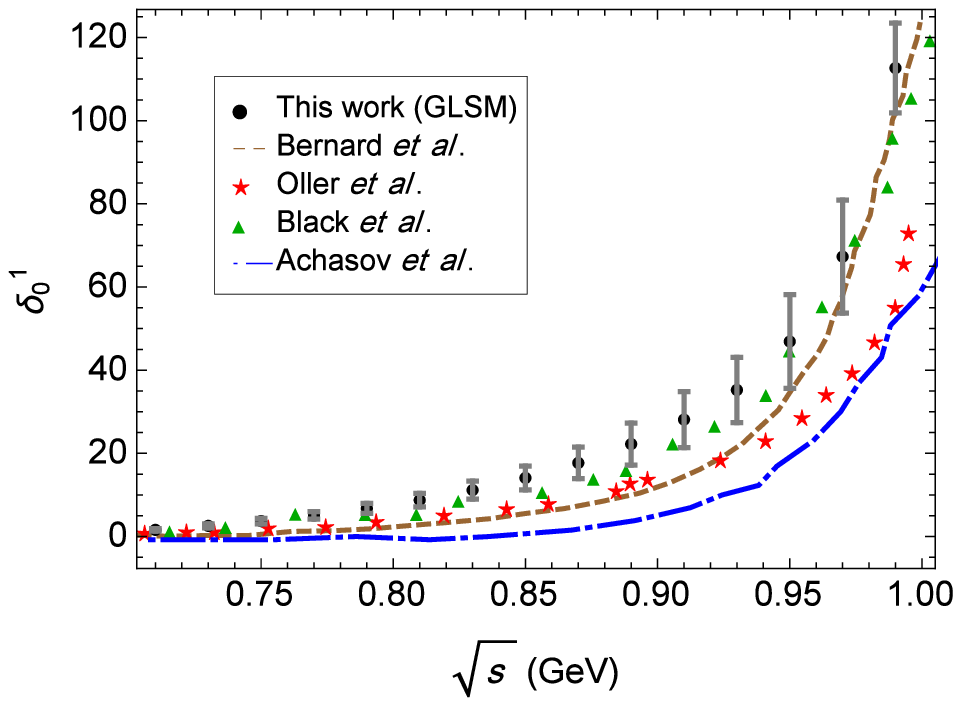}
\epsfxsize = 1 cm
\includegraphics[keepaspectratio=true,scale=0.64,width=5.75cm]{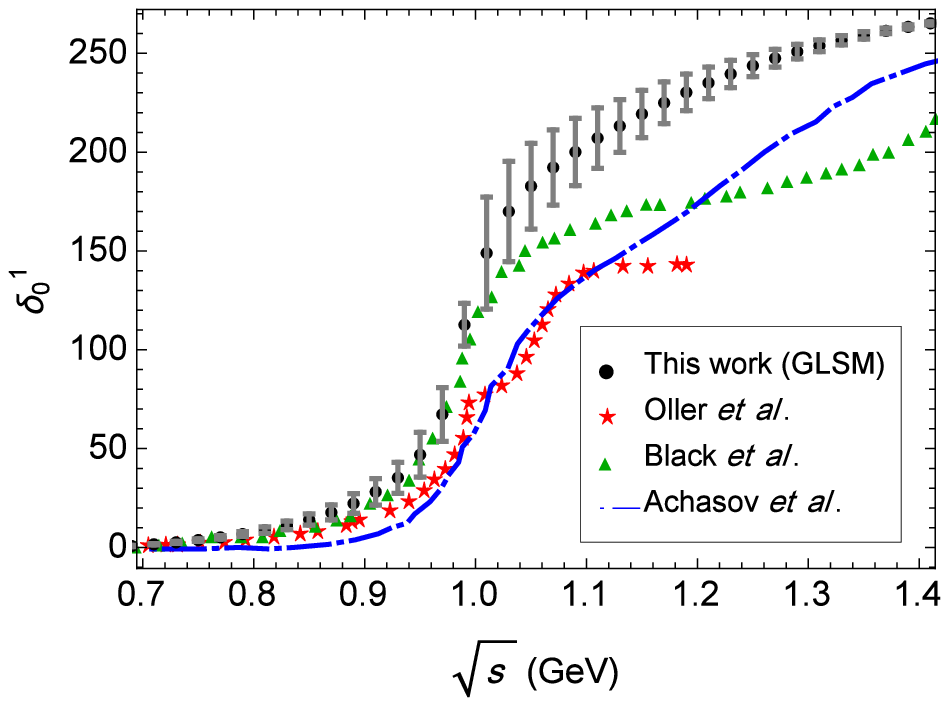}
\caption{Left figure shows phase shift computed (with the specific inputs $A_3/A_1$=30 and $m[\pi(1300)]$=1.38 GeV) 
from the K-matrix  unitarized s-wave $\pi\eta \rightarrow \pi \eta$ scattering amplitude in GLSM and compared with predictions by Bernard et al \cite{ChPT_bernard},  Oller et al \cite{oller2},  Black et al \cite{pieta} and Achasov et al  \cite{belledata}.   The averaged predictions of GLSM (circles) together with uncertainties stemming from variations of $A_3/A_1$ [27,30] and and $m[\pi(1300)]$ [1.22 GeV,1.38 GeV] estimated by one standard deviation around the averages (error bars) are compared with predictions by other works (middle and right figures). A reasonable qualitative agreement up to about 1 GeV,  particularly with the work of \cite{pieta},  is evident.}
\label{F_phase_shift}
\end{center}
\end{figure}

Similarly, the behavior of the imaginary part of the K-matrix unitarized amplitude can be traced to the structure of the bare amplitude.  In this case,  the poles and zeros in the bare amplitude  force the imaginary part of the unitarized amplitude to respectively become 1 and 0.    As a result,  the modulus of the K-matrix unitarized amplitude also becomes 1 and 0 at the location of poles and zeros in the bare amplitude.     We recognize that this behavior is partly enforced by the K-matrix unitarization method, which at first seems quite arbitrary, and in principle may or may not fetch any  nontrivial physics.
However, in practice the simple K-matrix unitarization has had reasonable success (at least up to about 1 GeV) for the cases of $\pi\pi$ and $\pi K$ scatterings studied in Refs. \cite{mixing_pipi} and \cite{mixing_piK}.   For the present case of $\pi\eta$ scattering, due to lack of experimental data,  is not immediately clear whether the K-matrix still gives a good description.  Comparison of GLSM (with specific choice of $m[\pi(1300)] =1.38$ GeV and $A_3/A_1=30$)  with the work of Achasov and Shestakov \cite{belledata} displayed in Fig. \ref{modT_GLSM} (left) only shows a similarity in mathematical form with their ``variant 2'' result (dashed-line), i.e. both raise to a maximum, then both fall to a local minimum and then again both rise to their global maximum.   
  We see that with the inclusion of chiral mixing, which naturally brings into the picture  the heavier $a_0(1450)$, the functional similarity with  ``variant 2'' of Ref. \cite{belledata} extends above 1 GeV (as we saw in the single nonet case in Fig. \ref{F_modT01_sn}, this similarity was limited to below 1 GeV). Also shown in Fig. \ref{modT_GLSM} (left), is a comparison with the prediction of Ref. \cite{pieta} within a nonlinear chiral Lagrangian model. The effects of simultaneous variations of $m[\pi(1300)]$ and $A_3/A_1$ (in ranges [1.22 GeV,1.38 GeV] and $[27,30]$, respectively) are given in Fig. \ref{modT_GLSM} (right), 
  which show a reasonable overlap with Ref. \cite{pieta} up to about $0.9$ GeV.
For further comparison,  Fig. \ref{F_phase_shift} (left) gives the $\pi\eta$ phase shift predicted in this work together with  those predicted by \cite{pieta,ChPT_bernard,oller2,belledata} with an overall qualitative agreement.     The sensitivity to variation of $m[\pi(1300)]$ is shown in the middle and right figures. 
The overall disagreement defined by Eq. (\ref{disag}) among different phase shift predictions of \cite{ChPT_bernard,oller2,pieta,belledata}  up to around $1$ GeV is about $75\%$; among the phase shift predictions of \cite{oller2,pieta,belledata} up to around $1.2$ GeV is  about $51\%$; and among the phase shift predictions of \cite{pieta,belledata} is about $33\%$.   To estimate the phase shift uncertainty in GLSM below $1$ GeV, we have looked at the distribution of $\delta _0^1$, its mean and its standard deviation at any given energy $\sqrt{s}$ (note that in GLSM there is some sensitivity to $m[\pi(1300)]$ as well as to $A_3/A_1$). The uncertainty reaches to about $12$ degrees or about $ 10\%  $ of our predicted phase shift at $1$ GeV. Mean disagreement of GLSM with the results of \cite{Albaladejo_ph} is about $ 20\% $, while the mean disagreements among different phase shift predictions given by different investigators range from $ 33\% $ to $ 75\% $.  Our uncertainty is not larger than the disagreement on the phase shift by other investigators \cite{ChPT_bernard,oller2,pieta,belledata}.

\subsection{The scattering lenghts}

\begin{figure}[!htbp]
\begin{center}
\epsfxsize = 1.5 cm
 \includegraphics[keepaspectratio=true,scale=0.57]{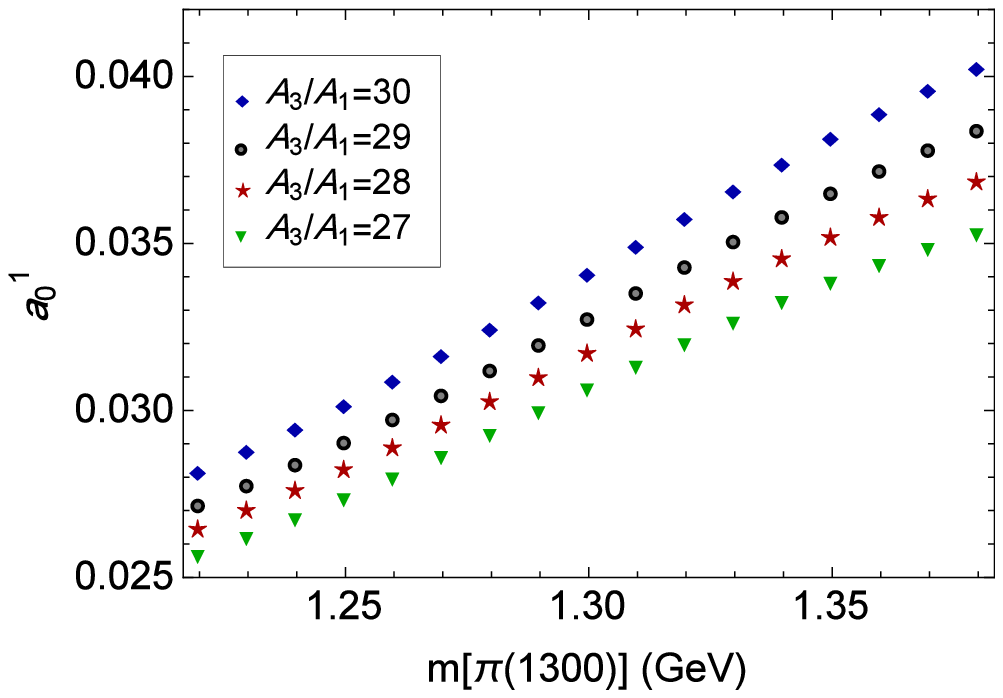}
\hskip .1 cm
\epsfxsize = 1.5 cm
\includegraphics[keepaspectratio=true,scale=0.57]{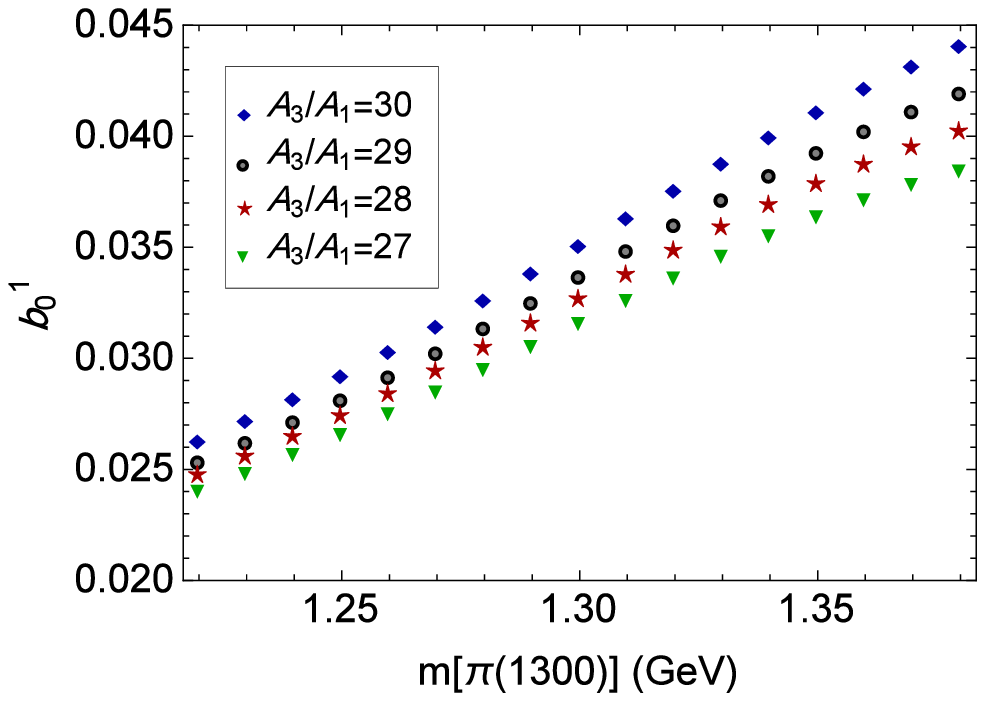}
\hskip .1 cm
\epsfxsize = 1.5 cm
\includegraphics[keepaspectratio=true,scale=0.57]{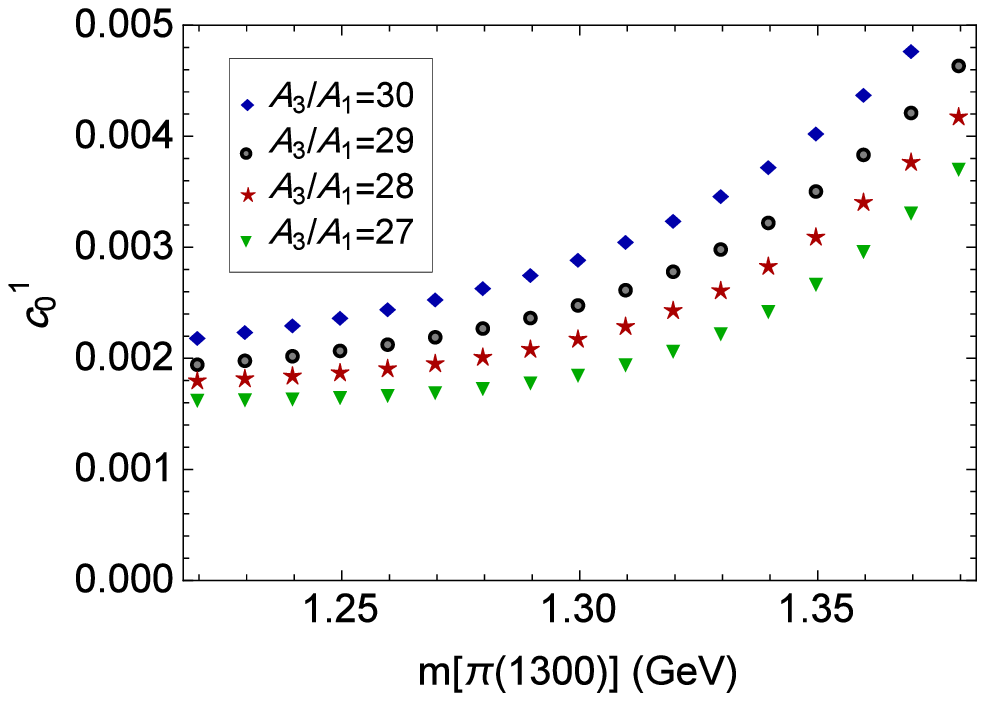}
\caption{The $J=0$, $I=1$, elastic $\pi\eta$  scattering lengths in GLSM for different values of $A_3/A_1$ and $m[\pi(1300)]$.}
\label{glsm_sl_fig}
\end{center}
\end{figure}

Similar to the discussion of scattering lengths in Sec. III,  here we compute these quantities within the GLSM and try to see whether there is any noticeable improvement compared to the single nonet predictions.    The dependency of the results on  $A_3/A_1$ and $m[\pi(1300)]$ are shown in Fig. \ref{glsm_sl_fig} and numerical values are given in Table \ref{sn_sl} and compared with chiral perturbation theory prediction.     As also noted in Sec. III, the effect of unitarization on these quantities is negligible (also see \cite{07_FJS3}).       Even though this computation is rather out of GLSM league,  nevertheless it maybe understandable why other models have a better overall agreement with  SNLSM than with GLSM.     In the latter case, the model aims to reach a wider energy range (by inclusion of chiral mixing with the next to lowest lying scalar and pseudoscalar nonets) and the price it pays is to loose further accuracy near the threshold.

\subsection{The physical poles}

\begin{figure}[!htb]
	\begin{center}
		\vskip 1cm
		\epsfxsize = 7.5cm
		\includegraphics[keepaspectratio=true,scale=0.63]{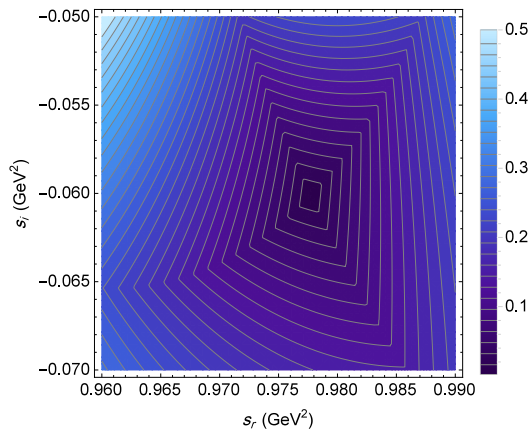}
		\caption{ Contour plot of function ${\cal F} \left(s_r, s_i\right)$ [defined in (\ref{F_srsi})]. The point at which ${\cal F}$=0 represents the first physical isovector scalar meson pole and provides  the properties of the $a_0(980)$.}
		\label{con_plot}
	\end{center}
\end{figure}

\begin{figure}[!htb]
\begin{center}
\vskip 1cm
\epsfxsize = 7.8cm
\includegraphics[keepaspectratio=true,scale=0.51]{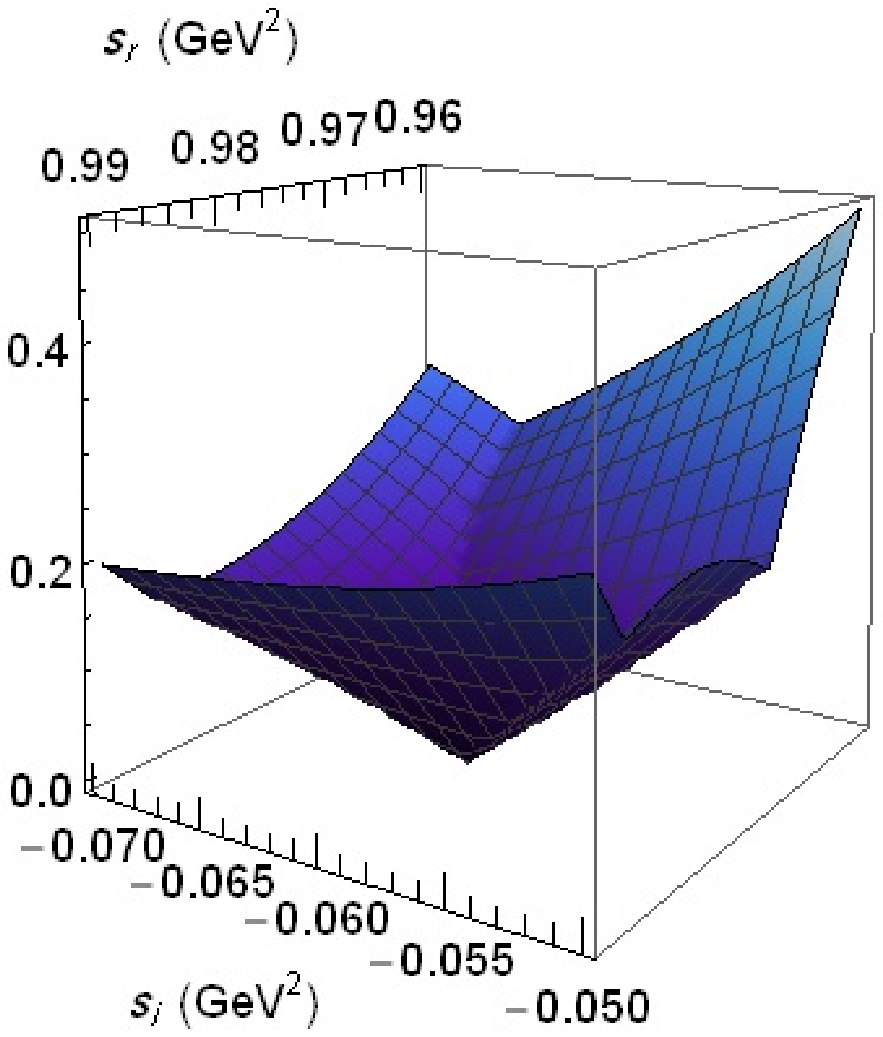}
\hskip 1cm
\epsfxsize = 7.5cm
\includegraphics[keepaspectratio=true,scale=0.44]{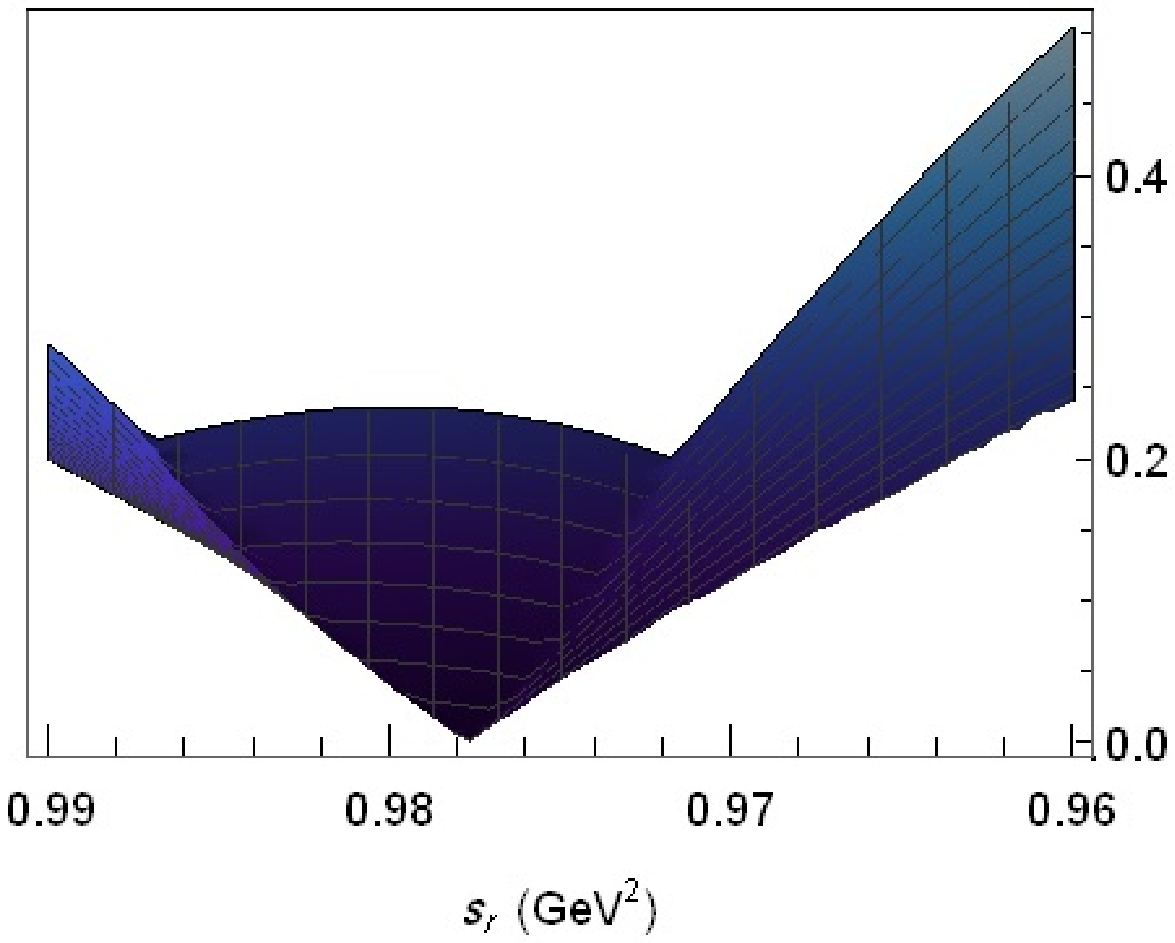}
 \hskip 0.2cm
\epsfxsize = 7.5cm
\includegraphics[keepaspectratio=true,scale=0.45]{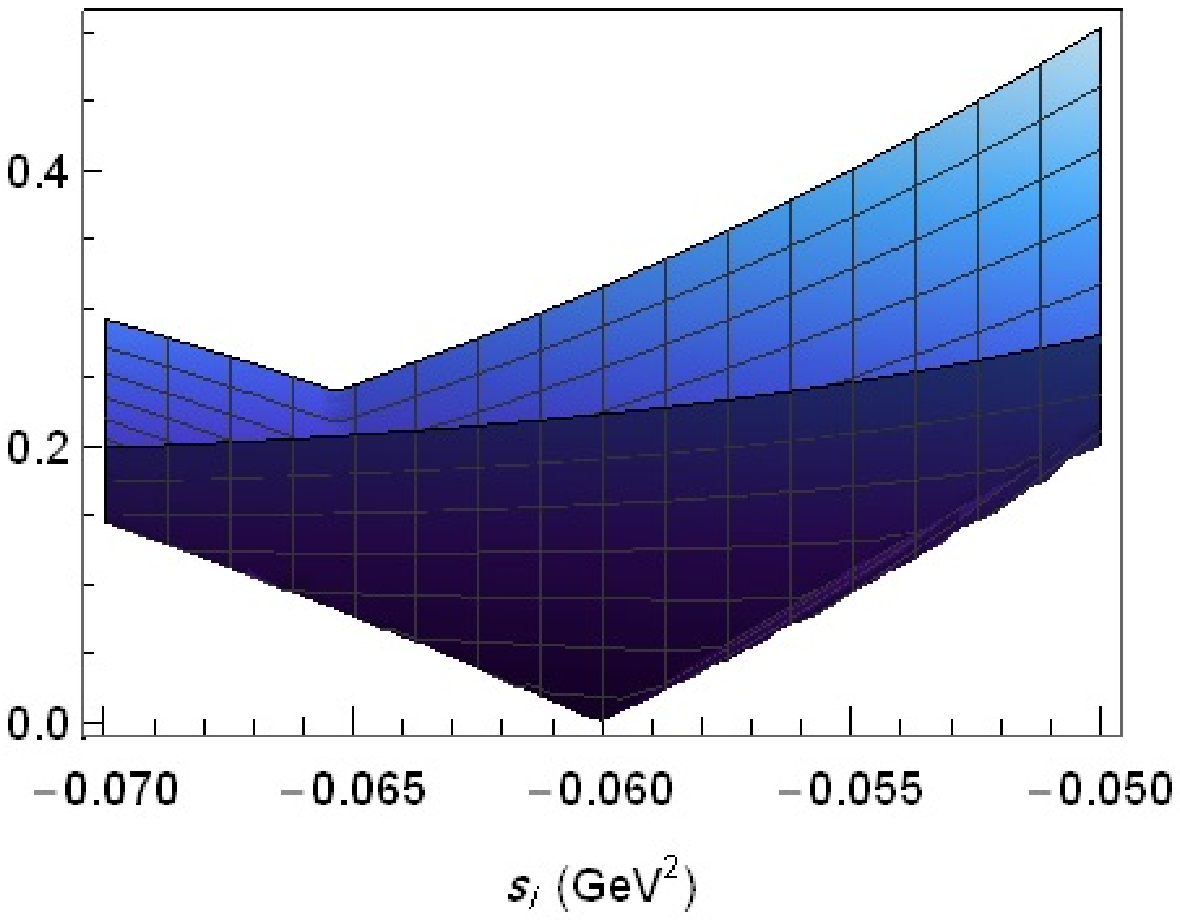}
 \caption{3D plot of function ${\cal F} \left(s_r, s_i\right)$ [defined in (\ref{F_srsi})] over complex plane (left), together with its projections onto  ${\cal F}-s_r$ plane (middle) and  ${\cal F}-s_i$ plane (right).   The point at which ${\cal F}$ touches down represents the first physical isovector scalar meson pole and provides  the properties of the $a_0(980)$.}
 \label{3D_plot}
\end{center}
\end{figure}

As stated in Sec. I, in the case of $\pi\pi$ scattering \cite{mixing_pipi} the first pole in the K-matrix unitarized scattering amplitude clearly captured the properties of light and broad sigma and the second pole resembled the $f_0(980)$.   Similarly, in the case of  $\pi K$ scattering \cite{mixing_piK} the first pole of the K-matrix unitarized amplitude was quite consistent with the properties of light and broad kappa meson.
In the present work too,   we find the pole positions in the complex plane of the
analytically continued expression for $T_0^{1}$.
We examine these physical pole positions by
solving for the complex roots of the
denominator of the  K-matrix unitarized amplitude
Eq. (\ref{T01_unitary}):
\begin{equation}
{\cal D} (s) = 1 - i\, T_0^{1B} = 0,
\label{pole_eq}
\end{equation}
with $T_0^{1B}$ given by Eq. (\ref{pietatem}).   We
search for solutions, $s^{(j)}= s_r^{(j)}+ i s_i^{(j)} = {\widetilde m}_j^2 - i {\widetilde m}_j {\widetilde \Gamma}_j$ of this equation, where
${\widetilde m}_j$ and ${\widetilde \Gamma}_j$ are interpreted as the
mass and decay width of the $j$-th physical
resonance. A
first natural attempt would be to try to  simultaneously solve the two equations:
\begin{eqnarray}
{\rm Re}{\cal D} \left(s_r, s_i\right) = 0, \nonumber \\
{\rm Im}{\cal D} \left(s_r, s_i\right) = 0,
\label{ReandIm}
\end{eqnarray}

for $s_i$ and $s_r$, however, this approach turns out to be rather tedious to be  implemented. A more efficient numerical approach, that was  first pointed out in \cite{mixing_pipi},  is to consider the positive function
\begin{equation}
{\cal F} \left(s_r, s_i\right) =
\left| {\rm Re}
\left(
{\cal D} (s_r, s_i)
\right)\right| +
\left| {\rm Im}
\left(
{\cal D} (s_r, s_i)
\right)\right|,
\label{F_srsi}
\end{equation}
which allows determination of poles by searching for the zeros of this function.  To illustrate the methodology,   Fig. \ref{con_plot} shows the contour plot of ${\cal F} (s_r, s_i)$ over the complex $s$-plane for the specific choice of $m[\pi(1300)]=1.3$ GeV and $A_3/A_1=30$. Also the function
${\cal F} (s_r, s_i)$ is plotted over the complex plane around the first pole in Fig. \ref{3D_plot}.
Clearly, the search of parameter space leads to two solutions for the pole positions which in turn result in the physical masses and decay widths for
the two isotriplets that the model predicts [to be identified with $a_0(980)$ and $a_0(1450)$].    We then zoom in on each pole and study the uncertainties.

\begin{figure}[h]
\begin{center}
\epsfxsize = 1.5cm
\includegraphics[keepaspectratio=true,scale=0.57]{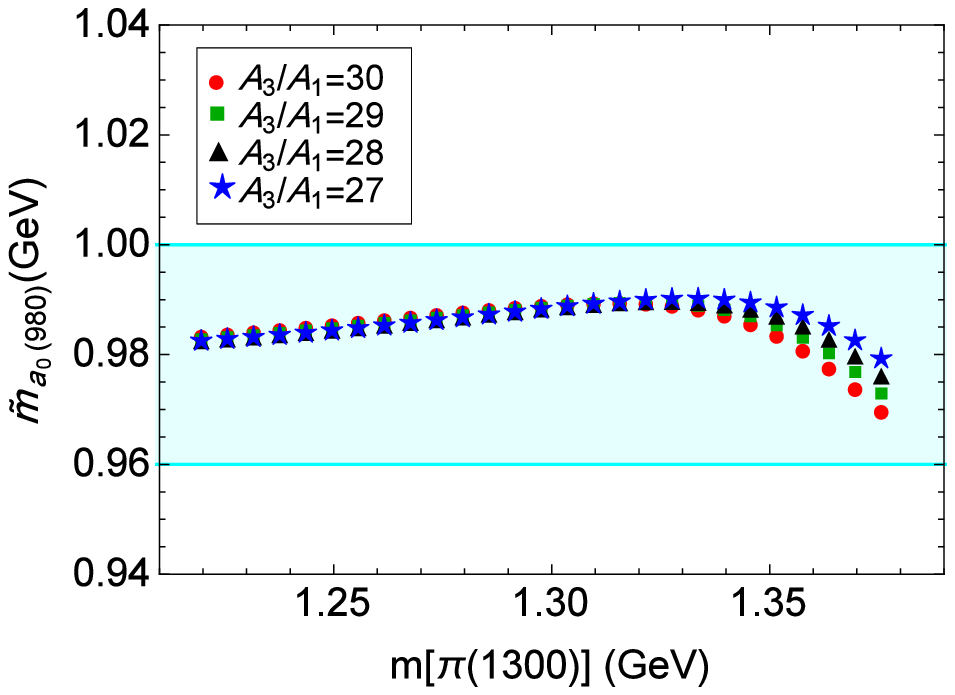}
\hskip .1 cm
\epsfxsize = 1.5cm
\includegraphics[keepaspectratio=true,scale=0.57]{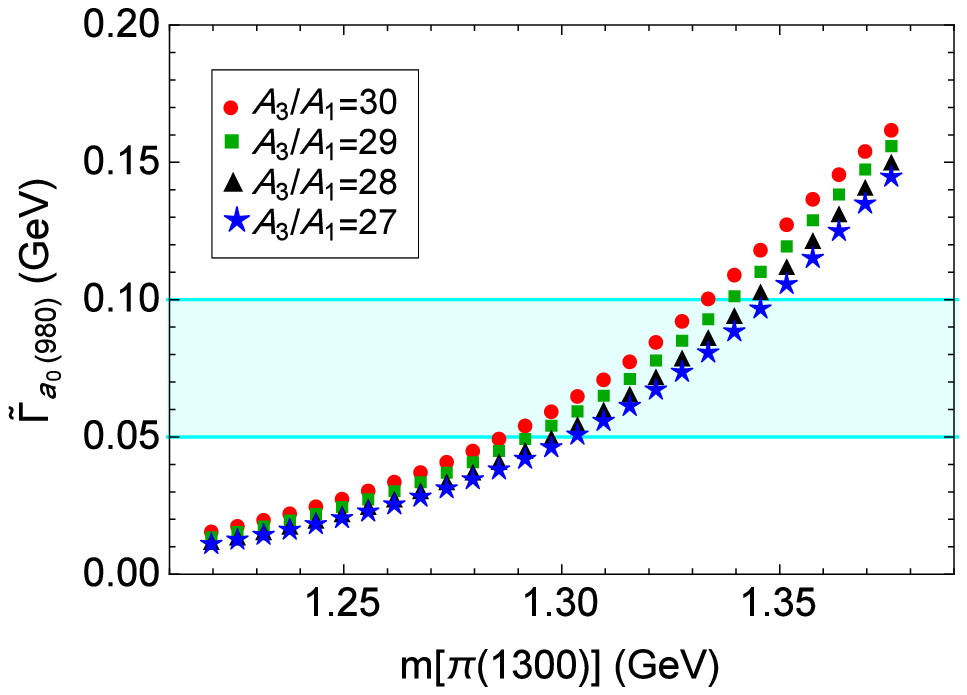}
 \epsfxsize = 1.5cm
 \includegraphics[keepaspectratio=true,scale=0.57]{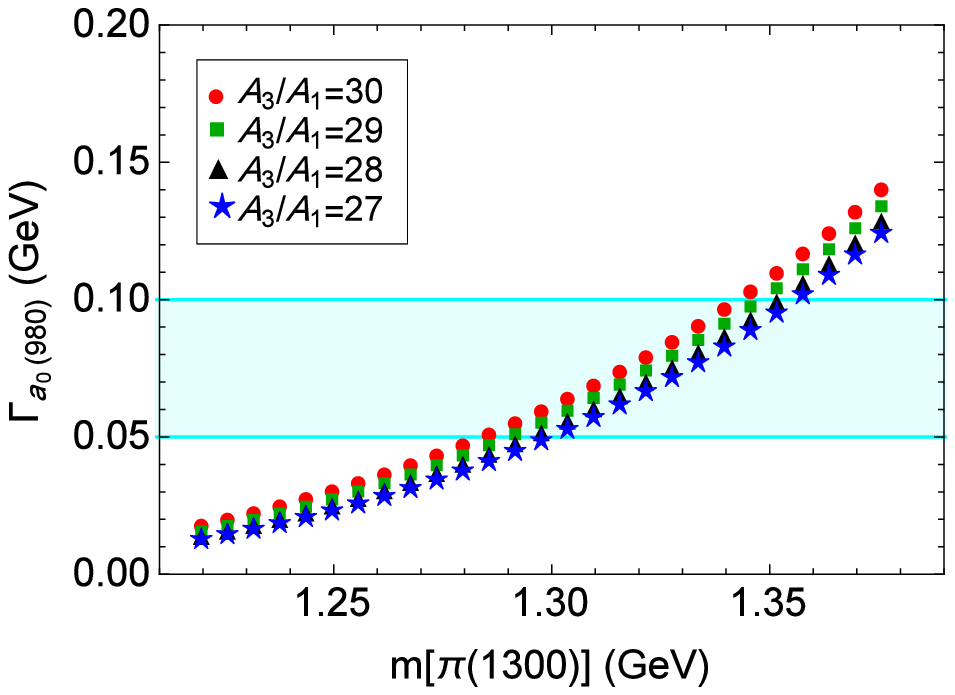}
\caption{Predicted physical mass (left) and decay width (middle) of the $a_0(980)$ resulted from the first pole position in the unitarized $\pi\eta$ elastic scattering amplitude are compared with the corresponding experimental ranges (shaded) reported by PDG \cite{pdg}.  The total decay width  is almost identical to the partial decay width to $\pi\eta$ (right),  which is consistent with the known fact that $\pi\eta$ is the  dominant decay channel of $a_0(980)$.}
\label{F_mass_width_1}
\end{center}
\end{figure}

\begin{figure}[h]
\begin{center}
\epsfxsize = 1.5cm
\includegraphics[keepaspectratio=true,scale=0.65]{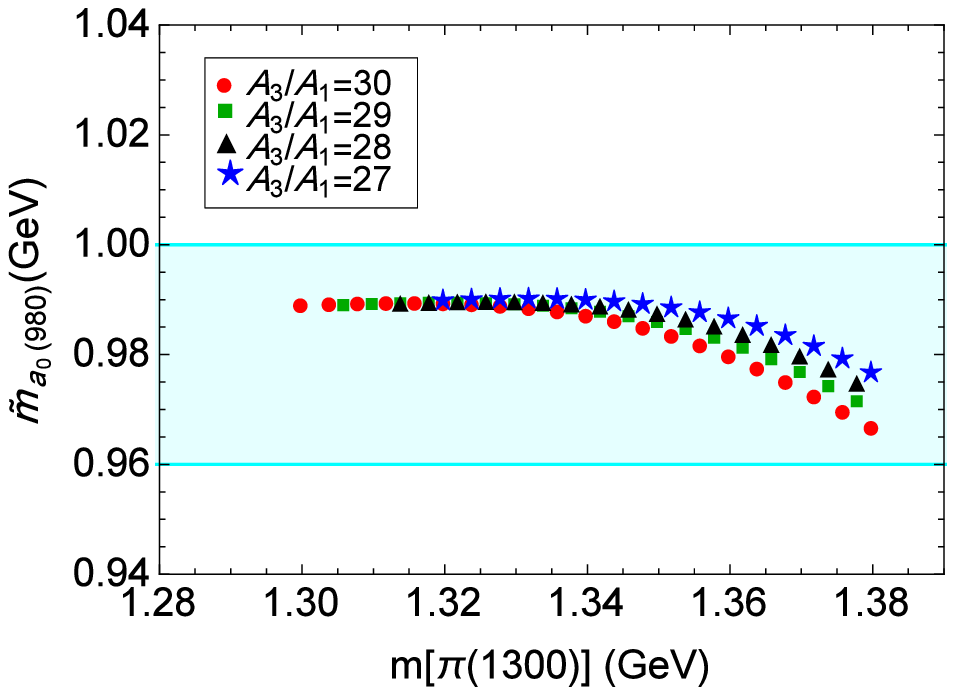}
\hskip .1 cm
\epsfxsize = 1.5cm
\includegraphics[keepaspectratio=true,scale=0.61]{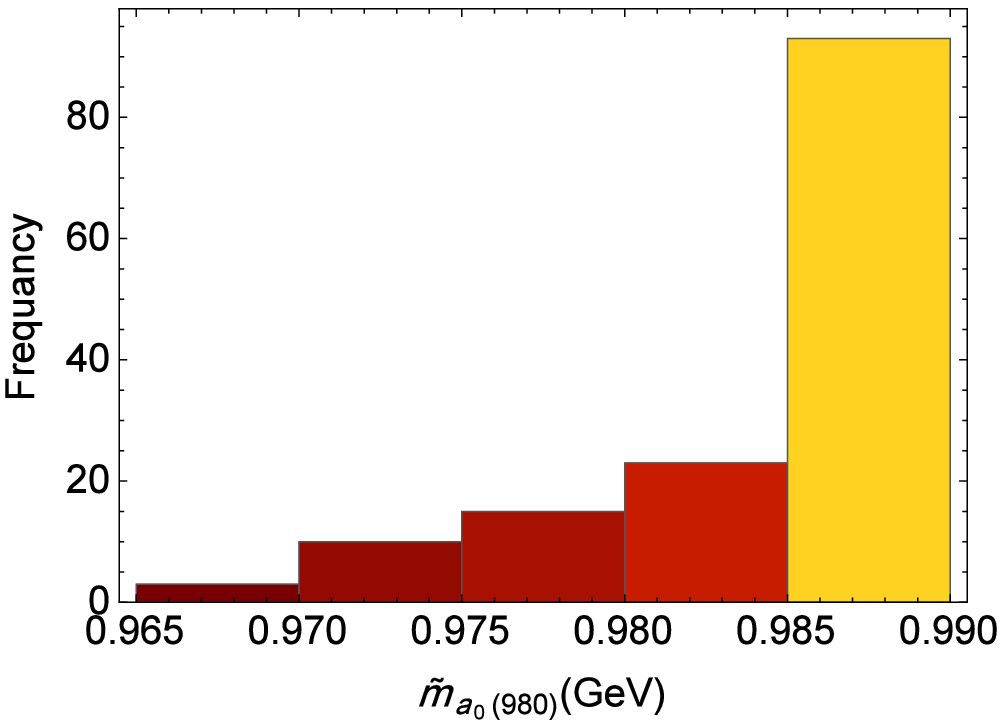}
\vskip 0.1cm
\epsfxsize = 1.5cm
\includegraphics[keepaspectratio=true,scale=0.65]{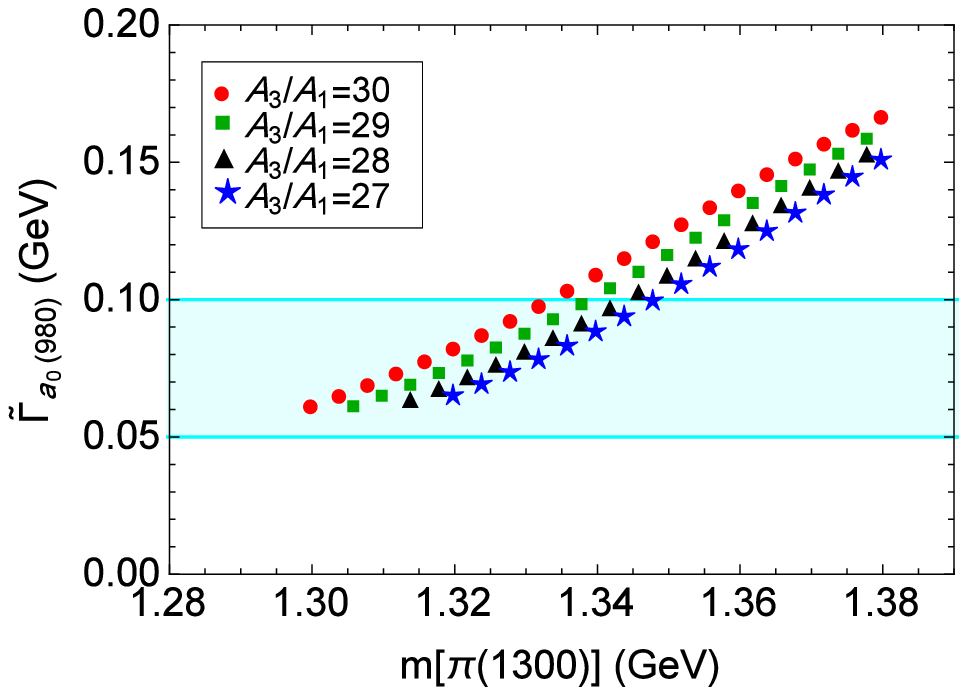}
\hskip .1 cm
\epsfxsize = 1.5cm
\includegraphics[keepaspectratio=true,scale=0.61]{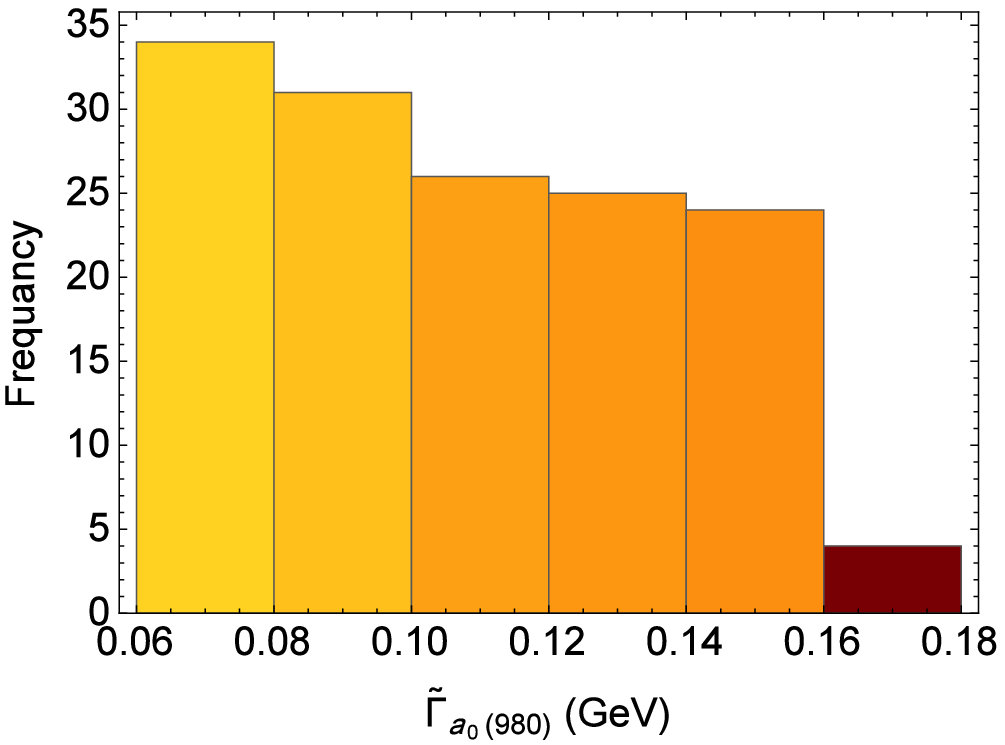}
\caption{The mass (top left) and decay width (bottom left) of the lightest isotriplet scalar state predicted in GLSM over the allowed range of $m[\pi(1300)]$ limited by condition (\ref{DR_positivity}).  The computed physical mass and decay width given in the left figures are organized into histograms in the right figures  representing the distribution of these quantities due to the variation of both $A_3/A_1$ as well as $m[\pi(1300)]$ .	
	}
\label{accepted_range}
\end{center}
\end{figure}

The first pole leads to the prediction of mass and decay width of the lighter isotriplet scalar as  displayed in Fig. \ref{F_mass_width_1} versus $m[\pi(1300)]$ for several values of $A_3/A_1$ (which are the main two experimental inputs in our model with largest uncertainties).  We recall that one of our experimental inputs is the $a_0(980)$ mass ($m[a_0(980)]=980 \pm 20 \: {\rm MeV}$) which is inputed for the bare mass (or Lagrangian mass) of the lighter isovector state in our GLSM.    Since $a_0(980)$ is a narrow state,  its interference  with background in $\pi\eta$ scattering is expected to be small,  and as a result,  the shifts in its mass and width due to the unitarization should be negligible compared to similar effects in $\pi\pi$ and $\pi K$ scatterings where broad states $\sigma$ and $\kappa$ are detected.   Fig. \ref{F_mass_width_1} shows that this is indeed the case and the properties of the first pole is clearly consistent with those of $a_0(980)$.  
While $\pi \eta$ is the dominant decay channel of $a_0(980)$, we expect  
\begin{equation}
 {\widetilde \Gamma}_{a_0(980)} \geq \Gamma[a_0(980)\rightarrow \pi\eta].
\label{DR_positivity}
\end{equation}
This further limits the range of variation of $m[\pi(1300)]$ to $1.3-1.38$ GeV
\footnote{
In addition, we can interpret the difference between the total decay width and partial decay width to $\pi\eta$ as a rough estimate of the partial decay width to $K{\bar K}$:
\[
\Gamma[a_0(980)\rightarrow K{\bar K}] \approx {\tilde \Gamma}_{a_0(980)} - \Gamma[a_0(980)\rightarrow \pi\eta].
\] 
and together with the condition (\ref{DR_positivity})  thereby estimate 
\[
{
	{\Gamma\left[a_0(980)\rightarrow K{\bar K}\right]}
		\over
	{\Gamma\left[a_0(980)\rightarrow \pi\eta\right]}	
	}
	= 0.105 \pm 0.056,
\]
which a value of $0.183 \pm 0.024$ reported by PDG \cite{pdg}. 
}
and in turn limits the predictions for the mass and decay width of $a_0(980)$ from those displayed in Fig. \ref{F_mass_width_1} to those displayed in Fig. \ref{accepted_range} (left) together with their histograms (right).

As a result,  the final predictions for the mass and decay width of  $a_0(980)$ in this work  (average $\pm$ STD) are:
\begin{eqnarray}
 m[a_0(980)] &=& 984 \pm 6 \hskip .125cm {\rm MeV},  \nonumber \\
\Gamma[a_0(980)] &=& 108 \pm 30 \hskip .125cm {\rm MeV},
\end{eqnarray}
to be compared with PDG values \cite{pdg}:
\begin{eqnarray}
m[a_0(980)] &=& 980 \pm 20 \hskip .125cm {\rm MeV} \hskip .5cm (\rm{PDG}), \nonumber \\
\Gamma[a_0(980)] &=& 50 \rightarrow 100 \hskip .125cm {\rm MeV} \hskip .5cm (\rm{PDG}).
\end{eqnarray}
This observation in turn persuades the appropriateness of the simple K-matrix method employed here.     The detection of $a_0(980)$ completes the lightest  nonet of scalar mesons below 1 GeV  predicted in the present order of the GLSM  (the cases of $\sigma$ and $\kappa$ were presented in Refs. \cite{mixing_pipi} and \cite{mixing_piK}).       This further reinforces the importance of the chiral mixing that underlies the properties of the scalar mesons according to which scalars below and above 1 GeV have considerable underlying mixings with those below 1 GeV being dominantly of four-quark nature and those above 1 GeV being closer to quark-antiquark states.

\begin{figure}[h]
\begin{center}
\epsfxsize = 6.5 cm
  \includegraphics[keepaspectratio=true,scale=0.57]{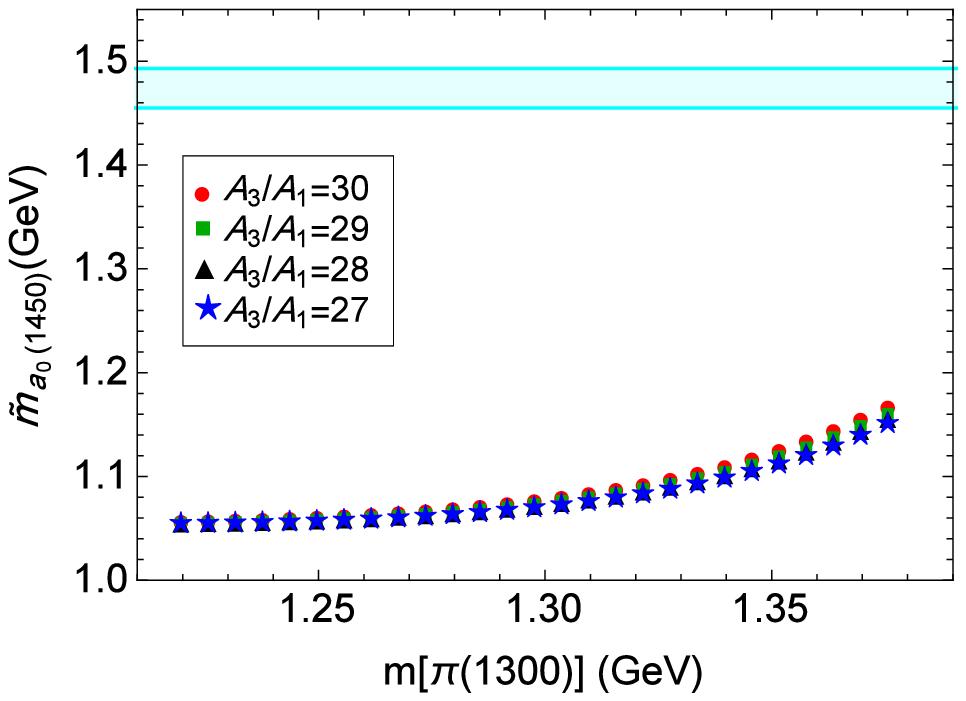}
  \hskip .1 cm
\epsfxsize = 6.5 cm
  \includegraphics[keepaspectratio=true,scale=0.57]{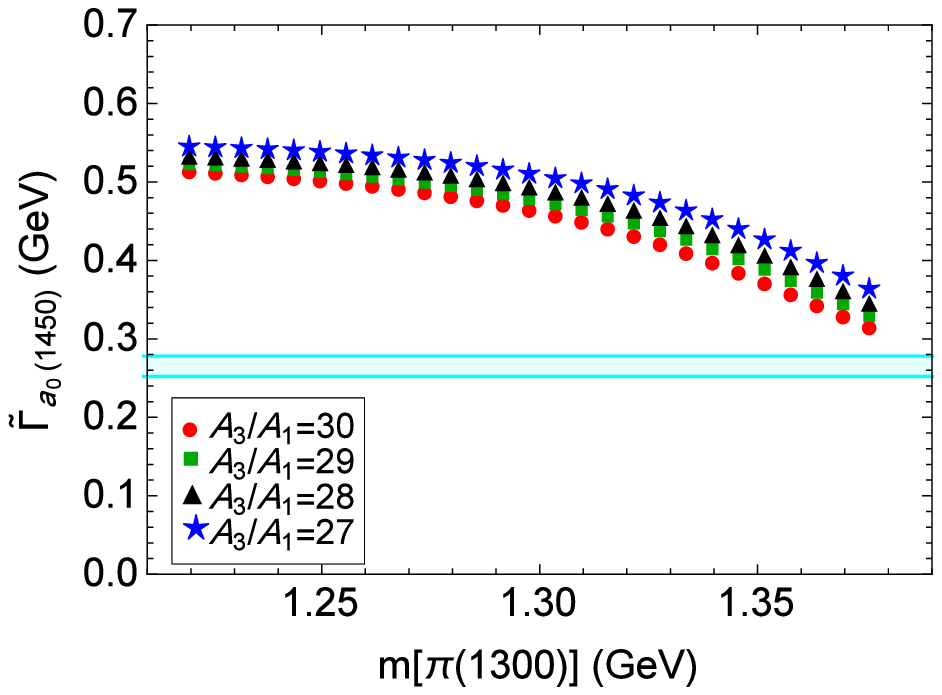}
\epsfxsize = 6.5 cm
  \includegraphics[keepaspectratio=true,scale=0.57]{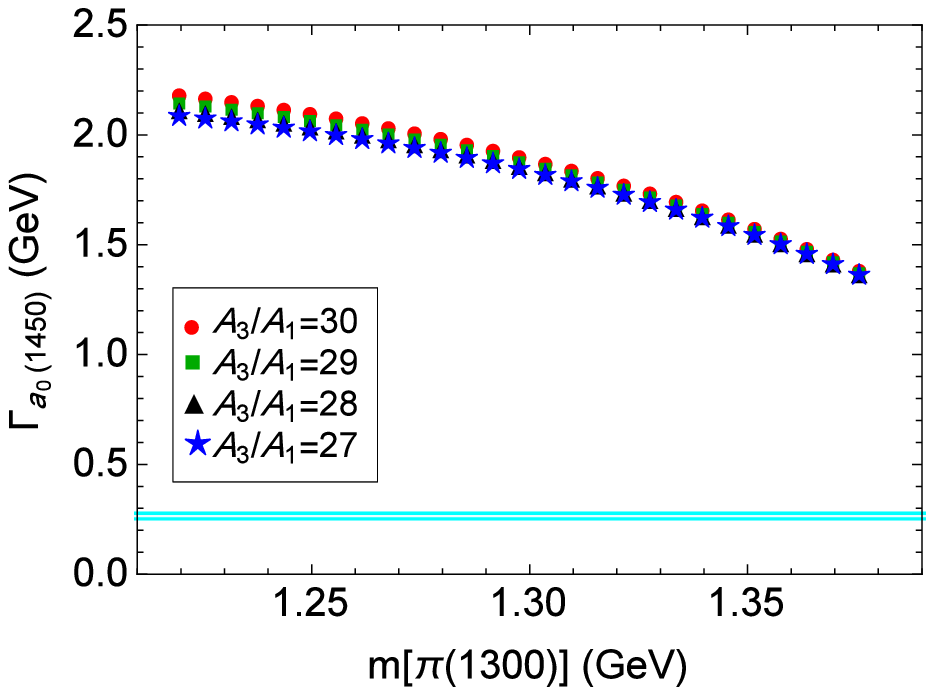}
\caption{Physical mass (left) and decay width (middle) of the second isovector scalar meson extracted from the second pole of
the unitarized $\pi\eta$ elastic scattering amplitude are compared with the corresponding experimental ranges (shaded) for $a_0(1450)$ reported by PDG \cite{pdg}.	
While unitarity corrections drop the mass of $a_0(1450)$  below its experimental range, they considerably improve the prediction of its decay width from an unphysical range (right) to a range that can get close to the experimental bound (middle, shaded). Inclusion of non-elastic channels are expected to improve the predictions.}
\label{F_mass_width_2}
\end{center}
\end{figure}

Although in this work we have studied the elastic $\pi\eta$ scattering in which a complete detection of  the $a_0(1450)$ is not expected to be possible, nevertheless, here we take a closer look at the second pole and try to see if it bears a  resemblance to  the $a_0(1450)$.    Extracting the mass and decay width from the second pole and including the effects of the uncertainties of the experimental inputs used to determine GLSM parameters,  the results are given in Fig. \ref{F_mass_width_2} and  show a large decrease in both mass and decay width of the second (heavier) isotriplet scalar state predicted by the  GLSM.   Again we recall that in this case too, the experimental mass of $a_0(1450)$ was  inputed for the bare mass of the heavier isovector state in our GLSM [see (\ref{inputs1})], but now we see that the unitarization lowers it considerably below the experimental mass of $a_0(1450)$.       The effect of unitarization on the decay width is shown in the same figure, but in this case the unitarization considerably improves the decay width, i.e.   compared to the bare decay width (computed before the unitarization) that was unphysically  large (see Fig. \ref{F_mass_width_2}),  the unitarization has improved the physical decay width and brought it to a comparable order of magnitude of the experimental decay width for the $a_0(1450)$ which is $265 \pm 13$ GeV \cite{pdg}.   Overall, in the elastic channel we can only partially probe the $a_0(1450)$.  For a complete probe of this state a full three coupled channel analysis is needed in which $K{\bar K}$ and ${\pi\eta'}$ channels are also included.   However, as far as our objective of exploring the quark substructure of the scalars below and above 1 GeV within the present order of GLSM is concerned,  the elastic channel provides enough insight since in this model the quark substructure of $a_0(980)$ and $a_0(1450)$ are reciprocal of each other and probing  the $a_0(980)$ in the elastic channel (which is a good approximation despite suffering from some inelasticities near the $K{\bar K}$ threshold) allows probing its substructure (an admixture of quark-antiquark and four-quark with the latter having an edge) which implies, within the context of GLSM, that $a_0(1450)$ is also an admixture of two- and four-quark combinations and that its  quark-antiquark component is favored.   The properties of the physical masses and decay widths presented in this section are further analyzed in Appendix D in which a pole expansion of the K-matrix unitarized scattering amplitude is given.

 \section{Summary and Conclusions}

The global study of scalar mesons below 2 GeV and their underlying two- and four-quark mixing patterns has been the paramount motivation for this work.       The framework for this global study was previously developed in \cite{global} in which an approximation scheme for limiting the (potentially very large) number of terms in the potential is defined in terms of the number of underlying quark and antiquark fields. When retaining terms in the potential with no more than eight quark and antiquark lines,  the model parameters are found by fits to mass spectrum of several scalar and pseudoscalar states, pion decay constant and the ratio of strange to non-strange quark masses.     Once these parameters are determined the model in turn provides the admixtures of the two- and four-quark components for each of the  members of its two scalar nonets (as well as its two pseudoscalar nonets).   The model shows a significant mixing among these components and favors larger four-quark components for the scalars below 1 GeV and larger two quark components for the scalars above 1 GeV (and a reverse situation for pseudoscalars below and above 1 GeV).    These predictions, while consistent with other investigations in the literature, need to be further tested and their robustness examined.   For this purpose,  the model predictions for other low-energy processes (that have not been used in the determination of the model parameters), have to be evaluated.     A delicate and important effect that can measure the effectiveness of the model is its predictions for the final-state interactions of Goldstone bosons.  Several prior works within the GLSM of ref. \cite{global} have examined the model predictions for the final-state interactions in $\pi\pi$ and $\pi K$ scatterings and $\eta'\rightarrow \eta\pi\pi$ decay.    In studies of $\pi\pi$ scattering the model agrees well with the experimental data up to about 1 GeV and predicts \cite{mixing_pipi} a  broad and light sima meson consistent with the PDG values.   Similarly,  in a recent work \cite{mixing_piK}, the model predictions for the $\pi K$ scattering amplitude showed a good prediction of the data as well as prediction of a light and broad kappa meson consistent with the PDG values.    The model also well predicts the experimental data on $\eta'\rightarrow\eta\pi\pi$ decay \cite{LsM_mmp_eta3p} in which the effects of the final-state interaction of pions are known to be important (see Sec. I).

In order to complete the probe of scalar mesons in Goldstone boson interactions,  in this work we applied the same order of the GLSM (with the same parameters fixed in \cite{global}) to predict the properties of $a_0(980)$ probed in $\pi\eta$ scattering.  Lack of experimental data does not allow testing  the model predictions for the scattering amplitude but the results showed a qualitative agreement with the recent work of Achasov and Shestakov in which they had examined the
the $\pi^0 \eta$ rescattering effects  in
$\gamma\gamma\rightarrow \pi^0 \eta$ data of Belle Collaboration \cite{belledata}.   More importantly,  when the $\pi\eta$ scattering amplitude was unitarized by K-matrix method and its poles were computed, it was shown that the lowest pole corresponds to an isovector scalar state with mass
 984 $\pm$ 6 MeV and decay width 108 $\pm$ 30 MeV, which is clearly in close agreement with the properties of $a_0(980)$ given by PDG \cite{pdg}.   The effects of the final-state interactions in the elastic $\pi\eta$ channel is not significant.    Therefore,   the GLSM provides a fairly coherent picture for the Goldstone boson interactions below 1 GeV in which the properties of the lowest scalar meson nonet is probed.  In this picture, the lowest lying scalar meson nonet is dominantly of four-quark type.

 The properties of the second isovector scalar meson  in this model (to be identified with the $a_0(1450)$) was also studied in this work within the elastic $\pi\eta$ channel.    It was shown that while K-matrix unitarization results is improving the overall properties of this second pole,  its identification with the $a_0(1450)$ requires the inelastic effects due to the opening of the $K{\bar K}$ and $\pi\eta'$ channels.
 \par
\begin{figure}[!htbp]
 	\begin{center}
 		\epsfxsize = 4 cm
 		\includegraphics[keepaspectratio=true,scale=0.62]{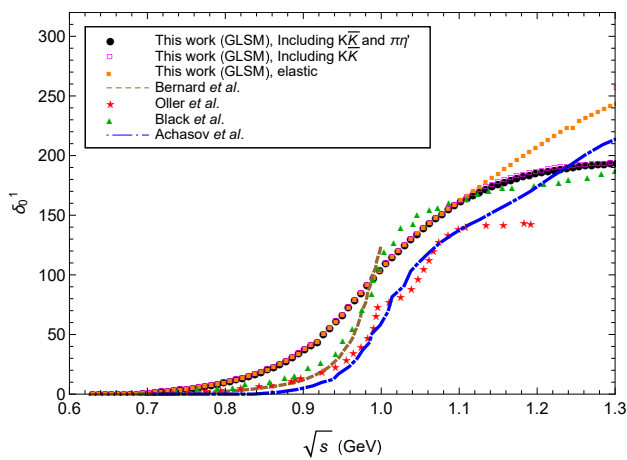}
 		\caption{The effect of the inelastic channels on the S-wave $\pi \eta$ scattering phase shift is compared with the phase shift computed in the elastic channel (with the specific inputs $A_3/A_1$=30 and $m[\pi(1300)]$=1.38 GeV),  as well as with the predictions by Bernard et al \cite{ChPT_bernard},  Oller et al \cite{oller2},  Black et al \cite{pieta} and Achasov et al  \cite{belledata}.  Including the inelastic channels moves the phase shift closer to the predictions of \cite{pieta,belledata,oller2} above $1.1$ GeV. }
  	\label{coupled_channel}
 	\end{center}
 \end{figure}
\begin{figure}[!htbp]
 	\begin{center}
 		\epsfxsize = 4 cm
 		\includegraphics[keepaspectratio=true,scale=0.62]{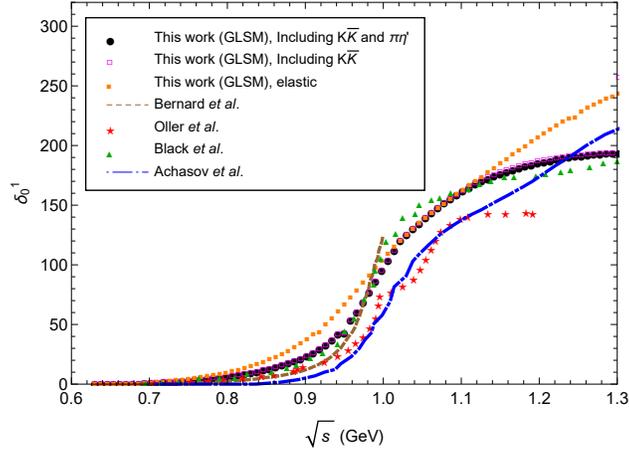}
 		\caption{The prediction of the model for the S-wave $\pi\eta$ phase shift using the  Flatt{\'e} parameterization (with the specific inputs $A_3/A_1$=30 and $m[\pi(1300)]$=1.38 GeV), is 
 	compared with predictions of \cite{ChPT_bernard},  \cite{oller2},  \cite{pieta} and \cite{belledata}.  Below the $K {\bar K}$ threshold, the Flatt{\'e} parameterization moves  the phase shift closer to the predictions of other models. }
 	 	\label{coupled_channel_flatte}
 	\end{center}
 \end{figure}

 Although the results presented in this work were for the elastic case, we have also done a preliminary  studies of the inelastic channels $K{\bar K}$ and $\pi\eta'$ and their effects on the $\pi \eta$ scattering amplitude, phase shift and the properties of $a_0(1450)$ which lies in the inelastic region. The full details are beyond the scope of this paper and will be presented in a follow up work.   As a comparison with our elastic results,  here we give the S-wave  phase shift in two coupled channel analysis of $\pi\eta$ and $K{\bar K}$ as well as a three coupled channel analysis of $\pi\eta$, $K{\bar K}$  and $\pi\eta'$. In Fig. \ref{coupled_channel} we see the effect of the inelastic channels on the S-wave $\pi \eta$ scattering phase shift compared with our elastic result,  and with other predictions  \cite{pieta,ChPT_bernard,belledata,oller2}.  It can be seen that inclusion of  the inelastic channels brings  the phase shift  closer to the predictions of \cite{pieta,belledata,oller2} above $1.1$ GeV. We find that this effect is driven by the $K{\bar K}$ channel, and that the  inclusion of the $\pi\eta'$ channel improves the properties of $a_0(980)$ and $a_0(1450)$.  
Moreover, we have also examined the effect of the  Flatt{\'e} parameterization \cite{flatte} on the K-matrix unitarized scattering amplitude 
by analytically continuing the center of mass momentum to the unphysical region below the threshold. This results in Fig. \ref{coupled_channel_flatte} in which we see that  the prediction of the model for the S-wave $\pi\eta$ scattering phase shift  gets  closer to the predictions of \cite{pieta,ChPT_bernard,belledata,oller2}.
 
There are several directions for future studies.   A full $3\times 3$ coupled channel scattering analysis of $\pi\eta$, $K{\bar K}$  and $\pi\eta'$ will further improve the probe of $a_0(1450)$.   Also it is important to examine the effects of terms beyond the present order of GLSM.   Our preliminary study  shows that these effects are not too significant but are likely to improve  the overall predictions.  It is also relevant to study the effects of glueballs in this framework (see \cite{Fariborz:2021gtc}-\cite{Fariborz:2018xxq} for some of the prior works) and particularly probe the mixing of pseudoscalar gluball with the eta system and determine what roles such mixings may play in $\pi\eta$ scattering.

\section*{Acknowledgments}
\vskip -.5cm
A.H.F. wishes to thank the Physics Dept. of Shiraz University for its hospitality in summer of 2012 where this work was initiated.
S.Z. gratefully acknowledges the support of the University of Sistan and Baluchestan Research Council.

\appendix
\section{Coupling Constants in the Single-Nonet Model}
The rotation matrices are
\begin{equation}
\left[
\begin{array}{c}
\pi^0 \\
\eta\\
\eta'
\end{array}
\right] =
R_\phi(\theta_p)
\left[
\begin{array}{c}
\phi_1^1 \\
\phi_2^2\\
\phi_3^3
\end{array}
\right]
=
\left[
\begin{array}{ccc}
{1\over \sqrt{2}} & -{1\over \sqrt{2}}      &   0\\
{a_p\over \sqrt{2}} & {a_p\over \sqrt{2}}  &  -b_p \\
{b_p\over \sqrt{2}} & {b_p\over \sqrt{2}}  &  a_p
\end{array}
\right]
\left[
\begin{array}{c}
\phi_1^1 \\
\phi_2^2\\
\phi_3^3
\end{array}
\right],
\label{Rp}
\end{equation}
with $a_p = ({{\rm cos} \theta_p - \sqrt{2} {\rm sin} \theta_p})/ {\sqrt{3}}$,
$b_p = ({\rm sin} \theta_p + \sqrt{2} {\rm cos} \theta_p) / {\sqrt{3}}$ and $\theta_p$ is the
pseudoscalar (octet-singlet) mixing angle.  Similarly,
 \begin{equation}
\left[
\begin{array}{c}
a_0^0 \\
\sigma\\
f_0
\end{array}
\right] =
R_s(\theta_s)
\left[
\begin{array}{c}
S_1^1 \\
S_2^2\\
S_3^3
\end{array}
\right]
=
\left[
\begin{array}{ccc}
{1\over \sqrt{2}} & -{1\over \sqrt{2}}      &   0\\
{a_s\over \sqrt{2}} & {a_s\over \sqrt{2}}  &  -b_s \\
{b_s\over \sqrt{2}} & {b_s\over \sqrt{2}}  &  a_s
\end{array}
\right]
\left[
\begin{array}{c}
S_1^1 \\
S_2^2\\
S_3^3
\end{array}
\right],
\label{Rs}
\end{equation}
with $a_s = ({\rm cos} \theta_s - \sqrt{2} {\rm sin} \theta_s)/\sqrt{3}$, $b_s = ({\rm sin} \theta_s + \sqrt{2}{\rm cos} \theta_s)
  / \sqrt{3}$ and $\theta_s$ is the
scalar (octet-singlet) mixing angle.
\\
The coupling constants are:
\begin{eqnarray}\label{coup}
 \gamma_{a_0\pi\eta} &=& \frac{\sqrt{2}}{F_\pi}a_p \left( m^2_{\rm BARE} (a_0) - m_\eta^2 \right),\nonumber\\
\gamma_{\sigma\pi\pi} &=& \frac{1}{F_\pi}a_s \left( m^2_{\rm  BARE} (\sigma) - m_\pi^2\right), \nonumber\\
\gamma_{f_0\pi\pi} &=& \frac{1 }{F_\pi}b_s\left( m^2_{\rm BARE} (f_0) - m_\pi^2 \right), \nonumber\\
\gamma_{\sigma\eta\eta} &=& \frac{a_s}{2\sqrt{2}}X - \frac{b_s}{2} Y,\nonumber\\
\gamma_{f_0\eta\eta} &=&
\frac{b_s}{2\sqrt{2}} X +\frac{a_s}{2} Y,
\label{trilinear-couplings}
\end{eqnarray}

where
\begin{eqnarray}
X = {\left(\frac{a_p}{\sqrt{2}} \right)}^2 \frac{2}{F_\pi} \left[ 2 a_s^2
m^2_{\rm  BARE} (\sigma) + 2 {b_s^2}m^2_{\rm  BARE}
(f_0) - m_\pi^2 - a_p^2 m_\eta^2 - b_p^2m_{\eta^\prime}^2 -
12 (2F_K - F_\pi) V_4 \right]  \nonumber \\
 + {b_p}^2 \frac{2}{2F_K - F_\pi} \left[ - \sqrt{2} a_s
b_s \left( m^2_{\rm  BARE} (\sigma) - m^2_{\rm  BARE} (f_0) \right) - 12 F_\pi V_4 \right]  +
\frac{48}{\sqrt{2}} a_p b_p V_4,  \nonumber\\
\end{eqnarray}
\begin{eqnarray}
Y= {\left(\frac{a_p}{\sqrt{2}} \right)}^2 \frac{2}{ F_\pi} \left[ -\sqrt{2}a_s
b_s \left( m^2_{\rm  BARE} (\sigma) - m^2_{\rm  BARE}
(f_0) \right) - 24 F_\pi V_4 \right] \nonumber \\
 + b_p^2
\frac{2}{2F_K - F_\pi} \left[ b_s^2 m^2_{\rm  BARE} (\sigma) +
a_s^2 m^2_{\rm  BARE} (f_0)  - b_p^2
m_\eta^2 - a_p^2 m_{f_0}^2 \right],
\end{eqnarray}
and
\begin{eqnarray}
\gamma_{\pi\eta}^{\left( 4 \right)}&=& \frac{4}{F_\pi} \left
[ \frac{a_s}{\sqrt{2}}\gamma_{\sigma\eta\eta}+
\frac{b_s}{\sqrt{2}}\gamma_{f_0\eta\eta}+\frac{a^2_p}{F_\pi}
\left(m^2_{\rm  BARE}(a_0)
- {m_\eta}^2 \right) \right].
\end{eqnarray}

\section{``Bare'' three- and four-point coupling constants}

\begin{eqnarray}
\left\langle\frac{\partial^{3}V}{\partial f_a\partial(\phi_{1}^{2})_{1}\partial(\phi_{2}^{1})_{1}}\right\rangle &=&
4 \sqrt{2} c_4^a\alpha_1,\\ \nonumber\\
\left\langle\frac{\partial^{3}V}{\partial f_b\partial(\phi_{1}^{2})_{1}\partial(\phi_{2}^{1})_{2}}\right\rangle &=&
\left\langle\frac{\partial^{3}V}{\partial f_b\partial(\phi_{1}^{2})_{2}\partial(\phi_{2}^{1})_{1}}\right\rangle=
\left\langle\frac{\partial^{3}V}{\partial f_d\partial(\phi_{1}^{2})_{1}\partial(\phi_{2}^{1})_{1}}\right\rangle=
4 e_3^a,\\ \nonumber\\
\left\langle\frac{\partial^{3}V}{\partial (S_{1}^{2})_{1}\partial(\phi_{2}^{1})_{1}\partial \eta_{a}}\right\rangle &=&
\frac{4 \sqrt{2}\Big (2 c_4^a \alpha _1^5 \beta _1+c_4^a \alpha _1^4 \alpha _3 \beta _3
+2 c_3 \alpha _3 \beta _3 \gamma
_1^2+2 c_3 \alpha _1 \beta _1 \gamma _1 (1+\gamma _1)\Big)}{\alpha _1^3 (2 \alpha _1 \beta _1+\alpha _3 \beta_3)},\\ \nonumber\\
\left\langle\frac{\partial^{3}V}{\partial (S_{1}^{2})_{1}\partial(\phi_{2}^{1})_{2}\partial \eta_{a}}\right\rangle&=&
-\frac{8 \sqrt{2} c_3 \left(-1+\gamma _1\right) \Big (\alpha _3 \beta _3 \gamma _1+\alpha _1 \beta _1 \left(1+\gamma _1\right)\Big )}{\alpha
_1 \left(2 \alpha _1 \beta _1+\alpha _3 \beta _3\right){}^2},\\ \nonumber\\
\left\langle\frac{\partial^{3}V}{\partial (S_{1}^{2})_{2}\partial(\phi_{2}^{1})_{1}\partial \eta_{a}}\right\rangle&=&
\frac{8 \sqrt{2} c_3\left(-1+\gamma _1\right) \Big (\alpha _3 \beta _3 \gamma _1+\alpha _1 \beta _1 \left(1+\gamma _1\right)\Big )}{\alpha
_1 \left(2 \alpha _1 \beta _1+\alpha _3 \beta _3\right){}^2},\\ \nonumber\\
\left\langle\frac{\partial^{3}V}{\partial (S_{1}^{2})_{1}\partial(\phi_{2}^{1})_{1}\partial \eta_{b}}\right\rangle&=&
\frac{8 c_3 \gamma _1 \left(\alpha _3 \beta _3+2 \alpha _1 \beta _1 \gamma _1\right)}{\alpha _1^2 \alpha _3 \left(2 \alpha _1
\beta _1+\alpha _3 \beta _3\right)},\\ \nonumber\\
\left\langle\frac{\partial^{3}V}{\partial (S_{1}^{2})_{1}\partial(\phi_{2}^{1})_{2}\partial \eta_{b}}\right\rangle&=&
4 e_3^a-\frac{8 c_3 \left(-1+\gamma _1\right) \left(\alpha _3 \beta _3+2 \alpha _1 \beta _1 \gamma _1\right)}{\alpha _3 \left(2
\alpha _1 \beta _1+\alpha _3 \beta _3\right){}^2},\\ \nonumber\\
\left\langle\frac{\partial^{3}V}{\partial (S_{1}^{2})_{2}\partial(\phi_{2}^{1})_{1}\partial \eta_{b}}\right\rangle&=&
4 e_3^a+\frac{8 c_3 \left(-1+\gamma _1\right) \left(\alpha _3 \beta _3+2 \alpha _1 \beta _1 \gamma _1\right)}{\alpha _3 \left(2
\alpha _1 \beta _1+\alpha _3 \beta _3\right){}^2},\\ \nonumber\\
\left\langle\frac{\partial^{3}V}{\partial (S_{1}^{2})_{1}\partial(\phi_{2}^{1})_{1}\partial \eta_{c}}\right\rangle&=&
\frac{8 \sqrt{2} c_3 \left(-1+\gamma _1\right) \gamma _1}{\alpha _1 \left(2 \alpha _1 \beta _1+\alpha _3 \beta _3\right)},\\ \nonumber\\
\left\langle\frac{\partial^{3}V}{\partial (S_{1}^{2})_{1}\partial(\phi_{2}^{1})_{2}\partial \eta_{c}}\right\rangle&=&
-\frac{8 \sqrt{2} c_3 \alpha _1 \left(-1+\gamma _1\right){}^2}{\left(2 \alpha _1 \beta _1+\alpha _3 \beta _3\right){}^2},\\ \nonumber\\
\left\langle\frac{\partial^{3}V}{\partial (S_{1}^{2})_{2}\partial(\phi_{2}^{1})_{1}\partial \eta_{c}}\right\rangle&=&
\frac{8 \sqrt{2} c_3 \alpha _1 \left(-1+\gamma _1\right){}^2}{\left(2 \alpha _1 \beta _1+\alpha _3 \beta _3\right){}^2},\\ \nonumber\\
\left\langle\frac{\partial^{3}V}{\partial (S_{1}^{2})_{1}\partial(\phi_{2}^{1})_{1}\partial \eta_{d}}\right\rangle&=&
\frac{8 e_3^a \alpha _1^3 \beta _1+4 e_3^a \alpha _1^2 \alpha _3 \beta _3+8 c_3 \alpha _3 \left(-1+\gamma _1\right)
\gamma _1}{\alpha _1^2 \left(2 \alpha _1 \beta _1+\alpha _3 \beta _3\right)},\\ \nonumber\\
\left\langle\frac{\partial^{3}V}{\partial (S_{1}^{2})_{1}\partial(\phi_{2}^{1})_{2}\partial \eta_{d}}\right\rangle&=&
-\frac{8 c_3 \alpha _3 \left(-1+\gamma _1\right){}^2}{\left(2 \alpha _1 \beta _1+\alpha _3 \beta _3\right){}^2},\\ \nonumber\\
\left\langle\frac{\partial^{3}V}{\partial (S_{1}^{2})_{2}\partial(\phi_{2}^{1})_{1}\partial \eta_{d}}\right\rangle&=&
\frac{8 c_3 \alpha _3 \left(-1+\gamma _1\right){}^2}{\left(2 \alpha _1 \beta _1+\alpha _3 \beta _3\right){}^2},\\ \nonumber\\
\left\langle\frac{\partial^{3}V}{\partial f_{a}\partial  \eta_ a \partial\eta_ a}\right\rangle &=&
\frac{1}{\alpha _1^3 \left(2 \alpha _1 \beta _1+\alpha _3 \beta _3\right){}^3}4 \sqrt{2} \Big(8 c_4^a \alpha _1^7 \beta _1^3+12
c_4^a \alpha _1^6 \alpha _3 \beta _1^2 \beta _3+6 c_4^a \alpha _1^5 \alpha _3^2 \beta _1 \beta _3^2\nonumber\\
&&+ c_4^a \alpha _1^4 \alpha _3^3\beta _3^3+4 c_3 \alpha _3^3 \beta _3^3 \gamma _1^2+24 c_3 \alpha _1^2 \alpha _3 \beta _1^2 \beta _3 \gamma _1 \left(1+\gamma _1\right)\nonumber\\
&&+8
c_3 \alpha _1^3 \beta _1^3 \left(1+\gamma _1\right){}^2+4 c_3 \alpha _1 \alpha _3^2 \beta _1 \beta _3^2 \gamma _1 \left(1+5 \gamma _1\right)\Big),\\ \nonumber\\
\left\langle\frac{\partial^{3}V}{\partial f_{a}\partial  \eta_ a \partial\eta_ b}\right\rangle &=&
\frac{8 c_3}{\alpha
_1^2 \alpha _3 \left(2 \alpha _1 \beta _1+\alpha _3 \beta _3\right){}^3} \Big(6 \alpha _1 \alpha _3^2 \beta _1 \beta _3^2 \gamma _1+\alpha _3^3 \beta _3^3 \gamma _1+4 \alpha _1^3 \beta _1^3
\gamma _1 \left(1+\gamma _1\right)\nonumber\\
&&+2 \alpha _1^2 \alpha _3 \beta _1^2 \beta _3 \left(2+\gamma _1+3 \gamma _1^2\right)\Big),\\ \nonumber\\
\left\langle\frac{\partial^{3}V}{\partial f_{a}\partial  \eta_ a \partial\eta_ c}\right\rangle &=&
\frac{8 \sqrt{2} c_3 \beta _1 \left(-1+\gamma _1\right) \Big(2 \alpha _1 \beta _1 \left(1+\gamma _1\right)+\alpha _3 \beta _3
\left(-1+3 \gamma _1\right)\Big)}{\left(2 \alpha _1 \beta _1+\alpha _3 \beta _3\right){}^3},\\ \nonumber\\
\left\langle\frac{\partial^{3}V}{\partial f_{a}\partial  \eta_ a \partial\eta_ d}\right\rangle &=&\frac{-4}{\alpha _1^2 \left(2 \alpha _1 \beta _1+\alpha _3 \beta _3\right){}^3}
 \bigg[8 e_3^a \alpha _1^5 \beta _1^3+12 e_3^a \alpha _1^4 \alpha _3 \beta _1^2 \beta _3+6 e_3^a \alpha _1^3
\alpha _3^2 \beta _1 \beta _3^2\nonumber\\
&&-12 c_3 \alpha _1 \alpha _3^2 \beta _1 \beta _3 \left(-1+\gamma _1\right) \gamma _1
-2 c_3 \alpha _3^3
\beta _3^2 \left(-1+\gamma _1\right) \gamma _1\nonumber\\
&&+\alpha _1^2 \alpha _3 \Big(e_3^a \alpha _3^2 \beta _3^3-8 c_3 \beta _1^2 (-1+\gamma
_1^2)\Big)\bigg],
\end{eqnarray}

\begin{eqnarray}
\left\langle\frac{\partial^{3}V}{\partial f_{a}\partial  \eta_ b \partial\eta_ b}\right\rangle &=&
-\frac{16 \sqrt{2} c_3 \beta _1 \beta _3 \left(-1+\gamma _1\right) \left(\alpha _3 \beta _3+2 \alpha _1 \beta _1 \gamma _1\right)}{\alpha
_3 \left(2 \alpha _1 \beta _1+\alpha _3 \beta _3\right){}^3},\\ \nonumber\\
\left\langle\frac{\partial^{3}V}{\partial f_{a}\partial  \eta_ b \partial\eta_ c}\right\rangle &=&\frac{- 4}{\left(2 \alpha _1 \beta _1+\alpha _3 \beta _3\right){}^3}
\bigg[  8 e_3^a\alpha _1^3 \beta _1^3+12 e_3^a \alpha _1^2 \alpha _3 \beta _1^2 \beta _3+\alpha _3 \beta _3^2 \Big(e_3^a
\alpha _3^2 \beta _3+2 c_3 \left(-1+\gamma _1\right)\Big)\nonumber\\
&&+\,2 \alpha _1 \beta _1 \beta _3 \Big(3 e_3^a \alpha _3^2 \beta _3+2 c_3
(1-3 \gamma _1+2 \gamma _1^2)\Big)\bigg] ,\\\nonumber\\
\left\langle\frac{\partial^{3}V}{\partial f_{a}\partial  \eta_ b \partial\eta_ d}\right\rangle &=&
\frac{8 \sqrt{2} c_3 \beta _1 \left(-1+\gamma _1\right) \Big(-\alpha _3 \beta _3 \left(-2+\gamma _1\right)+2 \alpha _1 \beta
_1 \gamma _1\Big)}{\left(2 \alpha _1 \beta _1+\alpha _3 \beta _3\right){}^3},\\ \nonumber\\
\left\langle\frac{\partial^{3}V}{\partial f_{a}\partial  \eta_ c \partial\eta_ c}\right\rangle &=&
-\frac{16 \sqrt{2} c_3 \alpha _1 \alpha _3 \beta _3 \left(-1+\gamma _1\right){}^2}{\left(2 \alpha _1 \beta _1+\alpha _3 \beta
_3\right){}^3},\\\nonumber\\
\left\langle\frac{\partial^{3}V}{\partial f_{a}\partial  \eta_ c \partial\eta_ d}\right\rangle &=&
-\frac{8 c_3 \alpha _3 \left(-2 \alpha _1 \beta _1+\alpha _3 \beta _3\right) \left(-1+\gamma _1\right){}^2}{\left(2 \alpha _1
\beta _1+\alpha _3 \beta _3\right){}^3},\\ \nonumber\\
\left\langle\frac{\partial^{3}V}{\partial f_{a}\partial  \eta_ d \partial\eta_ d}\right\rangle &=&
\frac{16 \sqrt{2} c_3 \alpha _3^2 \beta _1 \left(-1+\gamma _1\right){}^2}{\left(2 \alpha _1 \beta _1+\alpha _3 \beta _3\right){}^3},\\ \nonumber\\
\left\langle\frac{\partial^{3}V}{\partial f_{b}\partial  \eta_ a \partial\eta_ a}\right\rangle &=&
-\frac{32 c_3 \beta _1 \beta _3 \left(-1+\gamma _1\right) \Big(\alpha _3 \beta _3 \gamma _1+\alpha _1 \beta _1 \left(1+\gamma
_1\right)\Big)}{\alpha _1 \left(2 \alpha _1 \beta _1+\alpha _3 \beta _3\right){}^3},\\ \nonumber\\
\left\langle\frac{\partial^{3}V}{\partial f_{b}\partial  \eta_ a \partial\eta_ b}\right\rangle &=&
\frac{1}{\alpha_1 \alpha _3^2 \left(2 \alpha _1 \beta _1+\alpha _3 \beta _3\right){}^3}
\bigg[8 \sqrt{2} c_3 \Big(\alpha _3^3 \beta _3^3 \gamma _1+4 \alpha _1^3 \beta _1^3 \gamma _1 \left(1+\gamma _1\right)\nonumber\\
&&+\,6 \alpha
_1^2 \alpha _3 \beta _1^2 \beta _3 \gamma _1 \left(1+\gamma _1\right)+2 \alpha _1 \alpha _3^2 \beta _1 \beta _3^2 (1+2 \gamma _1^2)\Big)\bigg],\\ \nonumber\\
\left\langle\frac{\partial^{3}V}{\partial f_{b}\partial  \eta_ a \partial\eta_ c}\right\rangle &=&\frac{-4}{\left(2 \alpha _1 \beta _1+\alpha _3 \beta _3\right){}^3}
 \Bigg[8 e_3^a \alpha _1^3 \beta _1^3+12 e_3^a \alpha _1^2 \alpha _3 \beta _1^2 \beta _3
+2 \alpha _1 \beta _1 \beta
_3\nonumber\\
&& \Big(3 e_3^a  \alpha _3^2 \beta _3-4 c_3 \left(-1+\gamma _1\right)\Big)+\,\alpha _3 \beta _3^2 \Big(e_3^a  \alpha _3^2 \beta _3-4
c_3 \left(-1+\gamma _1\right) \gamma _1\Big)\Bigg],\\ \nonumber\\
\left\langle\frac{\partial^{3}V}{\partial f_{b}\partial  \eta_ a \partial\eta_ d}\right\rangle &=&
-\frac{8 \sqrt{2} c_3 \beta _1 \left(-1+\gamma _1\right) \Big(2 \alpha _1 \beta _1 \left(1+\gamma _1\right)+\alpha _3 \beta _3
\left(-1+3 \gamma _1\right)\Big)}{\left(2 \alpha _1 \beta _1+\alpha _3 \beta _3\right){}^3},\\ \nonumber\\
\left\langle\frac{\partial^{3}V}{\partial f_{b}\partial  \eta_ b \partial\eta_ b}\right\rangle &=&\frac{8}{\alpha _3^3 \left(2\alpha _1 \beta _1+\alpha _3 \beta _3\right){}^3}
 \bigg[\alpha _3^3 \left(2 c_3+c_4^a \alpha _3^4\right) \beta _3^3+6 \alpha _1 \alpha _3^2 \beta _1 \beta _3^2 \left(c_4^a
\alpha _3^4+2 c_3 \gamma _1\right)\nonumber\\
&&+\,8 \alpha _1^3 \beta _1^3 \left(c_4^a \alpha _3^4+2 c_3 \gamma _1^2\right)
+4 \alpha _1^2 \alpha _3 \beta _1^2 \beta _3 \Big(3 c_4^a \alpha _3^4
+\,2 c_3 \gamma _1 (1+2 \gamma _1)\Big)\bigg],\\ \nonumber\\
\left\langle\frac{\partial^{3}V}{\partial f_{b}\partial  \eta_ b \partial\eta_ c}\right\rangle &=&
\frac{16 \sqrt{2} c_3 \alpha _1 \left(-1+\gamma _1\right) \left(\alpha _3^2 \beta _3^2+2 \alpha _1^2 \beta _1^2 \gamma _1+3 \alpha
_1 \alpha _3 \beta _1 \beta _3 \gamma _1\right)}{\alpha _3^2 \left(2 \alpha _1 \beta _1+\alpha _3 \beta _3\right){}^3},\\ \nonumber\\
\left\langle\frac{\partial^{3}V}{\partial f_{b}\partial  \eta_ b \partial\eta_ d}\right\rangle &=&
\frac{8 c_3 \beta _3 \left(-1+\gamma _1\right) \Big(\alpha _3 \beta _3+2 \alpha _1 \beta _1 \left(-1+2 \gamma _1\right)\Big)}{\left(2
\alpha _1 \beta _1+\alpha _3 \beta _3\right){}^3},\\ \nonumber\\
\left\langle\frac{\partial^{3}V}{\partial f_{b}\partial  \eta_ c \partial\eta_ c}\right\rangle &=&
\frac{32 c_3 \alpha _1^2 \beta _3 \left(-1+\gamma _1\right){}^2}{\left(2 \alpha _1 \beta _1+\alpha _3 \beta _3\right){}^3},\\ \nonumber\\
\left\langle\frac{\partial^{3}V}{\partial f_{b}\partial  \eta_ c \partial\eta_ d}\right\rangle &=&
-\frac{8 \sqrt{2}c_3\alpha _1 \left(2 \alpha _1 \beta _1-\alpha _3 \beta _3\right) \left(-1+\gamma _1\right){}^2}{\left(2 \alpha
_1 \beta _1+\alpha _3 \beta _3\right){}^3},
\end{eqnarray}

\begin{eqnarray}
\left\langle\frac{\partial^{3}V}{\partial f_{b}\partial  \eta_ d \partial\eta_ d}\right\rangle &=&
-\frac{32 c_3 \alpha _1 \alpha _3 \beta _1 \left(-1+\gamma _1\right){}^2}{\left(2 \alpha _1 \beta _1+\alpha _3 \beta _3\right){}^3},\\ \nonumber\\
\left\langle\frac{\partial^{3}V}{\partial f_{c}\partial  \eta_ a \partial\eta_ a}\right\rangle &=&
\frac{16 \sqrt{2} c_3 \alpha _3 \beta _3 \left(-1+\gamma _1\right) \Big(\alpha _3 \beta _3 \gamma _1+\alpha _1 \beta _1 \left(1+\gamma
_1\right)\Big)}{\alpha _1 \left(2 \alpha _1 \beta _1+\alpha _3 \beta _3\right){}^3},\\ \nonumber\\
\left\langle\frac{\partial^{3}V}{\partial f_{c}\partial  \eta_ a \partial\eta_ b}\right\rangle &=&
-\frac{ 4 \Big(8 e_3^a \alpha _1^3 \beta c_3 \left(-1+\gamma _1\right)\Big)+\alpha _3 \beta _3^2 \Big(e_3^a \alpha _3^2 \beta _3+2
c_3 (1-3 \gamma _1+2 \gamma _1^2)\Big)}
{\left(2 \alpha _1 \beta _1+\alpha _3 \beta _3\right){}^3}
,\\ \nonumber\\
\left\langle\frac{\partial^{3}V}{\partial f_{c}\partial  \eta_ a \partial\eta_ c}\right\rangle &=&
\frac{8 \sqrt{2} c_3 \alpha _1 \left(-1+\gamma _1\right) \Big(2 \alpha _1 \beta _1 (1+\gamma _1)+\alpha _3 \beta _3
\left(-1+3 \gamma _1\right)\Big)}{\left(2 \alpha _1 \beta _1+\alpha _3 \beta _3\right){}^3},\\ \nonumber\\
\left\langle\frac{\partial^{3}V}{\partial f_{c}\partial  \eta_ a \partial\eta_ d}\right\rangle &=&
\frac{8 c_3 \alpha _3 \left(-1+\gamma _1\right) \Big(2 \alpha _1 \beta _1 \left(1+\gamma _1\right)+\alpha _3 \beta _3 \left(-1+3
\gamma _1\right)\Big)}{\left(2 \alpha _1 \beta _1+\alpha _3 \beta _3\right){}^3},\\ \nonumber\\
\left\langle\frac{\partial^{3}V}{\partial f_{c}\partial  \eta_ b \partial\eta_ b}\right\rangle &=&
-\frac{16 \sqrt{2} c_3 \alpha _1 \beta _3 \left(-1+\gamma _1\right) \left(\alpha _3 \beta _3+2 \alpha _1 \beta _1 \gamma _1\right)}{\alpha
_3 \left(2 \alpha _1 \beta _1+\alpha _3 \beta _3\right){}^3},\\ \nonumber\\
\left\langle\frac{\partial^{3}V}{\partial f_{c}\partial  \eta_ b \partial\eta_ c}\right\rangle &=&
\frac{16 c_3 \alpha _1^2 \left(-1+\gamma _1\right) \Big(-\alpha _3 \beta _3 \left(-2+\gamma _1\right)+2 \alpha _1 \beta _1 \gamma
_1\Big)}{\alpha _3 \left(2 \alpha _1 \beta _1+\alpha _3 \beta _3\right){}^3},\\ \nonumber\\
\left\langle\frac{\partial^{3}V}{\partial f_{c}\partial  \eta_ b \partial\eta_ d}\right\rangle &=&
\frac{8 \sqrt{2} c_3 \alpha _1 \left(-1+\gamma _1\right) \Big(-\alpha _3 \beta _3 \left(-2+\gamma _1\right)+2 \alpha _1 \beta
_1 \gamma _1\Big)}{\left(2 \alpha _1 \beta _1+\alpha _3 \beta _3\right){}^3},\\ \nonumber\\
\left\langle\frac{\partial^{3}V}{\partial f_{c}\partial  \eta_ c \partial\eta_ c}\right\rangle &=&
\frac{32 \sqrt{2} c_3 \alpha _1^3 \left(-1+\gamma _1\right){}^2}{\left(2 \alpha _1 \beta _1+\alpha _3 \beta _3\right){}^3},\\ \nonumber\\
\left\langle\frac{\partial^{3}V}{\partial f_{c}\partial  \eta_ c \partial\eta_ d}\right\rangle &=&
\frac{32 c_3 \alpha _1^2 \alpha _3 \left(-1+\gamma _1\right){}^2}{\left(2 \alpha _1 \beta _1+\alpha _3 \beta _3\right){}^3},\\ \nonumber\\
\left\langle\frac{\partial^{3}V}{\partial f_{c}\partial  \eta_ d \partial\eta_ d}\right\rangle &=&
\frac{16 \sqrt{2} c_3 \alpha _1 \alpha _3^2 \left(-1+\gamma _1\right){}^2}{\left(2 \alpha _1 \beta _1+\alpha _3 \beta _3\right){}^3},\\ \nonumber\\
\left\langle\frac{\partial^{3}V}{\partial f_{d}\partial  \eta_ a \partial\eta_ a}\right\rangle &=&\frac{-4}{\alpha _1 \left(2 \alpha _1 \beta _1+\alpha_3 \beta _3\right){}^3}
\bigg[  \bigg(8 e_3^a \alpha _1^4 \beta _1^3+12 e_3^a \alpha _1^3 \alpha _3 \beta _1^2 \beta _3+6 e_3^a \alpha _1^2
\alpha _3^2 \beta _1 \beta _3^2\nonumber \\
&&+8 c_3 \alpha _3^2 \beta _1 \beta _3 \left(-1+\gamma _1\right) \gamma _1
+\alpha _1 \alpha _3 \Big(e_3^a
\alpha _3^2 \beta _3^3+8 c_3 \beta _1^2 (-1+\gamma _1^2)\Big)\bigg)\bigg],\\ \nonumber\\
\left\langle\frac{\partial^{3}V}{\partial f_{d}\partial  \eta_ a \partial\eta_ b}\right\rangle &=&
\frac{8 \sqrt{2} c_3 \beta _1 \left(-1+\gamma _1\right) \Big(2 \alpha _1 \beta _1+\alpha _3 \beta _3 \left(-1+2 \gamma _1\right)\Big)}{\left(2
\alpha _1 \beta _1+\alpha _3 \beta _3\right){}^3},\\ \nonumber\\
\left\langle\frac{\partial^{3}V}{\partial f_{d}\partial  \eta_ a \partial\eta_ c}\right\rangle &=&
\frac{16 c_3 \alpha _3 \left(-1+\gamma _1\right) \left(2 \alpha _1 \beta _1+\alpha _3 \beta _3 \gamma _1\right)}{\left(2 \alpha
_1 \beta _1+\alpha _3 \beta _3\right){}^3},\\ \nonumber\\
\left\langle\frac{\partial^{3}V}{\partial f_{d}\partial  \eta_ a \partial\eta_ d}\right\rangle &=&
\frac{8 \sqrt{2} c_3 \alpha _3^2 \left(-1+\gamma _1\right) \left(2 \alpha _1 \beta _1+\alpha _3 \beta _3 \gamma _1\right)}{\alpha
_1 \left(2 \alpha _1 \beta _1+\alpha _3 \beta _3\right){}^3},\\ \nonumber\\
\left\langle\frac{\partial^{3}V}{\partial f_{d}\partial  \eta_ b \partial\eta_ b}\right\rangle &=&
\frac{32 c_3 \alpha _1 \beta _1 \left(-1+\gamma _1\right) \left(\alpha _3 \beta _3+2 \alpha _1 \beta _1 \gamma _1\right)}{\alpha
_3 \left(2 \alpha _1 \beta _1+\alpha _3 \beta _3\right){}^3},\\ \nonumber\\
\left\langle\frac{\partial^{3}V}{\partial f_{d}\partial  \eta_ b \partial\eta_ c}\right\rangle &=&
\frac{8 \sqrt{2} c_3 \alpha _1 \left(-1+\gamma _1\right) \Big(\alpha _3 \beta _3+2 \alpha _1 \beta _1 \left(-1+2 \gamma _1\right)\Big)}{\left(2
\alpha _1 \beta _1+\alpha _3 \beta _3\right){}^3},\\ \nonumber\\
\left\langle\frac{\partial^{3}V}{\partial f_{d}\partial  \eta_ b \partial\eta_ d}\right\rangle &=&
\frac{8 c_3 \alpha _3 \left(-1+\gamma _1\right) \Big(\alpha _3 \beta _3+2 \alpha _1 \beta _1 \left(-1+2 \gamma _1\right)\Big)}{\left(2
\alpha _1 \beta _1+\alpha _3 \beta _3\right){}^3},\\ \nonumber\\
\left\langle\frac{\partial^{3}V}{\partial f_{d}\partial  \eta_ c \partial\eta_ c}\right\rangle &=&
\frac{32 c_3 \alpha _1^2 \alpha _3 \left(-1+\gamma _1\right){}^2}{\left(2 \alpha _1 \beta _1+\alpha _3 \beta _3\right){}^3},
\end{eqnarray}

\begin{eqnarray}
\left\langle\frac{\partial^{3}V}{\partial f_{d}\partial  \eta_ c \partial\eta_ d}\right\rangle &=&
\frac{16 \sqrt{2} c_3 \alpha _1 \alpha _3^2 \left(-1+\gamma _1\right){}^2}{\left(2 \alpha _1 \beta _1+\alpha _3 \beta _3\right){}^3},\\ \nonumber\\
\left\langle\frac{\partial^{3}V}{\partial f_{d}\partial  \eta_ d \partial\eta_ d}\right\rangle &=&
\frac{16 c_3 \alpha _3^3 \left(-1+\gamma _1\right){}^2}{\left(2 \alpha _1 \beta _1+\alpha _3 \beta _3\right){}^3},\\ \nonumber\\
\left\langle\frac{\partial^{4}V}{\partial\eta_{a}\partial\eta_{a}\partial  (\phi_{1}^{2})_{1}\partial(\phi_{2}^{1})_{1}}\right\rangle &=&\frac{4 \Big(6 c_4^a \alpha _1^5 \beta _1+3 c_4^a\alpha _1^4 \alpha _3 \beta _3+8 c_3\alpha _3 \beta _3 \gamma _1^2+8
c_3 \alpha _1 \beta _1 \gamma _1 \left(1+\gamma _1\right)\Big)}{\alpha _1^4 \left(2 \alpha _1 \beta _1+\alpha _3 \beta _3\right)},\\ \nonumber\\
\left\langle\frac{\partial^{4}V}{\partial\eta_{a}\partial\eta_{a}\partial  (\phi_{1}^{2})_{1}\partial(\phi_{2}^{1})_{2}}\right\rangle
&=&\left\langle\frac{\partial^{4}V}{\partial\eta_{a}\partial\eta_{a}\partial  (\phi_{1}^{2})_{2}\partial(\phi_{2}^{1})_{1}}\right\rangle\nonumber\\
&=&-\frac{32 c_3 \beta _1 \left(-1+\gamma _1\right) \Big(\alpha _3 \beta _3 \gamma _1+\alpha _1 \beta _1 \left(1+\gamma _1\right)\Big)}{\alpha
_1 \left(2 \alpha _1 \beta _1+\alpha _3 \beta _3\right){}^3}
,\\ \nonumber \\
\left\langle\frac{\partial^{4}V}{\partial\eta_{a}\partial\eta_{b}\partial  (\phi_{1}^{2})_{1}\partial(\phi_{2}^{1})_{1}}\right\rangle &=&\frac{8 \sqrt{2} c_3 \gamma _1 \left(\alpha _3 \beta _3+2 \alpha _1 \beta _1 \gamma _1\right)}{\alpha _1^3 \alpha _3 \left(2 \alpha
_1 \beta _1+\alpha _3 \beta _3\right)},\\ \nonumber\\
\left\langle\frac{\partial^{4}V}{\partial\eta_{a}\partial\eta_{b}\partial  (\phi_{1}^{2})_{1}\partial(\phi_{2}^{1})_{2}}\right\rangle
&=&\left\langle\frac{\partial^{4}V}{\partial\eta_{a}\partial\eta_{b}\partial  (\phi_{1}^{2})_{2}\partial(\phi_{2}^{1})_{1}}\right\rangle\nonumber\\
&=&\frac{-8 \sqrt{2} c_3 \left(-1+\gamma _1\right) \Big(2 \alpha _1^2 \beta _1^2 \gamma _1+\alpha _3^2 \beta _3^2 \gamma _1+\alpha
_1 \alpha _3 \beta _1 \beta _3 \left(2+\gamma _1\right)\Big)}{\alpha _1 \alpha _3 \left(2 \alpha _1 \beta _1+\alpha _3 \beta _3\right){}^3},\nonumber\\\\
\left\langle\frac{\partial^{4}V}{\partial\eta_{a}\partial\eta_{c}\partial  (\phi_{1}^{2})_{1}\partial(\phi_{2}^{1})_{1}}\right\rangle &=&
\frac{16 c_3 \left(-1+\gamma _1\right) \gamma _1}{\alpha _1^2 \left(2 \alpha _1 \beta _1+\alpha _3 \beta _3\right)},\\ \nonumber\\
\left\langle\frac{\partial^{4}V}{\partial\eta_{a}\partial\eta_{c}\partial  (\phi_{1}^{2})_{1}\partial(\phi_{2}^{1})_{2}}\right\rangle &=&
\left\langle\frac{\partial^{4}V}{\partial\eta_{a}\partial\eta_{c}\partial  (\phi_{1}^{2})_{2}\partial(\phi_{2}^{1})_{1}}\right\rangle
=\frac{16 c_3 \left(-1+\gamma _1\right) \left(2 \alpha _1 \beta _1+\alpha _3 \beta _3 \gamma _1\right)}{\left(2 \alpha _1 \beta
_1+\alpha _3 \beta _3\right){}^3},\\ \nonumber\\
\left\langle\frac{\partial^{4}V}{\partial\eta_{a}\partial\eta_{d}\partial  (\phi_{1}^{2})_{1}\partial(\phi_{2}^{1})_{1}}\right\rangle &=&
\frac{8 \sqrt{2} c_3 \alpha _3 \left(-1+\gamma _1\right) \gamma _1}{\alpha _1^3 \left(2 \alpha _1 \beta _1+\alpha _3 \beta _3\right)},\\ \nonumber\\
\left\langle\frac{\partial^{4}V}{\partial\eta_{a}\partial\eta_{d}\partial  (\phi_{1}^{2})_{1}\partial(\phi_{2}^{1})_{2}}\right\rangle &=&
\left\langle\frac{\partial^{4}V}{\partial\eta_{a}\partial\eta_{d}\partial  (\phi_{1}^{2})_{2}\partial(\phi_{2}^{1})_{1}}\right\rangle
=\frac{8 \sqrt{2} c_3 \alpha _3 \left(-1+\gamma _1\right) \left(2 \alpha _1 \beta _1+\alpha _3 \beta _3 \gamma _1\right)}{\alpha
_1 \left(2 \alpha _1 \beta _1+\alpha _3 \beta _3\right){}^3},\\ \nonumber\\
\left\langle\frac{\partial^{4}V}{\partial\eta_{b}\partial\eta_{b}\partial  (\phi_{1}^{2})_{1}\partial(\phi_{2}^{1})_{2}}\right\rangle &=&\left\langle\frac{\partial^{4}V}{\partial\eta_{b}\partial\eta_{b}\partial  (\phi_{1}^{2})_{2}\partial(\phi_{2}^{1})_{1}}\right\rangle
=-\frac{16 c_3 \beta _3 \left(-1+\gamma _1\right) \left(\alpha _3 \beta _3+2 \alpha _1 \beta _1 \gamma _1\right)}{\alpha _3 \left(2
\alpha _1 \beta _1+\alpha _3 \beta _3\right){}^3},\\ \nonumber\\
\left\langle\frac{\partial^{4}V}{\partial\eta_{b}\partial\eta_{c}\partial  (\phi_{1}^{2})_{1}\partial(\phi_{2}^{1})_{2}}\right\rangle &=&
\left\langle\frac{\partial^{4}V}{\partial\eta_{b}\partial\eta_{b}\partial  (\phi_{1}^{2})_{2}\partial(\phi_{2}^{1})_{1}}\right\rangle\nonumber\\
&=&\frac{8 \sqrt{2} c_3 \alpha _1 \left(-1+\gamma _1\right) \Big(-\alpha _3 \beta _3 \left(-2+\gamma _1\right)+2 \alpha _1 \beta_1 \gamma _1\Big)}{\alpha _3 \left(2 \alpha _1 \beta _1+\alpha _3 \beta _3\right){}^3},\\ \nonumber\\
\left\langle\frac{\partial^{4}V}{\partial\eta_{b}\partial\eta_{d}\partial  (\phi_{1}^{2})_{1}\partial(\phi_{2}^{1})_{2}}\right\rangle &=&\left\langle\frac{\partial^{4}V}{\partial\eta_{b}\partial\eta_{d}\partial  (\phi_{1}^{2})_{2}\partial(\phi_{2}^{1})_{1}}\right\rangle\nonumber\\
&=&\frac{8 c_3 \left(-1+\gamma _1\right) \Big(-\alpha _3 \beta _3 \left(-2+\gamma _1\right)+2 \alpha _1 \beta _1 \gamma _1\Big)}{\left(2
\alpha _1 \beta _1+\alpha _3 \beta _3\right){}^3},\\ \nonumber\\
\left\langle\frac{\partial^{4}V}{\partial\eta_{c}\partial\eta_{c}\partial  (\phi_{1}^{2})_{1}\partial(\phi_{2}^{1})_{2}}\right\rangle &=&\left\langle\frac{\partial^{4}V}{\partial\eta_{c}\partial\eta_{c}\partial  (\phi_{1}^{2})_{2}\partial(\phi_{2}^{1})_{1}}\right\rangle
=\frac{32 c_3 \alpha _1^2 \left(-1+\gamma _1\right){}^2}{\left(2 \alpha _1 \beta _1+\alpha _3 \beta _3\right){}^3},\\ \nonumber\\
\left\langle\frac{\partial^{4}V}{\partial\eta_{c}\partial\eta_{d}\partial  (\phi_{1}^{2})_{1}\partial(\phi_{2}^{1})_{2}}\right\rangle &=&
\left\langle\frac{\partial^{4}V}{\partial\eta_{c}\partial\eta_{d}\partial  (\phi_{1}^{2})_{2}\partial(\phi_{2}^{1})_{1}}\right\rangle
=\frac{16 \sqrt{2} c_3 \alpha _1 \alpha _3 \left(-1+\gamma _1\right){}^2}{\left(2 \alpha _1 \beta _1+\alpha _3 \beta _3\right){}^3},\\ \nonumber\\
\left\langle\frac{\partial^{4}V}{\partial\eta_{d}\partial\eta_{d}\partial  (\phi_{1}^{2})_{1}\partial(\phi_{2}^{1})_{2}}\right\rangle &=&\left\langle\frac{\partial^{4}V}{\partial\eta_{d}\partial\eta_{d}\partial  (\phi_{1}^{2})_{2}\partial(\phi_{2}^{1})_{1}}\right\rangle
=\frac{16 c_3\alpha _3^2 \left(-1+\gamma _1\right){}^2}{\left(2 \alpha _1 \beta _1+\alpha _3 \beta _3\right){}^3}.
\end{eqnarray}

\section{Recovering current algebra limit}
As a consistency check, we  recover the  current algebra result for this scattering from GLSM.   To decouple the four-quarks the limit of $d_2, e_{3}^{a}\rightarrow 0$ and $\gamma_1 \rightarrow 1$ is imposed:

\begin{eqnarray}
m_{\pi}^2&=&-2 c_2 +4 c_4^a \alpha_1^2 \CL 0,\nonumber \\
m_{f_1}^2&=&m_{a}^2=-2 c_ 2 +12 c_4^a \alpha_1^2,\nonumber \\
m_{f_2}^2&=&-2 c_ 2 +12 c_4^a \alpha_3^2,\nonumber \\
m_{K}^2&=&-2 c_ 2 +4 c_4^a( \alpha_1^2-\alpha_1 \alpha_3 +\alpha_3^2)\CL 0,\nonumber \\
m_{\kappa}^2&=&-2 c_ 2 +4 c_4^a( \alpha_1^2+\alpha_1 \alpha_3 +\alpha_3^2),\nonumber \\
F_{\pi}&=& 2 \alpha _{1}\nonumber, \\
m_{\eta}^2+m_{\eta^{'}}^2&=&-4 c_2-\frac{16 c_3 }{\alpha_1^2}+4 c_4^a \alpha_1^2-\frac{8 c_3 }{\alpha_3^2}+4 c_4^a \alpha_3^2
\CL -{{16 c_3}\over \alpha_1^2} - {{8 c_3}\over \alpha_3^2}.
\end{eqnarray}

 As expected, in the chiral limit ($C.L.$) $m_\pi$ and $m_K$ vanish.
The five model parameters are then found to be

\begin{eqnarray}\label{lpa}
\alpha_1&=&\frac{F_{\pi}}{2},\nonumber \\
\alpha_3&=&F_{\pi}\sqrt{\frac{m_{f_1}^2+ 2 m_{f_2}^2-3 m_{\pi}^2 }{12(m_{f_{1}}^2-m_{\pi}^2)}} \CL
F_{\pi}\sqrt{\frac{m_{f_1}^2+ 2 m_{f_2}^2}{12 m_{f_{1}}^2}}
,\nonumber \\
c_2&=&\frac{1}{4}(m_{f_1}^2-3m_{\pi}^2) \CL \frac{1}{4} m_{f_1}^2,\nonumber \\
 c_3 &=& -\frac{ F_{\pi}^2(m_{f_1}^2+ 2 m_{f_2}^2-3 m_{\pi}^2)\Big(m_{f_{1}}^2-m_{f_2}^2+3 (m_{\eta}^2+m_{\eta ' }^2-2 m_{\pi}^2 )\Big) }{96(5 m_{f_{1}}^2 +4 m_{f_2}^2-9 m_{\pi}^2)}\nonumber\\
 &&
 \CL
 -\frac{ F_{\pi}^2(m_{f_1}^2+ 2 m_{f_2}^2)\Big(m_{f_{1}}^2-m_{f_2}^2+3 (-{{16 c_3}\over \alpha_1^2} - {{8 c_3}\over \alpha_3^2} )\Big) }{96(5 m_{f_{1}}^2 +4 m_{f_2}^2)},
  \nonumber \\
 c_4^a &=& \frac{m_{f_{1}}^2-m_{\pi}^2}{2 F_{\pi}^2}\CL  \frac{m_{f_{1}}^2}{2 F_{\pi}^2}.
\end{eqnarray}
Also note that:
\begin{eqnarray}
m_{\kappa}^2 &=& \frac{3}{2}(m_{f_1}^2-m_{\pi}^2)+\frac{1}{2}(-m_{f_1}^2+3 m_{\pi}^2) \CL m_{f_1}^2
\end{eqnarray}
In the limit that the scalar masses are very heavy, they decouple and we expect to recover the current algebra,  i.e. in the limit $m_{f_1}=m_{f_2}=m_{a}=m_{\kappa}=m\rightarrow \infty$, we have:
\begin{eqnarray}\label{lpb}
\lim _{m\rightarrow \infty} \alpha_3 &=& \frac{ F_{\pi}}{2},\nonumber \\
\lim _{m\rightarrow \infty} c_2&=&{m^2\over {4}},\nonumber \\
\lim _{m\rightarrow \infty} c_3 &=& \frac{-1}{96}F_{\pi}^2(m_{\eta}^2+m_{\eta ' }^2-2 m_{\pi}^2 ),\nonumber \\
\lim _{m\rightarrow \infty} c_4^a&=&{m^2\over {2 F_\pi^2}} .\nonumber \\
\label{par_decoupling}
 \end{eqnarray}
The scalar-pseudoscalar-pseudoscalar vertices (in the limit of $d_2,e_{3}^{a}\rightarrow 0$ and $\gamma_1 \rightarrow 1$) become:
\begin{eqnarray}\label{gf2pipi1}
\gamma^{(4)}_{\pi \eta}&=&12 \,c_4^a \cos^2 \phi+\frac{32\, c_3 \cos^2\phi}{\alpha_1^4}-\frac{8\sqrt{2}\sin(2 \phi)}{\alpha_1^3 \alpha_3},\nonumber \\
\gamma_{f_1 \pi \pi }&=&4 c_4^a \alpha_1,\nonumber \\
\gamma_{f_2 \pi \pi }&=&0,\nonumber \\
\gamma_{a_0 \pi \eta }&=&\frac{8\sqrt{2}\,c_3 \cos\phi}{\alpha_{1}^3}+4\sqrt{2}\,c_4^a \alpha_1 \cos\phi-\frac{8\,c_3\sin\phi}{\alpha_{1}^2\alpha_3},\nonumber \\
\gamma_{f_1 \eta \eta}&=&\frac{8 \sqrt{2}\, c_3\cos^2\phi}{\alpha_1^3}+2 \sqrt{2} \, c_4^a \alpha_1 \cos^2\phi-\frac{4\, c_3 \sin(2 \phi)}{\alpha_1^2 \alpha_3}\nonumber\\
\gamma_{f_2 \eta \eta}&=&\frac{8 \, c_3\sin^2\phi}{\alpha_3^3}+4 \, c_4^a \alpha_3 \sin^2\phi-\frac{4\sqrt{2}\, c_3 \sin(2 \phi)}{\alpha_1 \alpha_3^2},
\end{eqnarray}
where $\phi$ is the strange-non-strange mixing angle, $\cos\phi=(\cos \theta_p-\sqrt{2}\sin \theta_p)/\sqrt{6}$. Eq. (\ref{gf2pipi1}) together with (\ref{par_decoupling}) yield

\begin{eqnarray}\label{gf2pipi}
\gamma ^{(4)}_{\pi\eta} &=&\frac{1}{3 F^2_\pi}\bigg[4 \sqrt{2}\left(m^2_\eta+m^2_{\eta'} -2 m^2_\pi \right)\sin(2 \phi)  +2 \Big(9\,m^2-8(m^2_\eta+m^2_{\eta'})+7m^2_\pi\Big) \cos^2 \phi\bigg] \nonumber\\
&&\CL \frac{6 m^2}{ F^2_\pi}\cos^2 \phi,\nonumber \\
\gamma_{f_1 \pi \pi }&=&\frac{m^2-m_{\pi}^2}{ F_{\pi}^2} \CL \frac{m^2}{ F_{\pi}^2} ,\nonumber \\
\gamma_{f_2 \pi \pi }&=&0,\nonumber \\
\gamma _{a_0\pi \eta }&=&\frac{1}{3 F_\pi}\bigg[  \left(m^2_\eta+m^2_{\eta'} -2 m^2_\pi \right)\Big(-2 \sqrt{2} \cos\phi+2\sin \phi\Big) +3 \sqrt{2} \left(m^2-m^2_\pi \right) \cos \phi \bigg] \nonumber\\
&&\CL \frac{\sqrt{2}m^2}{ F_\pi}\cos \phi     ,\nonumber \\
\gamma_{f_1 \eta \eta}&=&\frac{1}{6 F_\pi}\bigg[2 \left(m^2_\eta+m^2_{\eta'} -2 m^2_\pi \right)\sin(2 \phi)+\Big(\sqrt{2}(3 m^2 -4(m^2_\eta+m^2_{\eta'})+5m^2_\pi\Big)\cos^2\phi\bigg]\nonumber\\
&&\CL \frac{\sqrt{2}m^2}{2 F_\pi}\cos^2\phi,\nonumber \\
\gamma_{f_2 \eta \eta}&=&\frac{1}{3 F_\pi}\bigg[\sqrt{2} \left(m^2_\eta+m^2_{\eta'} -2 m^2_\pi \right)\sin(2 \phi)+\Big(3 m^2 -2(m^2_\eta+m^2_{\eta'})+m^2_\pi\Big)\sin^2\phi\bigg] \nonumber \\
&&\CL \frac{m^2}{ F_\pi}\sin^2\phi,\nonumber \\
\end{eqnarray}

The dependence of the  four-point amplitude on scalar mass is
\begin{equation}
M_{4p}=\xi_0+\xi_1 m^2.
\end{equation}
The contribution of the isosinglet scalars is of the form
\begin{equation}
M_{f_i}= 2\sqrt{2}\,\gamma_{f_i\pi\pi}\gamma_{f_i \eta \eta}\times (\rm{propagator}),
\end{equation}
with
\begin{eqnarray}
2\sqrt{2}\,\gamma_{f_i\pi\pi}\gamma_{f_i \eta \eta}&=&\rho_0+\rho_1 m^2+\rho_2 m^4,\nonumber\\
\rm{propagator}&=&\frac{1}{m^2+x}\simeq \frac{1}{m^2}-\frac{x}{m^4}+ {\cal O} (\frac{1}{m^6}).
\end{eqnarray}
Thus
\begin{equation}
\lim _{m\rightarrow \infty} M_{f_i}=\rho_1 - x \rho_2 +\rho_2 m^2.
\end{equation}
Similarly for the $a_0$ contribution
\begin{equation}
M_{a_0}=\gamma^2_{a_0\pi\eta}\big[\frac{1}{m^2+y_1}+\frac{1}{m^2+y_2}\big],
\end{equation}
with
\begin{eqnarray}
\gamma_{a_0\pi\eta}^2=\delta_0 +\delta_1 m^2 +\delta_2 m^4, \nonumber\\
\frac{1}{m^2+y_i}
\simeq\frac{1}{m^2}-\frac{y_i}{m^4}+{\cal O} (\frac{1}{m^6}).
\end{eqnarray}
Thus
\begin{equation}
\lim _{m\rightarrow \infty} M_{a_0}=2\, \delta_1-\sum_i y_i\delta_2 +2\, \delta_2 m^2.
\end{equation}
Taking everything into account we expect:
\begin{equation}
\lim_{m\rightarrow\infty} M_{\rm{total}}=M_{\rm{C.A.}}
\end{equation}
which results in two sum rules:
\begin{eqnarray}
\xi_0+\rho_1-x \rho_2+2\,\delta_1-\sum _i y_i \delta _2&=&M_{\rm{C.A.}},\nonumber\\
\xi_1 +\rho_2 +2\,\delta_2 &=&0.
\end{eqnarray}
The second sum-rule is identically satisfied.  The first one is:
\begin{equation}
M_{\rm{C.A.}}=\frac{1}{3 F_\pi^2}\Bigg(2\, \left(2 m^2_{\eta}-4 m^2_{\eta'}+5 m^2_\pi\right)\cos^2  \phi +2 \sqrt{2}  \left(m^2_{\eta}+m^2_{\eta'}-2 m^2_\pi\right)\sin (2 \phi)\Bigg).
\end{equation}
Since in the decoupling limit
\begin{equation}
2m_{\pi}^{2} \rightarrow m_{\eta}^{2}+m_{\eta^{'}}^{2},
\end{equation}
then
\begin{equation}
M_{\rm{C.A.}}=\frac{2\,\left(2 m^2_{\eta}-4 m^2_{\eta'}+5 m^2_\pi\right)}{3 F_{\pi}^2}\cos^2 \phi \rightarrow
\frac{2\, m^2_\pi}{ F_{\pi}^2}\cos^2 \phi,
\label{con1}
\end{equation}
in agreement with the last term in Eq. (2.1) of \cite{pieta}.


\section{Pole expansion}
\begin{figure}[!htb]
	\centering
	\includegraphics[keepaspectratio=true,scale=0.67]{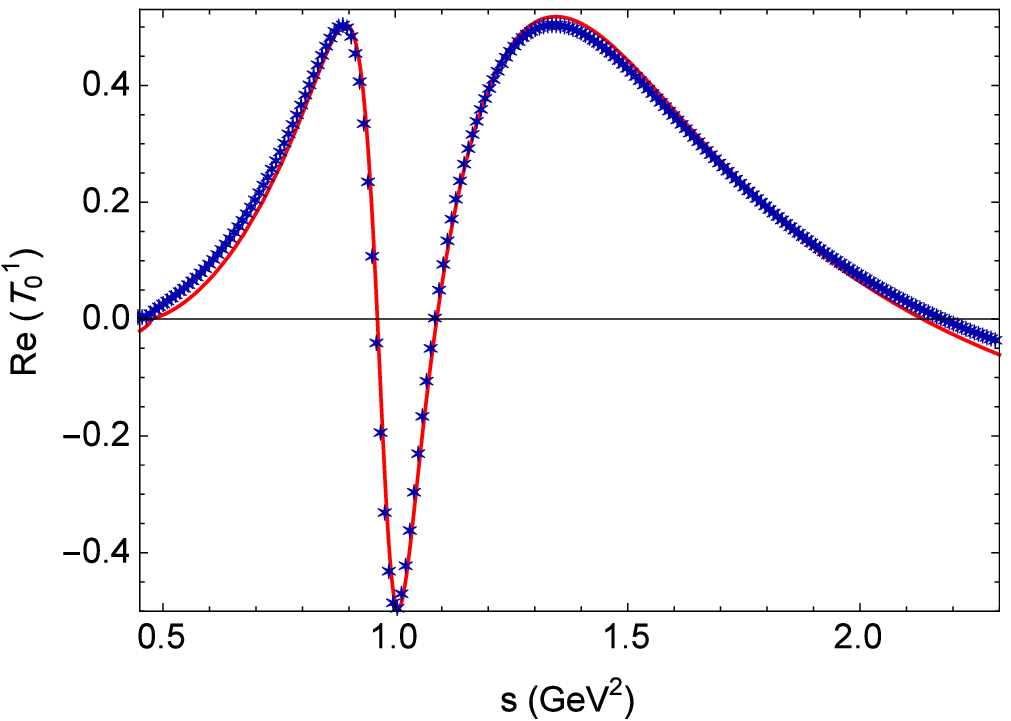}
	\hskip 0.1 cm
	\includegraphics[keepaspectratio=true,scale=0.67]{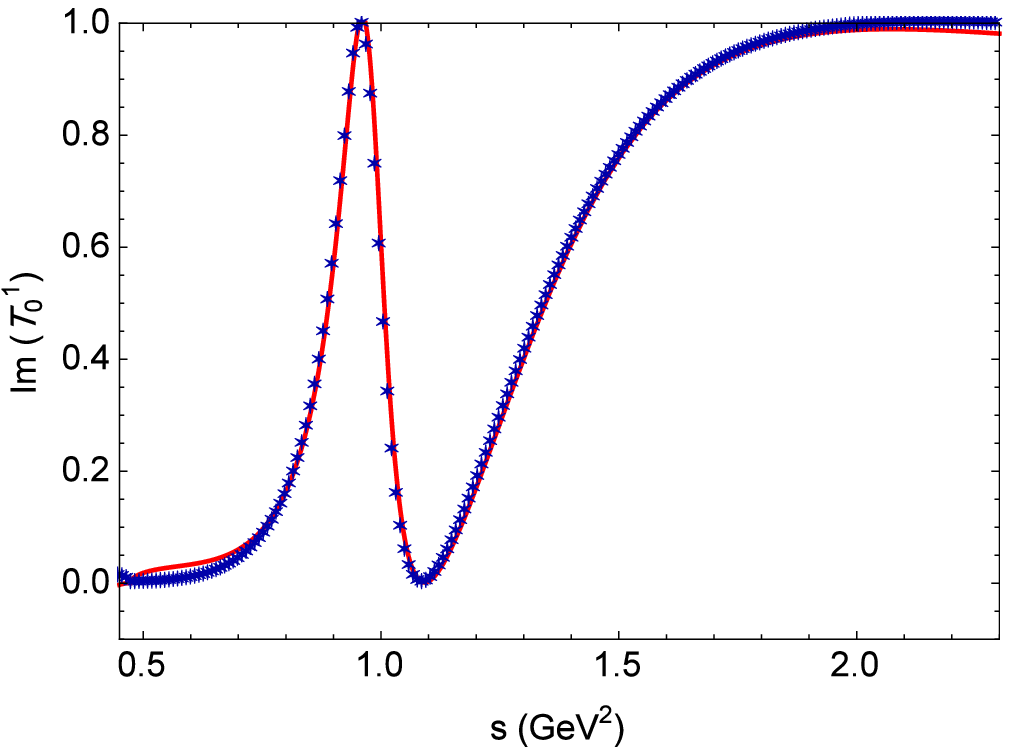}
	\caption{Close agreement of  the K-matrix unitarized $I=1$, $J=0$, $\pi \eta$ scattering amplitude with the expansion (\ref{pole_expan}) for $ A_3/A_1=30 $ and $ m[{\pi}(1300)]=1.3$ GeV.}
	\label{F_I12_compare}
\end{figure}

We highlight  an interesting property of the unitarization methodology applied in this work (the same observation was also made in the unitarization of $\pi\pi$ \cite{mixing_pipi} and $\pi K$ \cite{mixing_piK} scatterings).  Organizing the bare amplitude  in terms of the poles and a remaining background
\begin{eqnarray}
T_0^{1B}&=&T_{\alpha}+\sum_{j=1}^{n_{a}}{\frac{T_{\beta}^j}{m_{a_j}^2-s}} = \frac{\rho (s)}{2} \left[
T'_{\alpha} + \sum_{j=1}^{n_{a}}{\frac{{T'}_{\beta}^j}{m_{a_j}^2-s}}\right],
\end{eqnarray}
where
\begin{eqnarray}
T'_{\alpha}&=&-2\gamma_{\pi \eta}^{(4)}+\frac{1}{2q^2}\sum_{j=1}^{n_{a}} {\gamma_{a_j\pi \eta}^2}\ln{\left(\frac{(B_{\eta})_j+1}{(B_{\eta})_j-1}\right)}+\frac{\sqrt{2}}{q^2}\sum_{i=1}^{n_f} \gamma_{f_{i}\eta\eta}\gamma_{f_{i}\pi\pi}\ln{\left(1+\frac{4q^2}{m_{f_i}^2}\right)},  \\
{T'}_{\beta}^j&=&2 \gamma_{a_j\pi \eta}^2.
\label{T012_alpha_beta}
\end{eqnarray}
we can show that the K-matrix unitarized amplitude has a similar mathematical structure (in the complex plane) and can be written as a sum  of complex poles and a constant complex background
\begin{eqnarray}
T_0^{1}=\frac{T_0^{1B}}{1-i T_0^{1B}}
\approx
\widetilde{T}_{\alpha}+\sum_{j=1}^{n_{a}}{\frac{\widetilde{T}_{\beta}^j}{\widetilde{m}_{a_j}^2-s-i\widetilde{m}_{a_j}\widetilde{\Gamma}_{a_j}}}
\approx
\frac{\rho (s)}{2}
\left[
\widetilde{T}'_{\alpha}+\sum_{j=1}^{n_{a}}{\frac{\widetilde{T}_{\beta}^{\prime j}}{\widetilde{m}_{a_j}^2-s-i\widetilde{m}_{a_j}\widetilde{\Gamma}_{a_j}}}
\right]
\label{pole_expan}
\end{eqnarray}
which shows that the functional form of the K-matrix unitarized amplitde resembles the bare amplitude in which the bare masses are replaced by the physical poles in the complex $s$-plane.
The real part of the $I=1$, $J=0$ scattering amplitude obtained from the expansion (\ref{pole_expan}) is verified numerically in Fig. \ref{F_I12_compare}.

\begin{figure}[!htb]
	\begin{center}
		\epsfxsize = 3 cm
		\includegraphics[keepaspectratio=true,scale=0.67]{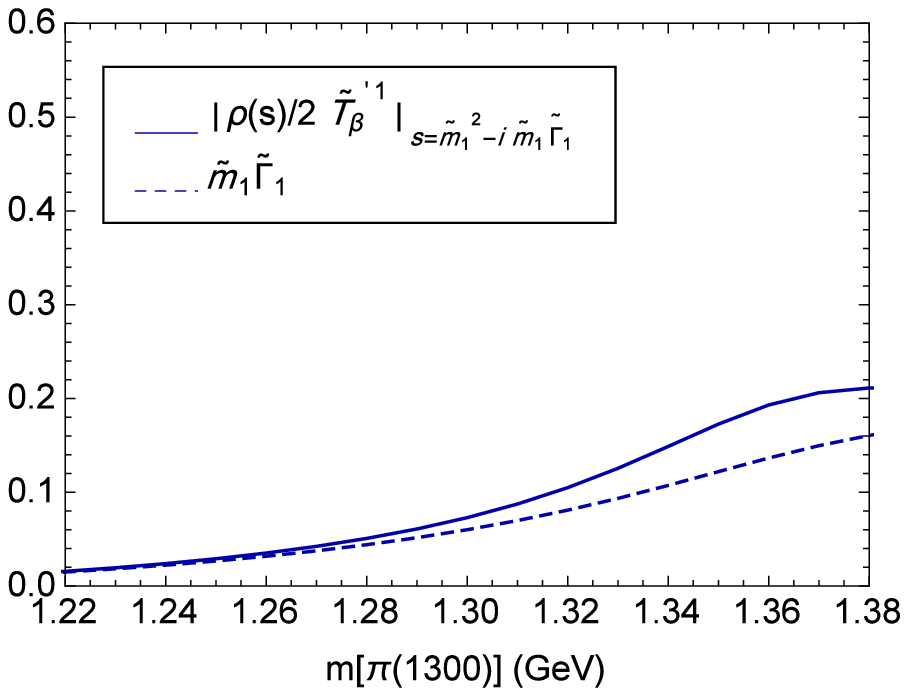}
		\hskip .1 cm
		\epsfxsize = 3 cm
		\includegraphics[keepaspectratio=true,scale=0.67]{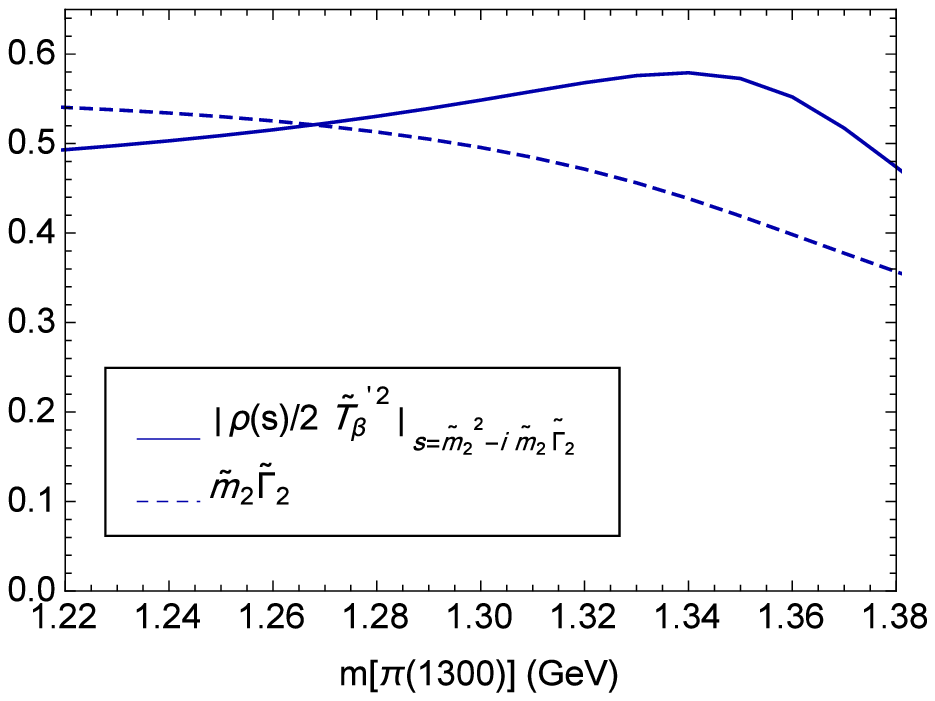}
		\caption{Comparison of $\widetilde{m}_{a_j}\widetilde{\Gamma}_{a_j}$ (dot-dashed line) with $ \left| \frac{\rho (s)}{2} \widetilde{T}^{'j}_{\beta}\right|_{s=\widetilde{m}_{a_j}^2 - i \widetilde{m}_{a_j} \widetilde{\Gamma}_{a_j}}$ (solid line).}
		\label{decay_width_compare_2}
	\end{center}
\end{figure}

Moreover,  the bare decay width  and mass of $a_0$'s satisfy
\begin{equation}
m_{a_j}\Gamma_{a_j} =
\left. \frac{\rho (s)}{2} {T}_{\beta}^{'j}\right|_{s=m_{a_j}^2}
\end{equation}
which is again in parallel with the physical decay width and mass of $a_0$'s
\begin{equation}
\widetilde{m}_{a_j} \widetilde{\Gamma}_{a_j} \approx \left|\widetilde{T}_{\beta}^j\right| \approx
\left| \frac{\rho (s)}{2} \widetilde{T}^{'j}_{\beta}\right|_{s=\widetilde{m}_{a_j}^2 - i \widetilde{m}_{a_j} \widetilde{\Gamma}_{a_j}}
\label{decay_width_compare_1}
\end{equation}
This relationship  is  numerically tested for  $A_3/A_1=30$ over the range of $m[\pi(1300)]$ in Fig. \ref{decay_width_compare_2}.

\begin{thebibliography}{10}

\bibitem{LsM_1960} M. Gell-Mann and M. Levy, Novo Cimento {\bf 16}, 705 (1960).

\bibitem{NLR_1969} W.A. Bardeen and B.W. Lee, Phys. Rev. {\bf 177}, 2389 (1969).

\bibitem{NJL_61}
Y. Nambu and G. Jona-Lasinio, Phys. Rev. {\bf 122}, 345 (1961).



\bibitem{lattice}
K.G. Wilson, Phys. Rev. D {\bf 10}, 2445 (1974); K.G. Wilson and J.B. Kogut, Phys. Rept. {\bf 12}, 75  (1974);
J. B. Kogut, D. K. Sinclair and L. Susskind, Nucl. Phys. B {\bf 114}, 199 (1976);
G. P. Lepage and P. B. Mackenzie, Phys. Rev. D {\bf 48}, 2250 (1993).



\bibitem{ChPT} S. Weinberg, Physica A {\bf 96},
327 (1979); J. Gasser and H. Leutwyler, Annalas
Phys. {\bf 158}, 142 (1984); Nucl. Phys. B {\bf
	250}, 465 (1985); H. Leutwyler, Annals of Physics 235, 165-203  (1994);
 I. Caprini, G. Colangelo and
H. Leutwyler, Phys. Rev. Lett. {\bf 96}, 132001 (2006).




\bibitem{ChUA1}
J.A. Oller and E. Oset, Nucl. Phys. A {\bf 620}, 438 (1997); Nucl. Phys. A {\bf 652}, 407 (1999).

\bibitem{ChUA2}
J.A. Oller and E. Oset,  Phys. Rev. D {\bf 60}, 074023 (1999).

\bibitem{ChUA3}
M. Jamin, J.A. Oller and A. Pich, Nucl. Phys. B {\bf 587}, 331 (2000).


\bibitem{ChUA4}
J.A. Oller,  Nucl.  Phys. A {\bf 727}, 353 (2003).

\bibitem{ChUA5}
J.A. Oller, Phys. Rev. D {\bf 71}, 054030 (2005).


\bibitem{ChUA6}
M. Albaladejo, J.A. Oller and L. Roca, Phys. Rev. D {\bf 82}, 094019 (2010).


\bibitem{ChUA7}
Z.H. Guo, J.A. Oller and J. Ruiz de Elvira, Phys. Rev. D {\bf 86}, 054006 (2012).


\bibitem{ChUA8}
M. Albaladejo and J.A. Oller, Phys. Rev. D {\bf 86}, 034003 (2012).

\bibitem{ChUA9}  J. Oller, Symmetry {\bf 12}, 1114 (2020).













\bibitem{IAM1} A. Dobado and J.R. Pelaez, Phys. Rev. D {\bf 56}, 3057 (1997).

\bibitem{IAM2} J.A. Oller, E. Oset and J. R. Pelaez,  Phys. Rev. Lett. {\bf 80}, 3452 (1998); 
Phys. Rev. D {\bf 59}, 074001 (1999);  Phys. Rev. D {\bf 60}, 099906
(1999);  Phys. Rev. D {\bf 75}, 099903 (2007).

\bibitem{IAM3} J.R. Pelaez, Phys. Rev. Lett. {\bf 92}, 102001 (2004).

\bibitem{IAM4}
A.~Salas-Bern\'ardez, F.~J.~Llanes-Estrada, J.~Escudero-Pedrosa and J.~A.~Oller,
SciPost Phys. \textbf{11}, no.2, 020 (2021).









\bibitem{Jaf} R.L. Jaffe, Phys. Rev. D {\bf 15}, 267
(1977).






\bibitem{sum_rules}
S. Narison, World Sci. Lect. Notes Phys. 26, 1 (1989);
P. Colangelo, A. Khodjamirian, arXiv:hep-ph/0010175 (2000).










\bibitem{Weinberg_13} S. Weinberg,   Phys. Rev. Lett. {\bf 110}, 261601 (2013).


\bibitem{Eli}
R.T. Kleiv, T.G. Steele, A. Zhang and I. Blokland, Phys. Rev. D {\bf 87}, 125018 (2013);
D. Harnett, R.T. Kleiv, K. Moats  and T.G. Steele,  Nucl. Phys. A {\bf 850}, 110 (2011);
J. Zhang, H.Y. Jin, Z.F. Zhang, T.G. Steele and D.H. Lu, Phys. Rev. D {\bf 79}, 114033 (2009);
Fang Shi, T.G. Steele, V. Elias, K.B. Sprague, Ying
Xue and A.H.  Fariborz, Nucl. Phys. A {\bf 671}, 416
(2000);
V. Elias, A.H. Fariborz, Fang Shi and
T.G. Steele, Nucl.  Phys.  A {\bf 633}, 279 (1998).

\bibitem{Lattice_scalars}
M. Wagner, et al., Acta Phys. Polon. Supp. {\bf 6} 847 (2013);
C. Alexandrou, et al,  JHEP {\bf 137}, 1304 (2013);
T. Kunihiro, S. Muroya, A. Nakamura, C. Nonaka, M. Sekiguchi, H. Wada, in proceedings of {\it International IUPAP Conference on Few-Body Problems in Physics (FB 19), Bonn, Germany, 31 Aug - 5 Sep 2009},  EPJ Web Conf.3:03010 (2010);
C. McNeile, in proceedings of {\it 11th Int. Conf. on Meson-Nucleon
Physics and
the Structure of the Nucleon}, 10-14 Sept. 2007, J\"ulich, Germany;
C. McNeile and C. Michael (UKQCD Collaboration), Phys. Rev. D {\bf 74},
014508 (2006);
N. Mathur et al, hep-ph/0607110;
A. Hart et al (UKQCD Collaboration),
Phys. Rev. D {\bf 74}, 114504 (2006);
H. Wada (SCALAR Collaboration), Nucl. Phys. Proc. Suppl. {\bf 129}, 432
(2004); T. Kunihiro et al (SCALAR Collaboration), Phys. Rev. D {\bf 70},
034504 (2003);
N. Ishii, H. Suganuma and H.
Matsufuru, Phys. Rev. D {\bf 66}, 014507 (2002);
Xi-Yan Fang, Ping Hui, Qi-Zhou Chen and D. Schutte,
Phys. Rev. D {\bf 65}, 114505 (2002);
M.G. Alford and R.L. Jaffe, Nucl. Phys. B {\bf
578}, 367 (2000);
C.J.
Morningstar and M. Peardon, Phys. Rev. D {\bf 60},
034509 (1999); J. Sexton, A. Vaccarino and D.
Weingarten, Phy. Rev. Lett. {\bf 75}, 4563 (1995);
G. Bali et al., Phys. Lett. B {\bf 309}, 378 (1993).



\bibitem{06_G}
 I. Eshraim, S. Janowski, F. Giacosa and  D.H. Rischke,
Phys. Rev. D {\bf 87}, 054036 (2013);
F. Giacosa, Phys. Rev. D {\bf 74}, 014028 (2006).



\bibitem{06_Pelaez}
J.R. Pelaez and A. Rodas Phys. Rev. D {\bf 93}, 074025 (2016); 
J.R. Pelaez, PoS CD12, 047 (2013);
R. Garcia-Martin, R. Kaminski, J.R. Pelaez, J. Ruiz de Elvira
Phys. Rev. Lett. {\bf 107}, 072001 (2011);
J.R. Pelaez, Phys. Rev. Lett. {\bf 97}, 242002 (2006).







\bibitem{generallag}
D. Parganlija, F. Giacosa and D. H. Rischke, Phys. Rev. D {\bf 82}, 054024 (2010);
S.Janowski, D. Parganlija, F. Giacosa and D. H. Rischke, Phys. Rev. D {\bf 84}, 054007 (2011).







\bibitem{08_tHooft} G. 't Hooft, G. Isidori, L.
Maiani, A.D. Polosa and V. Riquer,
arXiv: 0801.2288 [hep-ph].








\bibitem{06_MPPR}
L. Maiani, F. Piccinini, A.D. Polosa, V. Riquer, Eur. Phys. J. C {\bf 50},
609 (2007);  hep-ph/0604018.









\bibitem{05_N}
S. Narison, Phys. Rev. D {\bf 73}, 114024 (2006).



\bibitem{06_CCY} H.Y. Cheng, C.K. Chua and K.C. Yang, Phys. Rev. D {\bf
73}, 014017 (2006).

\bibitem{06_KKNHH} Yu. Kalashnikova, A. Kudryavtsev, A.V. Nefediev, J.
Haidenbauer and C. Hanhart, Phys. Rev. C {\bf 73}, 045203 (2006).

\bibitem{06_BCKR} E. van Beveren, J. Costa, F. Kleefeld and G. Rupp,
Phys. Rev. D {\bf 74}, 037501 (2006).

\bibitem{06_Aetal} M. Ablikim et al, Phys. Lett. B {\bf 633}, 681 (2006).

\bibitem{06_T}N.A. T\"ornqvist, hep-ph/0606041.








\bibitem{04_Ynd}
F.J. Yndurain, Phys. Lett. B {\bf 578}, 99
(2004); Phys. Lett. B {\bf
612}, 245 (2005).

\bibitem{05_TKM} T. Teshima, I. Kitamura and N. Morisita,  Nucl. Phys. A
{\bf 759}, 131 (2005).

\bibitem{05_GGF}
F. Giacosa, T. Gutsche, A. Faessler, Phys. Rev. C {\bf 71}, 025202
(2005).

\bibitem{05_VVFS}
J. Vijande, A. Valcarce, F. Fernandez, B. Silvestre-Brac,
Phys. Rev. D {\bf 72}, 034025 (2005).



\bibitem{05_BNNB} T.V. Brito, F.S. Navarra, M. Nielsen,
M.E. Bracco, Phys. Lett. B {\bf 608}, 69 (2005).

\bibitem{05_GGLF}
F. Giacosa, Th. Gutsche, V.E. Lyubovitskij, A. Faessler,
Phys. Lett. B {\bf 622}, 277 (2005).


\bibitem{JGR} S. Janowski, F. Giacosa, D.H. Rischke, Phys. Rev. D {\bf 90}, 114005 (2014).

\bibitem{15_BGB} W. Broniowski, F.  Giacosa and  V. Begun,  Phys. Rev. C {\bf 92}, 034905 (2015). 

\bibitem{16_WSG} T. Wolkanowski,  M. So?tysiak and  F. Giacosa, Nucl. Phys. B {\bf 909}, 418  (2016).

\bibitem{16_WGR} T. Wolkanowski, F. Giacosa  and D.H.  Rischke, Phys. Rev. D {\bf 93}, 014002  (2016). 







 

\bibitem{FJR_05} A.H. Fariborz, R. Jora and J. Schechter, Phys. Rev. D {\bf 72}, 034001 (2005);











\bibitem{04_KMNNSW} T. Kunihiro, S. Muroya, A. Nakamura,
C. Nonaka, M. Sekiguchi and H. Wada, Phys. Rev. D {\bf 70}, 034504 (2004).

\bibitem{04_UNOT} T. Umekawa, K. Naito, M. Oka and M. Takizawa, Phys. Rev.
C {\bf 70}, 055205 (2004).

\bibitem{04_MPPR} L. Maiani, F. Piccinini, A.D. Polosa and  V. Riquer,
Phys. Rev. Lett. {\bf 93}, 212002 (2004).

\bibitem{04_TKM} T. Teshima, I. Kitamura and N. Morisita, J. Phys. G.
{\bf 30}, 663 (2004).

\bibitem{04_NR}
M. Napsuciale and  S. Rodriguez, Phys. Rev. D {\bf 70}, 094043 (2004).




\bibitem{04_ACCGL}
A. Ananthanarayan, I. Caprini, G. Colangelo, J. Gasser and H. Leutwyler,
Phys. Lett. B {\bf 602}, 218 (2004).




\bibitem{E791} E.M. Aitala et al, Phys. Rev. Lett.
{\bf 89}, 121801 (2002).










\bibitem{01_CGL} G. Colangelo, J. Gasser and H.
Leutwyler, Nucl. Phys. B {\bf 603}, 125 (2001).







\bibitem{Ach} N.N. Achasov, Phys. Usp. {\bf 41}, 1149 (1999), hep-ph/9904223;
N.N. Achasov and G.N. Shestakov, hep-ph/9904254.


\bibitem{IH} K. Igi and K. Hikasa, Phys. Rev. {\bf
D59}, 034005 (1999).



\bibitem{Ishida_kappa} S.~Ishida, M.~Ishida, T.~Ishida,
K.~Takamatsu and T.~Tsuru, Prog. Theor. Phys. {\bf 98}, 621
(1997). See also M. Ishida and S. Ishida, Talk given at 7th
International Conference on Hadron Spectroscopy (Hadron
97), Upton, NY, 25-30 Aug. 1997, hep-ph/9712231.


\bibitem{AnSa}A.V. Anisovich and A.V. Sarantsev, Phys. Lett. {\bf B413},
137 (1997).

\bibitem{Ishida} S. Ishida, M.Y. Ishida, H. Takahashi, T. Ishida,
K. Takamatsu and T Tsuru, Prog. Theor. Phys. {\bf 95}, 745 (1996).


\bibitem{T}
{N.A.~T\"ornqvist} and M. Roos, Phys. Rev. Lett. {\bf 76}, 1575
(1996).

\bibitem{Sv} M. Svec, Phys. Rev. {\bf D53}, 2343 (1996).

\bibitem{Tor} N.A. T\"ornqvist, Z. Phys. C {\bf 68},
647 (1995).

\bibitem{JPHS}
{G.~Janssen, B.C.~Pearce, K.~Holinde and J.~Speth}, Phys. Rev. {\bf
D52},  2690  (1995).

\bibitem{DS} R. Delbourgo and M.D. Scadron, Mod. Phys. Lett. {\bf
A10}, 251 (1995).


\bibitem{AS94} N.N. Achasov and G.N. Shestakov, Phys. Rev. {\bf
D49}, 5779 (1994). A summary of the recent work of the Novosibirsk
group is given in N.N. Achasov, arXiv:0810.2601[hep-ph].

\bibitem{Kam94}{R. Kam\'inski}, {L. Le\'sniak} and J. P. Maillet,
Phys. Rev. {\bf D50}, 3145 (1994).


\bibitem{AS} N.N. Achasov and G.N. Shestakov,
Phys. Rev. D {\bf 49}, 5779 (1994).



\bibitem{MP} D. Morgan and M. Pennington, Phys.
Rev. {\bf D48}, 1185 (1993).

\bibitem{BMPV} A.A. Bolokhov, A.N. Manashov, M.V. Polyakov and
V.V. Vereshagin, Phys. Rev. {\bf D48}, 3090 (1993).




\bibitem{Isg} J. Weinstein and N. Isgur, Phys. Rev.
D {\bf 41}, 2236 (1990).


\bibitem{Aston} D. Aston et al., Nucl. Phys. B
{\bf 296}, 493 (1988).


\bibitem{vanBeveren} E. van Beveren, T.A. Rijken, K.
Metzger, C. Dullemond, G. Rupp and J.E. Ribeiro, Z. Phys.
C {\bf 30}, 615 (1986).

\bibitem{vanBev} E. van Beveren, T.A. Rijken,
K. Metzger, C. Dullemond, G. Rupp and J.E.
Ribeiro, Z. Phys. {\bf C30}, 615 (1986).




















\bibitem{3flavor}
A.H. Fariborz,  R. Jora, J. Schechter and M.N. Shahid,
Phys. Rev. D {\bf 83}, 034018 (2011).

\bibitem{Ds_decays}
A.H. Fariborz, R. Jora, J. Schechter and M.N. Shahid, Phys. Rev. D {\bf 84}, 094024 (2011);
arXiv:1108.3581 [hep-ph].







\bibitem{LsM_scatt_length}
D. Black, A.H. Fariborz, R. Jora, N.W. Park, J. Schechter and M.N. Shahid,  Mod. Phys. Lett. A {\bf 24}, 2285 (2009).


\bibitem{LsM_gauged}
A.H. Fariborz,  N.W. Park, J. Schechter and M.N. Shahid, Phys. Rev. D {\bf 80}, 113001 (2009).

\bibitem{bfjpss09}D. Black, A.H. Fariborz, R.
Jora, N.W. Park, J. Schechter and M.N. Shahid,
Mod. Phys. Lett. A {\bf 28}, 2285 (2009).





\bibitem{07_FJS2} A.H. Fariborz, R. Jora and J.
Schechter, Phys. Rev. D {\bf 77}, 034006
(2008).

\bibitem{07_FJS4} A.H. Fariborz, R. Jora and J.
Schechter, Phys. Rev. D {\bf 77}, 094004
(2008).




\bibitem{05_FJS} A.H. Fariborz, R. Jora and J.
Schechter, Phys. Rev. D {\bf 72}, 034001
(2005).

\bibitem{05_FJS2} A.H. Fariborz, R. Jora and J.
Schechter, Int. J. of Mod. Phys. A {\bf 20},
6178 (2005).





\bibitem{SU} J. Schechter and
Y. Ueda, Phys. Rev. D {\bf 4}, 733 (1971).






\bibitem{e3p} A. Abdel-Rehim, D. Black, A.H.
Fariborz and J. Schechter,  Phys. Rev. D
{\bf 67}, 054001 (2003).




\bibitem{Blk_rad} D. Black, M. Harada and J.
Shechter, Phys. Rev.  Lett. {\bf 88}, 181603 (2002).




\bibitem{99FS}A.H. Fariborz and J. Schechter, Phys.
Rev. D {\bf 60}, 034002 (1999).


\bibitem{BFSS2}D. Black, A.H. Fariborz, F. Sannino
and J. Schechter, Phys.  Rev. D {\bf 59}, 074026
(1999).

\bibitem{BFSS1}D. Black, A.H. Fariborz, F. Sannino
and J. Schechter, Phys.  Rev. D {\bf 58}, 054012
(1998).

\bibitem{HSS1}M. Harada, F. Sannino and J. Schechter, Phys. Rev. D {\bf
54}, 1991 (1996).


\bibitem{SS}F.~Sannino and J.~Schechter, Phys. Rev.  D {\bf 52},  96
(1995).





\bibitem{Far_IJMPA} A.H. Fariborz, Int. J. of
Mod. Phys. A {\bf 19}, 2095 (2004).


\bibitem{04_F} A.H. Fariborz, Int. J. of
Mod. Phys. A {\bf 19}, 5417 (2004).

\bibitem{06_F}
A.H. Fariborz,  Phys. Rev. D {\bf 74},
054030 (2006).









\bibitem{NR04}M. Napsuciale and S. Rodriguez, Phys. Rev. D {\bf 70},
094043 (2004).



\bibitem{mixing}T. Teshima, I. Kitamura and N. Morisita,
J. Phys. G
{\bf 28}, 1391 (2002); {\it ibid} {\bf 30}, 663 (2004).


\bibitem{close}F. Close and N.
Tornqvist, {\it ibid.}
{\bf 28}, R249 (2002).

\bibitem{Mec}D. Black, A.H. Fariborz and J.
Schechter, Phys. Rev. D {\bf 61}, 074001 (2000).

\bibitem{global}
A.H. Fariborz, R. Jora and J. Schechter,
Phys. Rev. D {\bf 79}, 074014 (2009).


\bibitem{mixing_pipi}
A.H. Fariborz, R. Jora, J. Schechter and M.N. Shahid, Phys. Rev. D {\bf 84}, 113004 (2011); arXiv:1106.4538 [hep-ph].


\bibitem{LsM_mmp_eta3p}
A.H. Fariborz, J. Schechter, S. Zarepour and S.M. Zebarjad,
Phys. Rev. D {\bf 90}, 033009 (2014).


\bibitem{pieta}D. Black, A.H. Fariborz and J.
Schechter, Phys. Rev.  D {\bf 61}, 074030 (2000).



\bibitem{LsM}D. Black, A.H. Fariborz, S. Moussa, S.
Nasri and J.  Schechter, Phys. Rev. D {\bf 64},
014031 (2001).

\bibitem{LsM_Maple}
A.H. Fariborz, Int. J. Mod. Phys. A {\bf 26}, 2327 (2011).

\bibitem{07_FJS1} A.H. Fariborz, R. Jora and J.
Schechter, Phys. Rev. D {\bf 76}, 014011
(2007).


\bibitem{07_FJS3} A.H. Fariborz, R. Jora and J.
Schechter, Phys. Rev. D {\bf 76}, 114001
(2007).





\bibitem{07_KZ}
E. Klempt and A. Zaitsev, Phys. Rept. {\bf 454},1 (2007).

\bibitem{Pelaez_Review}
J.R. Pelaez, Phys. Rept. {\bf 658}, 1 (2016).




\bibitem{pdg}
 P.A. Zyla et al. (Particle Data Group), Prog. Theor. Exp. Phys. \textbf{8},  083C01 (2020).


\bibitem{mixing_piK}
A. H. Fariborz, E. Pourjafarabadi, S. Zarepour and and S. M. Zebarjad, Phys. Rev. D {\bf 92}, 113002 (2015).








\bibitem{ChPT_bernard}
V. Bernard and N. Kaiser, Phys. Rev. D {\bf 44}, 3698 (1991).

\bibitem{ChPT_Novotny}
J. Novotny and M. Kolesar, arXiv:0212311 [hep-ph] (2003).

\bibitem{ChPT_kolesar}
M. Kolesar and J. Novotny, Eur. Phys. J. C {\bf 56}, 231 (2008).


\bibitem{belledata}
N. N. Achasov and G. N. Shestakov, Phys. Rev. D {\bf 81}, 094029 (2010).


\bibitem{Albaladejo_ph}
M. Albaladejo and B. Moussallam, Eur. Phys. J. C {\bf 75}, 488 (2015).


\bibitem{pieta_lattice}
J. J. Dudek, R. G. Edwards and D. J. Wilson, Phys. Rev. D {\bf 93}, 094506 (2016).

\bibitem{guo}
  Z. H.  Guo, L. Liu, U. G. Meißner, J. A. Oller and A. Rusetsky, Phys. Rev. D \textbf{95}, 054004 (2017).

\bibitem{Lu}
  J. Lu and B. Moussallam, Eur. Phys. J. C \textbf{80}, 436 (2020).

\bibitem{oller2}
J. A. Oller, E. Oset and J. R. Pelaez, Phys. Rev. D {\bf 59}, 074001 (1999).






\bibitem{kubis}
B. Kubis and S. P. Schneider, Eur. Phys. J. C {\bf 62}, 511 (2009).

\bibitem{bijnens}
J. Bijnens, G. Ecker and J. Gasser, arXiv:9411232 [hep-ph] (1994).

\bibitem{amoros}
G. Amoros, J. Bijnens and P. Talavera, Nucl. Phys. B {\bf 585}, 293 (2000); Erratum-ibid. B {\bf 598}, 665 (2001).


\bibitem{flatte}
 S.  Flatt{\'e}, Phys. Lett. \textbf{63B}, 224 (1976).



\bibitem{Fariborz:2021gtc}
A.~H.~Fariborz and M.~Lyukova,
Nucl. Phys. A \textbf{1015}, 122286 (2021).

\bibitem{Fariborz:2018tyi}
A.~H.~Fariborz, R.~Jora and M.~Lyukova,
Int. J. Mod. Phys. A \textbf{34}, 1950034 (2019).

\bibitem{Fariborz:2018unf}
A.~H.~Fariborz and R.~Jora,
Phys. Lett. B \textbf{790}, 410 (2019).

\bibitem{Fariborz:2018xxq}
A.~H.~Fariborz and R.~Jora,
Phys. Rev. D \textbf{98}, 094032 (2018).




\end{thebibliography}
\end{document}